\newif\ifarxiv
\arxivtrue 

\PassOptionsToPackage{table}{xcolor} 

\documentclass[12pt, hidelinks]{book}

\usepackage{hyperref}

\usepackage{algorithm}
\usepackage{algpseudocode}
\algrenewcommand\algorithmicrequire{\textbf{Input:}}
\algrenewcommand\algorithmicensure{\textbf{Output:}}
\algnewcommand\Not{\textbf{not} }

\usepackage{adjustbox}
\usepackage[headings]{fullpage}
\usepackage{xargs}
\usepackage{xspace}
\usepackage[doublespacing]{setspace}

\usepackage{microtype}
\usepackage[T1]{fontenc}

\usepackage{wrapfig}
\usepackage[font={sf}]{caption}
\usepackage[subrefformat=parens]{subcaption}

\usepackage{multirow}

\usepackage{xcolor}
\usepackage{graphicx}
\graphicspath{{_build/}{figures/}}

\usepackage{amsmath}
\usepackage{amssymb}

\newtheorem{theorem}{Theorem}[chapter]
\newtheorem{proof}{Proof}[chapter]

\usepackage[textsize=tiny]{todonotes}

\usepackage{libertine}
\usepackage[libertine]{newtxmath}
\usepackage[scaled=.85]{DejaVuSansMono}

\usepackage{listings}
\lstdefinelanguage{Rust}{
  sensitive,
  morecomment=[l]{//},
  morecomment=[s]{/*}{*/},
  moredelim=[s][{\itshape\color[rgb]{0,0,0.75}}]{\#[}{]},
  morestring=[b]{"},
  alsodigit={},
  alsoother={},
  alsoletter={!},
  otherkeywords={=>},
  morekeywords={break, continue, else, for, if, in, loop, match, return, while},
  morekeywords={as, const, let, move, mut, ref, static, unsafe},
  morekeywords={dyn, enum, fn, impl, Self, self, struct, trait, type, use, where},
  morekeywords={crate, extern, mod, pub, super},
  morekeywords={abstract, alignof, become, box, do, final, macro,
    offsetof, override, priv, proc, pure, sizeof, typeof, unsized, virtual, yield},
  morekeywords=[2]{Send},
  morekeywords=[3]{bool, char, f32, f64, i8, i16, i32, i64, isize, str, u8, u16, u32, u64, unit, usize, i128, u128},
}%

\def\mytitle{Generation of Compiler Backends\\ from Formal Models of Hardware}
\def\myauthor{Gus Henry Smith}
\def\year{2024}

\title{\mytitle}
\author{\myauthor}

\newcommand\thesisstmt{%
something something
}

\newcommand\Thesisstmt{\expandafter\MakeUppercase \thesisstmt}

\usepackage{mdframed}
\usepackage{glossaries}

\ifarxiv
\else

\usepackage[normalem]{ulem}

\fi

\makenoidxglossaries

\newglossaryentry{compiler}
{
    name=compiler,
    description={
A tool which converts between representations}
}

\newglossaryentry{target}
{
    name=target,
    description={
The platform which a \glstext{compiler} generates code for}
}

\newglossaryentry{compilerbackend}
{
    name=compiler backend,
    description={
The portion of a \glstext{compiler} that concerns the \glstext{target},
  e.g., target-specific optimizations and code generation}
}

\newglossaryentry{validation}
{
    name=validation,
    description={
The process of sanity-checking a program or hardware design using a
  limited, finite set
  of inputs, and judging the correctness 
  of the outputs.
Also referred to as \textit{testing.}
Validation should specifically be distinguished from
  \glstext{verification}.
Confusingly, 
  in the world of hardware design,
  validation is often referred to as
  verification,
  while verification (by our definition)
  is referred to as \textit{formal} verification}
}

\newglossaryentry{verification}
{
    name=verification,
    description={
The process of mathematically proving correctness
  about a program or hardware design.
While the proofs of correctness themselves
  can have limitations,
  verification is generally more thorough
  than \glstext{validation} or \textit{testing.}
In the world of hardware design,
  ``verification''
  generally refers to what we call validation or testing,
  while ``formal verification''
  refers to our verification}
}

\newglossaryentry{hardwaresynthesis}
{
    name={hardware synthesis},
    description={
Another term for hardware compilation.
The process of converting a hardware design
  captured in a high-level language
  to a low-level target implementation,
  for example,
  a \gls{netlist} which can be programmed onto an FPGA
  or a geometry file to be made into an ASIC}
}

\newglossaryentry{netlist}{
    name={netlist},
    description={
Much like \glstext{rtl} or \glstext{hls}, netlists are
  a way to capture a hardware specification.
Netlists are lower-level than RTL or HLS, however.
As their name implies,
  netlists are simply lists of ``nets'' (or wires),
  plus, importantly,
  the modules they are connected to}
}

\newglossaryentry{dsl}
{
    name={DSL},
    first={domain-specific language (DSL)},
    description={
A programming language designed for a narrow domain, generally for a specific purpose.
DSLs are generally smaller
  than general-purpose programming languages,
  with fewer complicated features
  like general control flow.
This makes them easier to reason about
  and optimize}
}

\newglossaryentry{mlkernel}
{
    name={machine learning kernel},
    description={
A core subroutine used within many machine learning workloads---for example,
  matrix multiplication.
Machine learning kernels are often highly optimized,
  often by building custom hardware to implement
  them efficiently}
}

\newglossaryentry{instruction-selection}
{
    name={instruction selection},
    description={
A core compiler algorithm
  in which higher-level,
  hardware-independent operations
  are lowered to
  hardware-specific instructions~\cite{blindell2016instruction}}
}

\newglossaryentry{technology-mapping}
{
    name={technology mapping},
    description={
A core compiler algorithm
  in \glstext{hardwaresynthesis} tools
  in which
  hardware-independent components of a design
  (e.g.~a Verilog multiply operator \texttt{*})
  are lowered to
  hardware-specific primitive instantiations
  (e.g.~a Xilinx DSP48E2 DSP)}
}

\newglossaryentry{hls}
{
    name={HLS},
    first={High-Level Synthesis (HLS)},
    description={
A set of tools~\cite{cong2011high}
  for compiling high-level implementations of algorithms
  (often written in C)
  into hardware designs.
HLS is often contrasted against
  \glstext{rtl},
  written in languages like Verilog or VHDL,
  which is a comparatively lower level
  of abstraction}
}

\newglossaryentry{rtl}
{
    name={RTL},
    first={Register Transfer Level (RTL)},
    description={
A specific level of abstraction for a hardware design specification,
  in which registers (hardware modules which hold state)
  are explicit in the design,
  but computation is still modeled at a high level.
This is in contrast to a higher level representation
  like \glstext{hls}, which does not include timing,
  or a lower, gate-level or netlist representation
  which may use specific gates or other hardware primitives}
}

\newglossaryentry{equality-saturation}
{
    name={equality saturation},
    description={
A term rewriting algorithm~\cite{tate2009equality,tate2011equality,willsey2021egg}
  which uses a specific data structure---%
  the \egr---%
  to capture a large space of equivalent programs}
}

\newglossaryentry{tensorization}
{
    name={tensorization},
    description={
An algorithm within machine learning compilers
  which uncovers places to invoke primitives
  operating over tensors---%
  often in the form of accelerator 
  invocations~\cite{tvmtensorization}.
Tensorization is analogous to vectorization
  in standard compilers,
  in which groups of instructions are translated
  into a single vector instruction
  to improve performance}
}

\newglossaryentry{accelerator}
{
    name={accelerator},
    description={
Custom hardware, specialized to a specific task
  or set of tasks.
In this dissertation,
  we are most frequently referring to
  accelerators for
  \glstext{mlkernel}s}
}

\newglossaryentry{program-synthesis}
{
    name={program synthesis},
    description={
A broad term capturing a set of 
  techniques for generating programs,
  often using automated reasoning tools~\cite{gulwani2017program}.
In this dissertation,
  we are generally referring to solver-aided synthesis,
  which formulates program generation as a constraint solving problem
  and applies solvers (e.g.~\glstext{smt} solvers).
Specifically, we are generally referring to
  \textit{sketch-guided} program synthesis,
  in which the final compiled program
  is captured as a sketch, with holes to be filled in by the solver~\cite{solar2008program}}
}

\newglossaryentry{primitive}
{
    name={primitive},
    description={
The atomic units of a language or representation.
In the context of this dissertation, we often mean
  either \glstext{accelerator} primitives
  or hardware primitives.
Accelerator primitives are operations
  implemented by accelerators which are, from the 
  view of software, black box operations
  which cannot be split into smaller sets of instructions.
They represent the smallest unit
  a compiler can target.
Hardware primitives are the modules
  provided for use by a hardware platform---%
  e.g. the gates and devices available for use
  on an FPGA.
A \glstext{netlist}
  is composed of primitive instantiations}
}

\newglossaryentry{fpga}
{
    name={FPGA},
    description={
Field Programmable Gate Arrays are
  a reprogrammable hardware platform
  allowing users to design hardware
  without fabricating an \glstext{asic}.
FPGAs are pre-fabricated chips
  composed of programmable \glstext{primitive}s.
The FPGA can be reprogrammed
  to configure and connect the various primitives,
  producing different hardware designs}
}

\newglossaryentry{asic}
{
    name={ASIC},
    first={Application-Specific Integrated Circuit (ASIC)},
    description={
A hardware chip produced for a specific purpose.
In contrast to FPGAs, which are reprogrammable,
  ASICs are static.
As a consequence, the design of ASICs is an intensive process
  involving significant \glstext{validation} and \glstext{verification}
  to ensure hardware correctness}
}

\newglossaryentry{automated-reasoning}
{
    name={automated reasoning},
    description={
A term to capture algorithms such as \glstext{smt} or \glstext{equality-saturation}.
In general, automated reasoning algorithms
  are able to solve
  complex constraint or optimization problems
  by applying mathematical and logical rules.
Automated reasoning algorithms
  are generally applicable to many tasks in computer science,
  as long as the can be captured
  in a way the algorithm can understand}
}

\newglossaryentry{smt}
{
    name={SMT},
    first={SMT (SAT modulo theories)},
    description={
A class of constraint problem.
SMT is simply SAT---Boolean satisfiability---%
  with extra ``theories'' added in,
  somewhat like libraries.
For example, the theory of fixed-width bitvectors,
  which includes operations over groups of bits.
SMT solvers are \glstext{automated-reasoning} algorithms
  which solve SMT problems}
}

\usepackage{soul}
\usepackage{ragged2e}
\usepackage{caption}
\usepackage{subcaption}
\usepackage[utf8]{inputenc}
\usepackage{tabularx}
\usepackage{listings}
\usepackage{amsmath}
\usepackage{xspace}
\usepackage{float}
\usepackage{enumitem}
\usepackage{microtype}
\microtypesetup{nopatch={footnote}}
\usepackage[T1]{fontenc}
\usepackage{graphicx}
\ifarxiv
\usepackage{grffile}
\fi
\usepackage{hyperref}
\usepackage[table]{xcolor}
\usepackage{wrapfig} 
\usepackage{cleveref}
\usepackage{bibentry}
\ifarxiv
\usepackage[frozencache,cachedir=.]{minted}
\else
\usepackage{minted}
\fi
\usepackage{tikz}
\usetikzlibrary{positioning, 
                quotes}

\newcommand{\g}{Glenside\xspace}

\newcommand{\accesspatternshape}[2]{$($$\left( #1 \right)$, $\left( #2 \right)$$)$}
\newcommand{\itc}{\texttt{im2col}\xspace}
\newcommand{\ctd}{\texttt{conv2d}\xspace}
\newcommand{\egr}{egraph\xspace}

\newcommand{\tcd}[1]{\texttt{#1}}
\newcommand{\mcd}[1]{\mathrm{\tcd{#1}}}

\newcommand\cse{Computer Science \& Engineering}

\newcommand{\TLA}{3LA\xspace}
\newcommand{\AppNum}{six\xspace}

\newcommand{\instrInText}[1]{\texttt{\small #1}}

\newcommand{\mapping}{IR-to-accelerator mapping\xspace}

\usepackage{amsfonts}
\usepackage{algorithm}
\usepackage{textcomp}
\usepackage{adjustbox}
\usepackage[htt]{hyphenat} 
\usepackage{semantic} 
\usepackage{pgfplots}
\usepackage{pgfplotstable}
\pgfplotsset{compat=1.17}
\usepackage{catchfile}
\usepackage{ragged2e}
\usepackage{subcaption} 
\usepackage[utf8]{inputenc}
\usepackage{tabularx}
\usepackage{listings}
\usepackage{booktabs}
\usepackage{color}
\usepackage{multirow}
\usepackage{syntax}
\usepackage[referable]{threeparttablex}
\usepackage[title]{appendix}
\usepackage{stmaryrd} 
\usepackage{siunitx} 
\usepackage{arydshln}
\usepackage{dashrule}
\usepackage{tikz-cd}
\usepackage{mathtools}
\usepackage{mathpartir}

\newcommand{\koika}{Kôika\xspace}

\newcommand{\lr}{Lakeroad\xspace}

\newcommand{\lrfn}{\text{$f_{\textsc{lr}}$}\xspace}   
\newcommand{\lrfnbmc}{\text{$f_{\textsc{lr}}^{*}$}\xspace}   

\newcommand{\SynProg}{\mathsf{Prog}\xspace}
\newcommand{\SynId}{\mathsf{Id}\xspace}
\newcommand{\SynBv}{\mathsf{BV}\xspace}
\newcommand{\SynVar}{\mathsf{Var}\xspace}
\newcommand{\SynNode}{\mathsf{Node}\xspace}

\newcommand{\Op}{\mathsf{OP}\xspace}
\newcommand{\OpBv}{\Op_{bv}}
\newcommand{\OpWire}{\Op_{w}}
\newcommand{\IRReg}{\lstinline[language=thelang]{Reg}\xspace}
\newcommand{\IRPrim}{\lstinline[language=thelang]{Prim}\xspace}

\newcommand{\Time}{\textsf{Time}\xspace}

\newcommand{\Sketch}{\textsc{Sketch}\xspace}

\definecolor{navy}{HTML}{0f1566}

\newcommand{\para}[1]{\paragraph{\hspace{-1em}#1}}

\newcommand{\norm}[1]{\left\lVert#1\right\rVert}

\newtheorem{property}{Property}
\definecolor{mygreen}{rgb}{0,0.6,0}
\lstdefinestyle{lispstyle}{
  backgroundcolor=\color{white},
  basicstyle=\ttfamily\footnotesize,
  breakatwhitespace=false,
  breaklines=true,
  captionpos=b,
  commentstyle=\color{mygreen},
  extendedchars=true,
  keepspaces=true,
  keywordstyle=\color{black},
  language=Lisp,
  morekeywords={*,...},
  numbers=none,
  numbersep=5pt,
  numberstyle=\tiny\color{mygray},
  rulecolor=\color{black},
  showspaces=false,
  showstringspaces=false,
  showtabs=false,
  stringstyle=\color{black},
  tabsize=2,
  title=\lstname
}
\lstdefinestyle{pystyle}{
  backgroundcolor=\color{white},
  basicstyle=\ttfamily\footnotesize,
  breakatwhitespace=false,
  breaklines=true,
  captionpos=b,
  commentstyle=\color{mygreen},
  extendedchars=true,
  keepspaces=true,
  keywordstyle=\color{blue},
  language=Python,
  morekeywords={*,...},
  numbers=none,
  numbersep=5pt,
  numberstyle=\tiny\color{mygray},
  rulecolor=\color{black},
  showspaces=false,
  showstringspaces=false,
  showtabs=false,
  stringstyle=\color{black},
  tabsize=2,
  title=\lstname
}

\newcommand\YAMLcolonstyle{\color{black}\mdseries}
\newcommand\YAMLkeystyle{\color{black}\bfseries}
\newcommand\YAMLvaluestyle{\color{black}\mdseries}

\makeatletter

\newcommand\language@yaml{yaml}

\expandafter\expandafter\expandafter\lstdefinelanguage
\expandafter{\language@yaml}
{
  keywords={true,false,null,y,n},
  keywordstyle=\color{darkgray}\bfseries,
  basicstyle=\color{black}\linespread{0.8}\ttfamily\footnotesize,
  comment=[l]{\#},
  morecomment=[s]{/*}{*/},
  commentstyle=\color{purple}\ttfamily,
  stringstyle=\YAMLvaluestyle\ttfamily,
  moredelim=[l][\color{orange}]{\&},
  moredelim=[l][\color{magenta}]{*},
  moredelim=**[il][\YAMLcolonstyle{:}\YAMLvaluestyle]{:},   
  morestring=[b]',
  morestring=[b]",
  literate =    {---}{{\ProcessThreeDashes}}3
                {>}{{\textcolor{red}\textgreater}}1     
                {|}{{\textcolor{red}\textbar}}1 
                {\ -\ }{{\mdseries\ -\ }}3,
}

\lst@AddToHook{EveryLine}{\ifx\lst@language\language@yaml\YAMLkeystyle\fi}
\makeatother

\newcommand\ProcessThreeDashes{\llap{\color{cyan}\mdseries-{-}-}}

\DeclareMathOperator{\defn}{\Coloneqq}

\lstdefinelanguage{thelang}{
  basicstyle=\ttfamily,
  keywordstyle=\color{black}\bfseries,
  morekeywords=[1]{let,in,:=,Reg,Prim,Op},
  morekeywords=[2]{},
  morekeywords=[3]{},
  alsoletter={:=},
  morestring=[b]",
  morecomment=[l]{\#},
  morecomment=[s]{(*}{*)},
  moredelim=**[is][\color{white}]{(&}{&)},
}

\newcommand{\seq}[1]{\langle#1\rangle}

\newcommand{\UberLang}{\ensuremath{\altmathcal{L}_\textsc{lr}}\xspace}
\newcommand{\SpecLang}{\ensuremath{\altmathcal{L}_\textsc{beh}}\xspace}
\newcommand{\ImplLang}{\ensuremath{\altmathcal{L}_\textsc{struct}}\xspace}
\newcommand{\SketchLang}{\ensuremath{\altmathcal{L}_\textsc{sketch}}\xspace}

\newcommand{\tighten}{\looseness=-1}

\DeclareMathAlphabet{\altmathcal}{OMS}{cmsy}{m}{n}

\newtheorem{lemma}[theorem]{Lemma}

\newcommand{\thesisubiquitylabel}{\textbf{\textsc{Ubiquity}}}
\crefformat{thesis:ubiquity}{#2\thesisubiquitylabel#3}

\newcommand{\thesisoptimizationslabel}{\textbf{\textsc{Optimizations}}}
\crefformat{thesis:optimizations}{#2\thesisoptimizationslabel#3}

\newcommand{\thesisdevtimelabel}{\textbf{\textsc{Devtime}}}
\crefformat{thesis:devtime}{#2\thesisdevtimelabel#3}

\newcommand{\thesiscorrectnesslabel}{\textbf{\textsc{Correctness}}}
\crefformat{thesis:correctness}{#2\thesiscorrectnesslabel#3}

\newcommand{\thesisalgorithmslabel}{\textbf{\textsc{Algorithms}}}
\crefformat{thesis:algorithms}{#2\thesisalgorithmslabel#3}
\newcommand{\thesismodelslabel}{\textbf{\textsc{Models}}}
\crefformat{thesis:models}{#2\thesismodelslabel#3}
\newcommand{\mythesis}{
Automatically generating
  compiler backends
  from explicit formal models
  of the hardware they target
  enables optimizations,
  improves correctness,
  and reduces development time.

}

\newlist{inlinelist}{enumerate*}{1}
\setlist*[inlinelist,1]{%
  label=(\arabic*)
}

\begin{document}












\pagestyle{empty}

\begin{center}
  {\huge \mytitle}
  \vfill

  {\Large \myauthor}
  \vfill

  \begin{spacing}{1}
    A dissertation \\
    submitted in partial fulfillment of the \\
    requirements for the degree of
  \end{spacing}
  \vfill

  Doctor of Philosophy
  \vfill

  University of Washington \\
  \year
  \vfill

  Reading Commitee: \\
  Zachary Tatlock, Chair \\
  Luis Ceze \\
  Michael Taylor
  \vfill

  \begin{spacing}{1}
    Program Authorized to Offer Degree: \\
    \cse
  \end{spacing}
  \clearpage

  \textcopyright{} Copyright \year\\
  \myauthor
  \clearpage
\end{center}

\pagestyle{plain}
\setcounter{page}{1}
\pagenumbering{roman}

\begin{center}
  University of Washington \\[1em]
  \textbf{Abstract}        \\[1em]
  Generation of Compiler Backends from Formal Models of Hardware
                   \\[1em]
  \myauthor                \\[1em]

  Chair of the Supervisory Committee: \\[-0.5em]
  Zachary Tatlock \\[-0.5em]
  Computer Science \& Engineering
  \\[2em]
\end{center}
Compilers convert
  between representations---%
  usually, from 
  higher-level, human writable code
  to lower-level,
  machine-readable code.
A compiler backend is
  the portion of the 
  compiler containing
  optimizations
  and code generation routines
  for a specific hardware target.
%
In this dissertation,
  I advocate for a specific way of
  building compiler backends:
  namely, by automatically generating them
  from explicit, formal models of hardware
  using automated reasoning algorithms.
I describe how automatically generating compilers
  from formal models of hardware
  leads to increased optimization ability,
  stronger correctness guarantees,
  and reduced development time
  for compiler backends.
As evidence, I present two
  case studies:
  first, \g,
  which uses
  equality saturation
  to increase the \TLA compiler's
  ability to offload operations
  to machine learning accelerators,
  and second,
  \lr,
  a technology mapper for FPGAs
  which uses program synthesis
  and semantics extracted from Verilog
  to map hardware designs
  to complex, programmable hardware primitives.

\setcounter{tocdepth}{1}
\begin{spacing}{1.5}
  \tableofcontents
\end{spacing}

\pagestyle{headings}
\setcounter{page}{1}
\pagenumbering{arabic}
\renewcommand{\chaptermark}[1]{\markboth{\sc{\chaptername\ \thechapter.\ #1}}{}}
\renewcommand{\sectionmark}[1]{\markright{\sc{\thesection.\ #1}}{}}

\chapter*{Acknowledgments}




First, allow me to thank
  everyone in my life 
  \textit{not} mentioned in the following paragraphs.
My PhD was far more than just a collection
  of projects;
  for the past six years,
  it was my life.
Anyone who has contributed to my life
  has contributed to
  this dissertation.

Thanks to my early research mentors.
To Vijay Narayanan and Jack Sampson at Penn State.
Without their early mentorship,
  I would have never considered
  the PhD a viable option.
To my initial research mentors in SAMPL:
  Thierry Moreau, Tianqi Chen, Jared Roesch, and Luis Vega.
Thank you for bringing me on
  to TVM
  and introducing me to compilers research.
To my advisor Luis Ceze,
  who hired me on at UW
  and changed
  the course of my entire life.
I could never thank Luis enough
  for the impact that single decision
  had on me.

Thanks to everyone I've worked with in industry.
To Aart Bik, Mangpo Phothilimthana,
  and Penporn Koanantakool
  at Google
  for their mentorship throughout my 
  extended 2021 internship.
To Aart and Penporn
  for bringing me onto the MLIR team,
  and to Mangpo for her patient tutelage
  in applying machine learning to compilers.
To Noah Evans and everyone at Sandia,
  for bringing me on
  and being interested in my research.
To Nina, Jannis, Claire, and the entire YosysHQ team
  for their support.
To Jin Yang and Jeremy Casas at Intel.

Thanks to every research team
  I've worked with.
To the Glenside team: Andrew Liu, Steven Lyubomirsky, Scott Davidson, Joseph McMahan, Michael Taylor, Zach Tatlock, and Luis Ceze.
To the 3LA team: Bo-Yuan Huang, Steven, Yi Li, Mike He, Thierry Tambe, Akash Gaonkar, Vishal Canumalla, Andrew Cheung, Gu-Yeon Wei, Aarti Gupta, Zach, and Sharad Malik.
To the Lakeroad team: Ben Kushigian, Vishal, Andrew, Steven, Sorawee Porncharoenwase, Ren\'e Just, Gilbert Bernstein, and Zach.
To the Churchroad team:
  Dan Petrisko,
  Colin Knizek,
  Chandra Nandi, Zach,
  Jonathan Balkind, Zach Sisco and Thanawat Techaumnuaiwit.

Thanks to those who volunteered to be on my committee---%
  to Mike Taylor,  Scott Hauck, 
  and Gilbert Bernstein.
Thanks especially to Gilbert,
  who invested time and effort
  into making me a sharper, more precise
  PL researcher
  in my last year at UW.
Gilbert was brave enough to
  attempt to teach me about coinduction---%
  and persistent enough to (mostly) succeed.
I will always appreciate his willingness to
  take the time and effort
  required
  to explain difficult concepts.

Thanks to the MISL lab
  for letting me hang around
  all these years,
  despite my utter lack of any formal
  training in biology.
Thanks to Chris Thachuk
  and Lancelot Wathieu
  for taking the
  Fridge Compiler
  from half-working course project
  to published paper.
Thanks to Jeff Nivala,
  Gwen Roote, Carina Imburgia,
  and the whole
  FPCA team
  for letting me tag along
  on the project.
A more down-to-earth lab I may never find.

To the entire Cornell Capra lab.
To Rachit Nigam, who
  served as my model of 
  success
  in programming languages and hardware design research.
To Adrian Sampson,
  who I can always look to for advice.
To Priya Srikumar,
  whose brief presence in PLSE
  in the summer of 2023
  was a blessing.
Priya's passing was felt
  throughout the Programming Languages
  research world,
  into their communities in Ithaca and Seattle,
  and beyond.
I will always remember Priya sharing
  a Tom Wayman poem with me,
  titled
  ``Did I Miss Anything?'',
  an excerpt of which
  will always remind me of them:
\begin{quote}
Nothing. When you are not present\\
how could something significant occur?
\end{quote}

\clearpage
To the PLSE lab.
It is harder to imagine a
  better lab
  exists anywhere on the planet.
The energy of the lab
  can be felt
  even before stepping through the door.
Whether it's laughter,
  or spirited research discussion,
  or someone playing one of the lab's many instruments,
  or even just the peaceful silence
  of face-in-laptop
  research progress,
  the energy of the lab is one of a kind.
Without realizing it,
  the lab has become my family;
  leaving it
  has been one of the most challenging things
  I've ever done.
I can only hope to start something
  half
  as vibrant, half as connected,
  half as supportive,
  wherever I end up.

Thanks to my mentee Hannah Leung,
  whose unyielding positivity
  and bravery
  in the face of challenges
  the likes of which I'll never 
  myself experience
  was a constant source of inspiration.
Though she may not know it,
  I've learned more from her
  than she has from me.

To my mentees
  Vishal Canumalla
  and Andrew Cheung.
Thanks for trusting me
  to mentor you in research and in life,
  and thanks for your friendship
  along the way.
I am so proud of you guys,
  and cannot wait to see what you do next.
Furthermore, thanks to Vishal's family,
  the Canumallas,
  who brought me in.
Thank you Anu and Sridhar
  for making me feel at home
  and making sure I'm well fed.

Thanks to Race Condition Running
  for the miles and miles of fun
  over the years.
Through my best and worst times
  in the PhD and in my personal life,
  the Saturday run and brunch
  were there for me
  as a consistent source of companionship
  and connection.
Thanks to Zach, Nick Walker, Ellis,
  Chandra, Max,
  and everyone else who
  kept the club going.
Thanks to everyone else who kept me active
  during my PhD---%
  thanks to Eric Zeng and Chien-Yu Lin
  for getting me out in the backcountry
  when the snow was good
  and to Gwen Roote for keeping me humble
  on the BJJ mats.
  
Thanks to the lab jam crew---%
  Ben, Carina, Rohan, and everyone else---%
  for providing a safe space
  to be a musical beginner.

To Ben Kushigian---a wonderful coauthor and friend.
  
To Nick and Danielle.
To Holly and Peter.
To Jared.

To Anjali, for all of the time you've given me.

To \texttt{\#binary-blobs}, \texttt{\#spicelords}, and associated friends:
  Pratyush Patel, Chien-Yu Lin,
  Jacob Van Geffen,
  Dan Petrisko, 
  Luzdary Ruelas,
  Ellis Michael,
  Max Ruttenberg,
  and Katie Lim.
Thanks to the countless hours
  of Twilight Imperium.

To Max Willsey and Sami Davies.
Without Max's urging,
  I may have never joined UW;
  even after I joined UW, without Max's urging,
  I may have never joined PLSE.
Max's quiet, confident intelligence
  served as a model for the entire PLSE lab.
I was lucky to witness
  a golden age of PLSE research,
  with Max's \texttt{egg} at the forefront.
Max showed me that a new project,
  an empty buffer,
  a clean slate,
  was nothing to be scared of
  if you face it with confidence and excitement.
And to Sami,
  whose intimidating intelligence
  is softened by her infectious smile
  and sincere desire for connection.

To Steven Lyubomirsky---%
  \textit{Lex Lyubomiricus} will forever guide my research.
Thank you for your constant mentorship
  at every stage of my research career.
Steven helped me present my first Programming Languages
  Reading Group paper in 2018---%
  ``Build Systems \`a la Carte'' by Mokhov et al.---%
  before I had any
  formal Programming Languages research
  under my belt.
Six years later, in 2024,
  Steven helped me publish the final
  first-author publication of my PhD,
  the Lakeroad paper.
  
To Chandra Nandi---%
  I could not have asked for a better
  research mentor and friend.
Chandra helped
  in every single one of my paper pushes
  in grad school,
  simply out of the love
  of camaraderie.
She taught me how
  to feel the unabashed joy
  that comes 
  from working hard
  on cool problems
  with your closest friends.

To Jeff's sons,
  Christian, Nick, and Aaron.
You all are so much like your father.
Thanks for 
  bucking the Seattle Freeze
  and making me feel at home here
  in my first 
  few years.
Here's to 
  many, many more rounds
  of pitch and putt.

To Eric, Rick, Zach,
  and the other men in the Montlake Men's Group.
Your support
  and mentorship
  helped me navigate the most challenging chapters
  of my life to date.
  
To my friends
  from the previous chapters of my life
  that I've been lucky enough
  to bring into this one:
  Spencer Norris, Imaz Athar,
  Anthony Marucci, Tim Lagnese, Jacob Brunette,
  and Drew Abbott.
Thank you to my oldest 
  and best friends,
  Spencer and Imaz.
There is something
  bittersweet about how,
  more than almost anything else,
  it's simply \textit{time} 
  that is the key ingredient
  for a deep relationship---%
  I'm so lucky to have invested
  my time in you guys.
You two are
  and will always be
  my brothers.
It has been an honor and a joy
  to grow with you
  through life's changes,
  to have you guys help me
  through my lowest lows
  and celebrate with me during my highest highs.
HSNE, boys.
Thanks to
  Anthony, Tim, and Jacob.
I could not have picked a better crew
  of friends to move to Seattle with,
  to spend my 20s with.
May our caps always fly true.
To Drew and Em---%
  we've rarely lived in the same city,
  but you two are family to me.
  
Boundless love and thanks to my older brother Simon.
In all my 29 years,
  he has never
  been far from my mind.

Thank you to Michaela,
  whose five years of love and loyalty transformed me
  into the man I am.
Her support carried me through the single
  continuous paper push
  that was the two years
  of my PhD.
Her patience with the all-consuming nature
  of my PhD
  was more than I could have asked for.
Her genuine care
  for everyone around her
  showed me that there is nothing more to life
  than the people in it.
There is no one more deserving
  of happiness.
Furthermore, thanks to the entire Brock and Lowe family,
  who brought me in
  without question,
  who made me a part of their family.
To Lisa Lowe, whose unhesitating love
  from the moment we met
  taught me the meaning of
  ``unconditional''.
To Ray Brock,
  who was
  a role model for men
  in a world where they are severely lacking.
I am immensely lucky to have known him.
It is through the Lowes
  that I came to understand
  the meaning of family---%
  a gift I could never
  repay.

Finally, to my advisor, Zach Tatlock.
Zach is the kind of researcher
  all should strive to be.
Intelligent, hardworking,
  and selfless to a fault.
Zach brought me on
  when he didn't need to.
He has a steady supply
  of top-tier PL students.
Bringing me on, as
  a student with zero formal programming languages
  research experience,
  was in many ways a questionable decision.
But, like Luis's decision to admit me,
  that single decision altered the course
  of my entire life.
Everything I know of research
  I learned first from Zach.
But Zach is far more than just a good researcher.
Zach is, in fact,
  the only thing I really strive to be---%
  a good man.
I've had very few role models in life,
  and Zach is first among them.
Thank you, Zach---%
  none of this could have happened
  without you.
\printnoidxglossaries
\chapter{Introduction}

A \gls{compiler}
  is a tool which
  converts from one representation
  to another---%
  usually, from a
  higher-level,
  human-writeable
  representation
  to a lower-level,
  machine-readable representation.
A classic example
  is the clang compiler
  from the LLVM suite~\cite{lattner2004llvm},
  which
  compiles programs written in 
  processor-independent C code
  into processor-specific
  machine code,
  which the hardware understands how
  to execute:
\begin{figure}[!h]
\centering
\begin{tikzpicture}
  \draw[inner sep = 5pt] (0,0) node [draw, thin] (c) \bgroup%
\begin{minipage}{10em}
\begin{minted}[fontsize=\small,baselinestretch=1]{c}
int square(int num) {
    return num * num;
}
\end{minted}
\end{minipage}
  \egroup;
  
\draw[inner sep = 5pt] (8.75,0) node [draw, thin] (asm) \bgroup%
\begin{minipage}{7em}
\begin{minted}[fontsize=\small,baselinestretch=1]{asm}
square:
 imul edi, edi
 mov eax, edi
 ret
\end{minted}
\end{minipage}
  \egroup;

\draw[inner sep = 5pt] (4.5,0) node [draw, thin] (compiler) {clang};

  \draw[->, very thick] (c) edge (compiler) (compiler) edge (asm);
\end{tikzpicture}
\caption{clang compiling high-level C code to target-specific x86 assembly.}
\label{fig:clang-c-to-x86}
\end{figure}
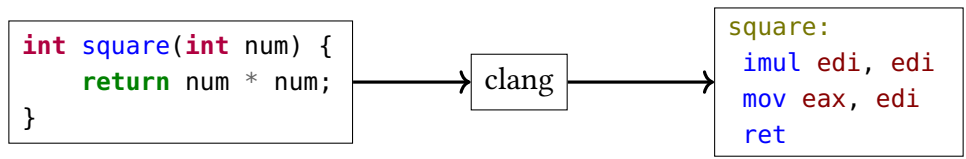

\noindent
In this case,
  clang is generating code 
  for an x86 processor;
  we refer to 
  the platform
  which the compiler is compiling for
  as the \gls{target}.
Though compilers perform many target-agnostic
  transformations and optimizations
  (modifications of the high-level code
    which are useful regardless of the target),
  a compiler's fundamental purpose
  is to produce a program
  in the target's language.
The portion of the compiler
  which handles target-specific optimizations
  and code generation is called the
  \gls{compilerbackend}---%
  for example,
  it is clang's x86 backend which is responsible for
  performing x86-specific optimizations
  and eventually producing
  x86 assembly code.
Compiler backends will be the focus
  of this dissertation;
  we will largely ignore other compiler components
  (i.e.~frontends and target-independent optimizations).

Throughout this intro,
  we will use three different compilers
  as our running examples.
As we have already seen, we will
  consider the
  general-purpose C compiler
  clang
  as our ``gold standard'' example
  of a commonly-known
  and understood compiler.
However, we will also consider
  two compilers from more specialized
  domains,
  which will be key to
  \cref{part:glenside-and-3la}
  and
  \cref{part:lakeroad}
  of this dissertation,
  respectively.
First,
  the TVM compiler~\cite{chen2018tvm},
  which will be a main focus
  of \cref{part:glenside-and-3la},
  is a compiler for deep learning
  programs
  which compiles code written in a
  high-level
  Python
  \gls{dsl}
  into optimized code
  for CPUs, GPUs,
  and custom \glspl{accelerator}.
Second,
  the open-source \gls{hardwaresynthesis}
  tool Yosys
  is a compiler for hardware designs
  supporting hardware targets such as
  \glspl{fpga}
  and will be a focus of
  \cref{part:lakeroad}.

Compiler backends are composed
  of multiple stages,
  and each stage is implemented with
  one or a number of core
  \textbf{algorithms.}
For example, a key stage in clang's x86 backend
  when compiling our example in
  \cref{fig:clang-c-to-x86}
  is
  \gls{instruction-selection},
  in which clang decides how to implement
  each operation in the C program
  using actual instructions
  provided by the processor~\cite{llvminstructionselection}.
It is in this stage
  where clang decides to
  implement
  C's \texttt{*} operator
  using x86's
  \texttt{imul} instruction.

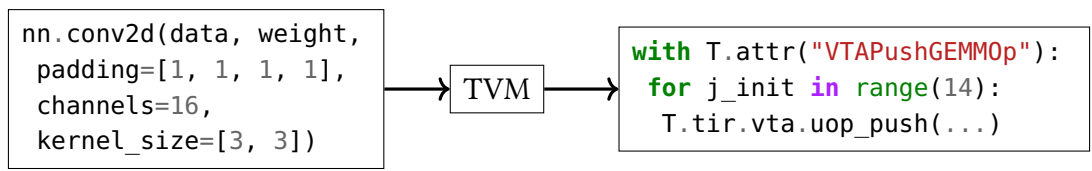
\begin{figure}
\centering
\begin{tikzpicture}
  \draw[inner sep = 5pt] (0,0) node [draw, thin] (c) \bgroup%
\begin{minipage}{11em}
\begin{minted}[fontsize=\small,baselinestretch=1]{python}
nn.conv2d(data, weight, 
 padding=[1, 1, 1, 1], 
 channels=16, 
 kernel_size=[3, 3])
\end{minted}
\end{minipage}
  \egroup;
  
\draw[inner sep = 5pt] (8.75,0) node [draw, thin] (asm) \bgroup%
\begin{minipage}{14em}
\begin{minted}[fontsize=\small,baselinestretch=1]{python}
with T.attr("VTAPushGEMMOp"):
 for j_init in range(14):
  T.tir.vta.uop_push(...)
\end{minted}
\end{minipage}
  \egroup;

\draw[inner sep = 5pt] (4,0) node [draw, thin] (compiler) {TVM};

  \draw[->, very thick] (c) edge (compiler) (compiler) edge (asm);
\end{tikzpicture}
\caption{Tensorizing a 2D convolution to VTA~\cite{moreau2018vta} accelerator calls.}
\label{fig:intro:tensorization}
\end{figure}

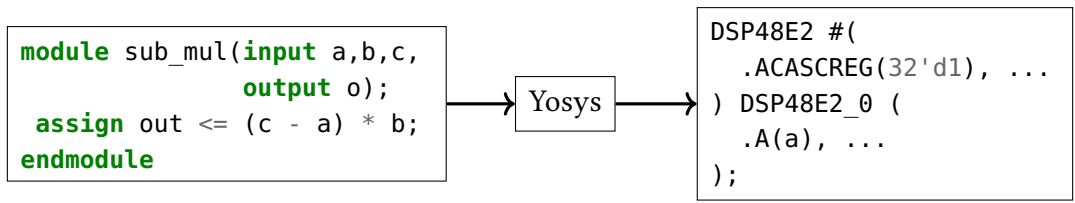
\begin{figure}[]
\centering
\begin{tikzpicture}
  \draw[inner sep = 5pt] (0,0) node [draw, thin] (c) \bgroup%
\begin{minipage}{13em}
\begin{minted}[fontsize=\small,baselinestretch=1]{verilog}
module sub_mul(input a,b,c,
               output o);
 assign out <= (c - a) * b;
endmodule
\end{minted}
\end{minipage}
  \egroup;
  
\draw[inner sep = 5pt] (8.75,0) node [draw, thin] (asm) \bgroup%
\begin{minipage}{11em}
\begin{minted}[fontsize=\small,baselinestretch=1]{verilog}
DSP48E2 #(
  .ACASCREG(32'd1), ...
) DSP48E2_0 (
  .A(a), ...
);
\end{minted}
\end{minipage}
  \egroup;

\draw[inner sep = 5pt] (4.5,0) node [draw, thin] (compiler) {Yosys};

  \draw[->, very thick] (c) edge (compiler) (compiler) edge (asm);
\end{tikzpicture}
\caption{Technology mapping a high-level
  hardware design
  to an instantiation of a specific 
  hardware primitive.}
\label{fig:intro:techmapping}
\end{figure}

Our other two compiler examples,
  TVM and Yosys,
  also rely on a few core algorithms
  to implement their backends.
TVM implements a step
  called 
  \gls{tensorization}~\cite{tvmtensorization},
  which, among other things, maps high-level 
  \glspl{mlkernel}
  to target-specific implementations,
  including invocations
  of specialized hardware \glspl{accelerator}.
\Cref{fig:intro:tensorization}
  shows an example of tensorizing 
  a 2-dimensional convolution
  to general matrix multiplication (GEMM) instructions
  for a specific accelerator backend, VTA.
Similarly,
  a core step in hardware compilation
  for
  Yosys and other hardware synthesis tools 
  is \gls{technology-mapping}
  in which the tool determines
  how to implement the high-level
  hardware design
  using the hardware primitives
  available on the hardware
  platform.
An example of technology mapping is shown in
  \cref{fig:intro:techmapping},
  in which a high-level, architecture-independent
  hardware module 
  is implemented using an
  architecture-specific hardware primitive
  (in this case, a DSP48E2 primitive
    present on Xilinx \glspl{fpga}).
Both tensorization
  and technology mapping
  will be key focuses of this dissertation,
  in \cref{part:glenside-and-3la,part:lakeroad}
  respectively.

To implement their core algorithms,
  compiler backends 
  employ \textbf{models} of hardware.
Following our running examples, 
  clang's instruction selection algorithm
  directly utilizes a model
  of the x86 architecture's
  instructions
  to determine what instructions
  are available for use.
The model is explicit,
  built into clang's x86 backend itself.
The following snippet is
  taken from the x86 instruction model,
  and is the declaration
  of the \texttt{imul} instruction
  used to implement our \texttt{square}
  function above:
  
\vspace{4mm}
\begin{figure}[H]
    \centering
\begin{minted}[baselinestretch=1]{cpp}
defm IMUL : Mul<0xF7, "imul", MRM5r, MRM5m, null_frag>;
\end{minted}
\caption{
Declaration of x86's
  \texttt{imul}
  instruction in LLVM's x86 
  backend~\cite{llvmx86tablegen}.
}
    \label{fig:intro:llvm-tablegen}
\end{figure}
\noindent
This instruction declaration
  tells clang's instruction selector
  that there is an instruction,
  \texttt{imul},
  available for use;
  other parts of the model (omitted)
  describe the functionality of the instruction,
  which helps the instruction selector
  decide when to use the \texttt{imul} instruction.

Not all models within compilers
  are made the same, however.
To contrast the explicit model in
  \cref{fig:intro:llvm-tablegen},
  consider this snippet from Yosys's technology mapper
  for Xilinx FPGAs:

\vspace{4mm}


\begin{figure}[H]
    \centering
\begin{minted}[baselinestretch=1]{c}
subpattern in_dffe
arg argQ clock
code
  dff = nullptr;
  if (argQ.empty())
    reject;
  for (const auto &c : argQ.chunks()) {
    if (!c.wire)
      reject;
    ...
\end{minted}
    \caption{
Snippet of code
  from Yosys's pmgen framework~\cite{yosysxilinxpmgen}
  attempting to map hardware designs
  to specific FPGA hardware primitives.
    }
    \label{fig:intro:yosys-pmgen}
\end{figure}


\noindent
This code is an imperative pattern matching
  algorithm
  written in Yosys's pmgen \gls{dsl}
  which searches for a specific pattern 
  in the hardware design.
Unlike \cref{fig:intro:llvm-tablegen},
  which is an explicit hardware model
  \textit{used by} clang's instruction selector algorithm,
  the above example is \textit{both} algorithm
  \textit{and} model:
  encoded implicitly
  within this algorithm
  is a model of the underlying hardware.%
\footnote{In fact, it would be quite difficult
  to build a compiler
  \textit{without} encoding some kind 
  of model of the underlying hardware.
That is, any compiler which generates code
  for a hardware target
  will encode facts about the hardware
  which amount to a model of the hardware.
The less those facts are explicitly separated out,
  the more implicit the model.}
Soon, I will argue why this method
  of entwining algorithm
  and model
  is disadvantageous;
  before I do that, however,
  I will introduce some terminology
  to make it easier to discuss 
  the properties
  of algorithms and models.

To better discuss compiler backends'
  algorithms
  and the hardware models on which
  they depend,
  I introduce two terms:
  \textit{model explicitness}
  and
  \textit{algorithm adaptability.}
  
\paragraph{Model explicitness.}
Model explicitness
  captures
  how overtly
  a model is encoded
  into a compiler backend.
For example,
  \cref{fig:intro:llvm-tablegen}
  presents an overt, 
  explicit model of hardware
  in the form of a list
  of instructions
  implemented on x86.
In contrast,
  \cref{fig:intro:yosys-pmgen}
  presents an implicit model
  embedded within a pattern matching
  algorithm.
We consider
  \cref{fig:intro:llvm-tablegen} more explicit
  as the model is easier to identify
  and interpret.
Model explicitness is also highly correlated
  to whether or not the model
  is captured in a non-executable, \textit{declarative} form
  such as the static list of instructions
  in \cref{fig:intro:llvm-tablegen},
  or in an executable, \textit{imperative} form
  such as the pattern matching
  algorithm in \cref{fig:intro:yosys-pmgen}.

\paragraph{Algorithm adaptability.}
Algorithm adaptability captures the
  ability of a particular compiler backend
  algorithm
  (e.g.~an instruction selection
    algorithm
    or a technology mapping algorithm)
  to adapt
  to new hardware
  with minimal modification.
For example,
  because
  clang's instruction selection algorithm
  reads its instructions from
  declarative models
  such as the one presented in
  \cref{fig:intro:llvm-tablegen},
  it can easily adapt to new instructions
  and hardware targets
  by simply being supplied
  a new list of instructions~\cite{llvminstructionselection}.
The snippet of 
  Yosys's technology mapping algorithm
  presented in 
  \cref{fig:intro:yosys-pmgen},
  on the other hand,
  encodes a model of the 
  target FPGA
  implicitly within the algorithm;
  thus, adapting it to a new FPGA
  would involve entirely rewriting
  the algorithm.

\vspace{10mm}

\begin{figure}
    \centering
    \includegraphics[width=.7\textwidth]{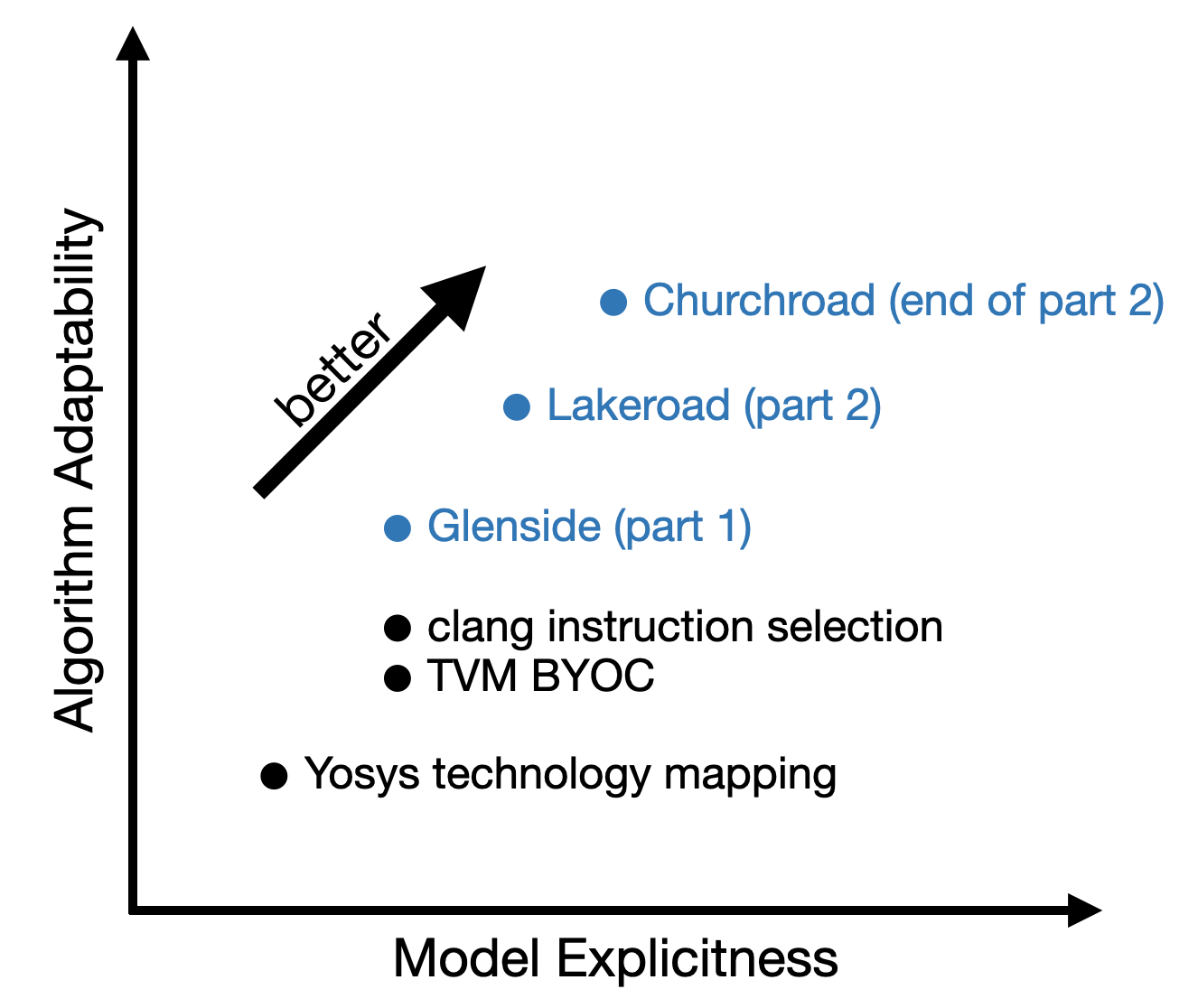}
    \caption{
A visualization of where various compiler backend
  components
  fall
  on the model explicitness--algorithm automation
  spectrum.
Glenside, Lakeroad, and Churchroad
  are the contributions of this
  dissertation.
    }
    \label{fig:intro:model-alg-spectrum}
\end{figure}

Now that we've introduced these terms,
  let's 
  reconsider the two examples we have discussed so far:
  clang's instruction selector
  and Yosys's technology mapper.
clang's instruction selector is powered by
  an \textit{explicit} model of the x86 ISA,
  part of which is presented in
  \cref{fig:intro:llvm-tablegen}.
Furthermore, its underlying algorithm
  is \textit{adaptable} to new models;
  the user simply needs to update
  the model, and the algorithm will adapt.
On the other hand,
  Yosys's technology mapper utilizes
  \textit{implicit} models,
  and its algorithm is \textit{inflexible}.
To visualize this, we
  we plot Yosys's techology mapper
  and clang's instruction selector
  on a 2D plane
  with model explicitness on the horizontal axis
  and algorithm adaptability on the vertical axis,
  shown in \cref{fig:intro:model-alg-spectrum}.
Yosys's technology mapper,
  being built upon implicit models
  and inflexible algorithms,
  is towards the bottom left.
Meanwhile, clang's instruction selection
  is further up and to the right.


\textbf{The core claim of this dissertation,
  put informally,
  is that pushing up
  and to the right
  on our model explicitness--algorithm adaptability
  spectrum (\cref{fig:intro:model-alg-spectrum})
  produces better compiler backends.}
By ``pushing up'',
  I mean
  using
  more adaptable
  \gls{automated-reasoning} algorithms
  to automatically generate
  compiler backends
  from explicit, formal models of hardware,
  as opposed to 
  hardcoding backends
  using inflexible algorithms
  and implicit models.
By ``better compiler backends'',
  I focus on three primary classes
  of improvements:
  correctness,
  optimization,
  and development time improvements.
  
Next, I present my formal
  thesis statement.
Afterwards, I will break down the statement
  and discuss each part.

\mdfsetup{
    frametitle={\colorbox{white}{\space{Thesis Statement}\space}},
    innertopmargin=10pt,
    frametitleaboveskip=-9,
    frametitlealignment=\center
    }
\begin{mdframed}
\mythesis
\end{mdframed}

\vspace{10mm}

I will now explain and define
  the individual components
  of this thesis.
I will use a set of five keywords
  to refer back to 
  to the primary components of my
  thesis statement:
  its two ``inputs'',
  \cref{thesis:algorithms} and
  \cref{thesis:models},
  and its three ``outputs'',
  \cref{thesis:optimizations},
  \cref{thesis:correctness}, and
  \cref{thesis:devtime}.

\paragraph{Automatically generating compiler 
  backends.}
  \label[thesis:algorithms]{thesis:algorithms}
  \label[thesis:models]{thesis:models}
At the highest level,
  my thesis advocates for
  automatically generating compiler backends.
By this, I mean
  utilizing automated reasoning techniques
  to automatically implement
  compiler backend tasks like
  \gls{tensorization} in ML compilers
  and \gls{technology-mapping} in \gls{fpga} \gls{hardwaresynthesis} tools.
At the highest level,
  automatically generating compiler backends
  takes the form of
  specializing an automated reasoning algorithm
  to a specific compilation task
  by feeding it
  a hardware model.
Thus, I break down the automated generation
  of compiler backends
  into two primary design decisions:
  choosing an automated reasoning algorithm
  (hence referred to with the keyword
    \cref{thesis:algorithms})
  and choosing the formal hardware models
  to feed into the algorithms
  (hence referred to with the keyword
    \cref{thesis:models}).
\cref{thesis:algorithms}
  refers
  to the automated reasoning algorithms
  we use
  to automatically generate our backends.
\cref{thesis:models}
  refers to the hardware models
  which the automated reasoning algorithms
  consume
  to generate the compiler backend.
I thus consider   
  the
  \cref{thesis:algorithms}
  and
  \cref{thesis:models}
  as ``inputs''
  or the independent variables
  in my research.
In \cref{part:glenside-and-3la,part:lakeroad},
  I will show how
  different choices of 
  \cref{thesis:algorithms}
  and
  \cref{thesis:models}
  produce different results.
In
  \cref{part:glenside-and-3la},
  we focus on an algorithm called
  \gls{equality-saturation},
  and
  our models take the form
  of program rewrites
  capturing the high-level 
  functional behavior
  of hardware accelerators.
In
  \cref{part:lakeroad},
  we focus on an algorithm called
  \gls{program-synthesis},
  and
  we directly utilize
  vendor-provided
  Verilog simulation models
  as our models of hardware.

\vspace{10mm}

In my thesis statement,
  I claim that automatically generating
  compiler backends
  benefits compiler
  \cref{thesis:optimizations},
  \cref{thesis:correctness}, and
  \cref{thesis:devtime}.
We consider these the ``outputs''
  or dependent variables
  in our experiments
  in \cref{part:glenside-and-3la,part:lakeroad}.
I now describe each in detail.

\paragraph{\cref{thesis:optimizations}.}
  \label[thesis:optimizations]{thesis:optimizations}
A key task of a compiler backend
  is to utilize the target hardware
  efficiently
  to produce optimized programs.
Compiler backends which rely on poor,
  implicitly encoded models
  and inflexible algorithms
  leave key optimizations on the table.
In \cref{part:glenside-and-3la},
  we demonstrate how inflexible algorithms
  lead to missed accelerator mapping opportunities
  in deep learning compilers.
In \cref{part:lakeroad},
  we show how inflexible algorithms
  and implicit models
  lead to poor utilization
  of specialized FPGA primitives
  during FPGA compilation.
Both of these cases correspond to missed opportunities for optimization.

\paragraph{\cref{thesis:correctness}.}
  \label[thesis:correctness]{thesis:correctness}
Compiler backends are expected to produce correct code.
However,
  backends which are built on
  implicit models of hardware
  which are deeply integrated into
  the algorithms themselves
  can have hard-to-find bugs.
A cleaner division between hardware model
  and algorithm
  makes it easier to find bugs on either side.
Furthermore, many of the
  more adaptable algorithms
  I advocate for in this dissertation
  (equality saturation, program synthesis)
  have open-source, highly-used, well-tested
  implementations
  which are likely more trustworthy
  than a hand-implemented
  algorithm used only within
  a single compiler backend.
In \cref{part:glenside-and-3la},
  I demonstrate how
  the difficulty of building compilers
  leads to a lack of validation
  in machine learning accelerators.
I show how automatically generating compilers
  for accelerators aids in rapid
  testing and validation,
  and directly leads to uncovering bugs
  in real hardware designs.
In \cref{part:lakeroad},
  I demonstrate how
  utilizing automated reasoning methods,
  we can generate correctness
  guarantees
  stronger than the guarantees
  provided by any
  existing hardware synthesis tool.

\paragraph{\cref{thesis:devtime}.}
  \label[thesis:devtime]{thesis:devtime}
And lastly,
  more explicit models
  and more adaptable algorithms
  can ease compiler development.
Implicit models,
  especially models deeply integrated
  into the algorithms themselves
  (such as the Yosys example
    in \cref{fig:intro:yosys-pmgen})
  are generally harder to comprehend
  and thus harder to update.
Understanding implicit,
  imperative models
  is often more challenging
  than understanding explicit,
  declarative models.
More flexible algorithms
  reduce development time in a number of ways.
First, the algorithms
  this dissertation promotes
  all have free-to-use open source 
  implementations,
  thus alleviating the need
  for the compiler engineer
  to write their own algorithm by hand.
Second, the greater flexibility
  of the algorithms
  allows them to adapt to new
  hardware
  with less engineering effort.
In both \cref{part:glenside-and-3la,part:lakeroad},
  we demonstrate
  how compiler backends
  generated with automated reasoning algorithms
  are more easily extensible.
In both cases, to target a new hardware platform,
  users simply need to provide
  models of the target hardware.
In \cref{part:glenside-and-3la},
  these models come in the form of
  rewrites capturing accelerator functionality.
In \cref{part:lakeroad},
  we use simulation models of FPGA primitives
  provided by the FPGA vendors.

\vspace{10mm}

I will demonstrate this thesis
  in two parts.
These parts are visualized on
  our model explicitness--algorithm adaptability
  spectrum in
  \cref{fig:intro:model-alg-spectrum}.
In \cref{part:glenside-and-3la},
  I introduce Glenside~\cite{smith2021pure}
  and \TLA~\cite{huang2024application},
  which demonstrate how
  a more adaptable algorithm
  can increase a compiler's
  ability to offload operations
  to machine learning accelerators.
As is shown in
  \cref{fig:intro:model-alg-spectrum},
  \cref{part:glenside-and-3la}
  only pushes along one axis 
  of our spectrum:
  namely, algorithm adaptability.
In \cref{part:lakeroad}, I
  more fully realize
  my thesis statement
  via Lakeroad~\cite{smith2024fpga}:
  a technology mapper for FPGAs
  which utilizes both more
  adaptable algorithms
  and
  more explicit models.
In the end of \cref{part:lakeroad},
  I also describe Churchroad~\cite{smith2024there},
  which seeks to extend the power of \lr
  to larger hardware designs.
Glenside, Lakeroad, and Churchroad
  demonstrate
  how,
  by automatically generating
  portions of compiler backends
  using more adaptable algorithms
  and more explicit models of hardware,
  we we improve their
  optimization ability and
  correctness, while
  easing development effort.

Before jumping into the content
  of this dissertation, though,
  let me first take the time
  to situate this work
  in the existing literature
  and explain my novel contributions.

\section*{Situating this Dissertation: What's New?}

The automatic generation
  of compilers
  is not a new idea;
  in fact, it has been somewhat
  of a holy grail
  for decades.
So what does this 
  dissertation bring to the table?
Before elaborating on that question,
  I will briefly chronicle
  related work from the past five decades
  related to the top-level topic
  of compiler generation.
Then, I will discuss
  the novel insights of this dissertation:
  namely, 
  (1) taking advantage of the new wave of 
  powerful, off-the-shelf
  automated reasoning tools,
  (2) constraining ourselves
  to domain-specific tasks
  to limit the size of the problem, and
  (3) directly generating
  compiler backends from 
  externally-supplied
  (i.e., not written by us)
  models of hardware.

The earliest citations
  for automated compiler generation
  go back to the late 1960s and early 1970s.
However, 
  though it was often referred to as
  \textit{compiler} generation,
  much of this work focused solely on
  compiler \textit{frontends:}
  parsers and lexers~\cite{
    Newcomer1975MachineindependentGO,
    mosses1975mathematical,
  Jones1980CompilerGF,
  Sethi1981CircularEE,
  Smith2005SemanticsDirectedCG,
  Feldman1968TranslatorWS}.
A common task was,
  given a grammar for a new programming language,
  could you generate the parser and lexer
  for that language.
This work culminated in 
  industry-standard tools
  like Yacc and GNU Bison.
Even just Yacc's name, which stands for
  ``yet another compiler--compiler''
  shows how our terminology may have changed;
  while Yacc might have been considered
  a compiler--compiler in the past,
  now it only handles the very frontend
  of compiler tasks: the parser and lexer.

However, not all of the initial wave of research
  focused on compiler frontends.
There was also work on generating components
  of compiler \textit{backends}---%
  often referred to as 
  ``code generation.''
In a 1977 survey of code generators 
  from R.~G.~Cattell~\cite{cattell1977survey},
  he states:
  \begin{quote}
Traditionally, compiler-generation systems have
been weak on automating the later stages of compilation, specifically code generation.
But as the formal methods and grammars applied have become better understood and
more powerful, their scope has gradually been evolving towards the later stages of
compilation.
  \end{quote}
Certainly this entire dissertation
  can simply be seen as one more step
  in this gradual evolution.
As we will discuss later,
  this dissertation benefits from
  the nearly five decades
  of formal methods and automated reasoning research
  since the time of Cattell's writing.

Nonetheless,
  researchers did approach the topic
  of backend generation~\cite{
  snyder1975portable,
  fraser1988automatic}.
Interestingly,
  the spectrum I identify
  earlier in this chapter---%
  the model explicitness--algorithm adaptability spectrum---%
  seems to apply even in the early years
  of automated backend generation research.
In his 1977 survey, Cattell
  describes a very similar spectrum
  in methods of 
  automatically producing code generators:
\begin{quote}
In general, there have been two kinds of approaches to more automatic production of
code generators. The first is the development of a specialized language for code
generators, with built-in machinery for dealing with common details of the process.
The second extreme is the development of a program to build a code generator for a
language from a purely structural and behavioral machine description. Rather than
being mutually exclusive, these procedural and descriptive language approaches,
respectively, represent points in a continuum of degrees of automatic programming.
\end{quote}
The two extremes on Cattell's spectrum
  have analogous points on my
  model explicitness--algorithm adaptability spectrum.
At one extreme
  on his spectrum,
  he is essentially describing
  code generator \glspl{dsl}:
  specialized languages for
  building procedural (in his words)
  or imperative (in mine)
  code generators.
We've already seen a
  modern example of this---%
  Yosys's pmgen DSL
  in \cref{fig:intro:yosys-pmgen}.
At the other extreme of his spectrum
  are programs which produce code generators from
  descriptive/structural (in his words)
  or declarative (in mine)
  machine descriptions and hardware models.
Again, we have seen a modern example of this
  in \cref{fig:intro:llvm-tablegen}: 
  clang's instruction selector
  and the declarative model of the ISA
  which it consumes.

%

Much of the early work on
  the automated generation of code generators
  (i.e.~compiler backends)
  was based on intermediate representations.
To generate compilers,
  these works introduce an intermediate
  representation;
  then, those wishing to generate a compiler
  would then specify how to compile
  that intermediate representation
  to their target machine.
A prime example of this 
  was Perry Miller's DMACS system~\cite{miller1971automatic}
  where DMACS stood for
  ``Descriptive MACro code generating System.''
This
  system introduced
  macros, which were 
  machine-independent
  operations
  which could be implemented
  differently
  for each target machine.
Much of this work can be seen
  as the predecessors
  of the 
  now-standard method of building compilers
  using one or a number of
  intermediate representations,
  most recently made popular
  by the
  foundational LLVM compiler 
  toolchain~\cite{lattner2004llvm}.

While it is further
  from the work of this dissertation,
  it is also worth mentioning work on
  partial evaluation~\cite{consel1993tutorial,futamura1999partial}.
Futamura projections
  specifically
  are of interest,
  as they describe how compilers can
  be viewed as partial evaluations
  of interpreters.
I do not use any partial evaluation techniques
  in this dissertation.

In the ensuing five decades,
  the topic of automated backend
  generation
  has seen steady interest~\cite{
  buchwald2018synthesizing,
  dias2010automatically,
  brandner2007compiler,
  daly2022synthesizing,
  leupers1997retargetable,
  brandner2013automatic}.
In the next few paragraphs,
  I highlight notable trends
  and important projects.
  
A common pattern in recent research
  is the application
  of more and more powerful
  \gls{automated-reasoning} algorithms,
  especially those employed in this dissertation:
  term rewriting~\cite{Richards2006VerificationOC,Despland1990UsingRT,Emmelmann1991CodeSB,Despland1990PAGODEAB,Daly2024EfficientlySL} and 
  synthesis based on \gls{smt}~\cite{Daly2024EfficientlySL}.
One of the most notable projects
  in this space
  is SPIRAL~\cite{franchetti2018spiral}. 
Historically,
  SPIRAL is an umbrella
  over many related grants,
  researchers, and projects,
  but at its core is the
  SPIRAL compiler.
The SPIRAL compiler's goal
  is to enable performance
  portability
  of specialized kernels
  across a wide range of architectures.
They use many techniques
  also used in this dissertation,
  such as
  capturing programs in a high-level,
  backend-nonspecific language
  and
  utilizing automated reasoning 
  (in their case, term rewriting systems)
  to adapt their compiler
  to different backends.

Another point worthy of note
  is the work of Norman Ramsey
  and his student Jo\~{a}o Dias,
  whose work on generating
  instruction selectors
  (among other compiler components)
  is very reminiscent
  of the work in this dissertation~\cite{
  ramsey2003pragmatic,
  ramsey2011resourceable,
  dias2010automatically}.
As is the general pattern
  in the work in this area,
  Ramsey and Dias
  focus on utilizing automated reasoning algorithms
  and high-level machine descriptions
  to implement compiler backend components,
  saving compiler development time.

If this dissertation focuses on
  generating software (i.e.~compilers)
  from hardware,
  it is important to also mention
  the parallel line of research
  which attempts to generate hardware from software.
Programs are a literal description
  of what needs to be computed;
  why not use them to determine
  what hardware we should make?
A perfect example of this is
  the concept of \gls{hls}~\cite{cong2011high,cong2022fpga}
  which allows hardware designers to produce hardware
  from
  software algorithms
  written in high-level languages like C.\footnote{
Note that the reality of modern HLS
  is a little messier than described,
  but this is the intention.
}
There is also an entire literature on
  hardware--software codesign~\cite{teich2012hardware,kokila2016survey,schaumont2012practical,wolf2003decade,Gupta1993HardwaresoftwareCF},
  which, while rich and varied,
  generally centers on the idea
  of exploring the hardware design space
  using representative software workloads
  as a starting place.
In fact,
  \g (presented in \cref{part:glenside-and-3la})
  was originally a
  hardware--software codesign tool
  which aimed to use equality saturation
  to rewrite machine learning models
  into potential accelerator designs.

The ideas presented in this dissertation
  are complementary
  to the ideas of hardware--software codesign,
  and
  I believe we should pursue both directions.
The core idea presented in
  this dissertation---%
  generating compiler backends
  from formal models of hardware---%
  targets the lower layers of the compiler stack.
In general, this dissertation answers the question of,
  given low-level primitives,
  how do we find places 
  to use those primitives 
  in programs or hardware designs?
Hardware--software codesign, on the other hand,
  targets higher levels of the stack.
Codesign seeks to answer the question,
  what hardware \textit{should} we design,
  given the software we need to run?
Thus, the methods presented here
  will only benefit codesign;
  better technology mapping from \lr, for example,
  will only benefit the designs
  produced by HLS tools.

I claim there are three specific features
  of this dissertation
  which set it apart from existing literature.
First is our focus on
  using
  \textit{off-the-shelf}
  automated reasoning tools,
  rather than developing our own.
Second is constraining ourselves
  to domain-specific tasks
  to limit the size of the 
  generation problem.
And last is the generation of
  compiler backends from 
  \textit{externally-supplied}
  (i.e., not written by us)
  models of hardware.
I will now discuss each of these in detail.

First, since the inception of
  automated compiler backend generation
  as a research topic,
  there have been significant
  advances in
  \gls{automated-reasoning} techniques,
  e.g.~%
  equational reasoning via \gls{equality-saturation}~\cite{tate2009equality,willsey2021egg},
  \gls{program-synthesis}~\cite{solar2008program,torlak2013growing},
  and machine learning for program generation~\cite{alon2019code2vec,austin2021program}.
Many of these advances
  have made automated reasoning techniques
  accessible to a broader audience.
Whereas previous work
  often may need to build 
  automated reasoning techniques
  by hand,
  in this dissertation I show how
  off-the-shelf tools
  have become powerful enough
  to use for tasks of this magnitude.
Specifically, I utilize
  the
  equality saturation
  library \texttt{egg}~\cite{willsey2021egg}
  in \cref{part:glenside-and-3la}
  and the program synthesis library
  Rosette~\cite{
  torlak2013growing,torlak2014lightweight}
  in \cref{part:lakeroad}.

Second, we focus on
  compiler generation
  for specialized hardware backends.
Compiler backend generation projects
  of the past often focused on
  generating compiler components
  for general-purpose processors~\cite{
  fauth1995describing,
  leupers1997retargetable,
  brandner2007compiler,
  brandner2009automatic,
  brandner2013automatic,
  ramsey2003pragmatic,
  ramsey2011resourceable,
  dias2010automatically}.
As hardware becomes more heterogeneous,
  however,
  the variety of hardware targets
  needing compilers
  has increased.
In this dissertation,
  we focus instead on 
  building compiler components
  for new, specialized platforms---%
  machine learning accelerators
  in \cref{part:glenside-and-3la}
  and specialized \gls{fpga} \glspl{primitive}
  like DSPs in
  \cref{part:lakeroad}.
Not only are compilers
  needed for these new specialized targets,
  but their specialization
  also constrains the search space,
  making automated reasoning
  far more tractable.

Lastly,
  while previous works have often leaned
  on machine descriptions~\cite{
  ramsey2011resourceable}
  when generating compiler components,
  these machine descriptions
  generally must be handwritten
  by the compiler engineer
  utilizing the compiler generation
  framework.
In \cref{part:lakeroad}
  we introduce a new technique---%
  semantics extraction from Verilog---%
  which directly leverages vendor-supplied
  Verilog.
Many of these models are
  built to be used with
  automated reasoning tools,
  but are currently only utilized for
  \textit{post-compilation verification,}
  rather than in compilation itself~\cite{sisco2022position,sisco2022synthesis}.
I demonstrate how these models
  can be used
  to generate more correct, more complete compilers
  for specialized hardware.

\part{Compilation to Machine Learning Accelerators}
\label{part:glenside-and-3la}
\chapter*{\Cref{part:glenside-and-3la} Abstract}

In \cref{part:glenside-and-3la},
  I describe an application of my underlying thesis
  to the generation of compilers
  for machine learning \glspl{accelerator}.
Specialized hardware, especially for high-performance fields
  such as machine learning,
  have only grown in importance
  over the last decade.
Despite this increased importance,
  it remains surprisingly difficult
  to build compilers for specialized hardware.
Existing approaches
  require significant
  developer effort (\cref{thesis:devtime}),
  and 
  often leave
  \cref{thesis:optimizations} on the table.
Because building a satisfactory compiler
  remains so challenging,
  many hardware designers choose not to.
Without a compiler,
  designers are unable to test
  their hardware on full workloads,
  leaving crucial bugs undiscovered
  (\cref{thesis:correctness}).
In the following chapters,
  I describe how,
  responding to the lack of testing
  in the accelerator design
  community,
  we applied
  my thesis
  to automatically generate compiler backends
  targeting machine learning accelerators.
Specifically, we utilized
  \gls{equality-saturation}
  (\cref{thesis:algorithms})
  driven by rewrites capturing
  the functional behavior of accelerators
  (\cref{thesis:models})
  to produce a backend
  which 
  requires little developer effort to use
  (\cref{thesis:devtime})
  but finds more mapping to accelerators
  than existing work
  (\cref{thesis:optimizations})
  and enables hardware designers to run
  crucial end-to-end testing
  (\cref{thesis:correctness}).
This is wrapped up inside a language and tool called
  \g;
  \g was then integrated into a larger compiler
  called 3LA.
\Cref{part:glenside-and-3la} draws from both the 
  \g paper,
  ``Pure Tensor Rewriting via Access Patterns''~\cite{smith2021pure},
  and the 
  3LA paper,
  ``Application-level Validation of Accelerator Designs Using a Formal Software/Hardware Interface''~\cite{huang2024application}.

\chapter{Introduction and Motivation}
\label{sec:part1-motivation}

Hardware acceleration has powered significant advances
  in subfields like artificial intelligence, image processing, and graph analysis~\cite{han2016eie,chen2016eyeriss,reagen2016minerva,zhang2016cambricon,hameed2010understanding,ham2016graphicionado,jouppi2017tpu, krizhevsky2012conv, reuther2019survey}.
This trend has highlighted the need
  for flexible \gls{accelerator} support
  in domain-specific compilers like
  Halide~\cite{halide},
  TVM~\cite{chen2018tvm},
  TensorFlow/MLIR~\cite{abadi2016tensorflow,mlir}, and
  PyTorch~\cite{pytorch}.

Despite our increasing dependence
  on accelerators,
  building compilers
  for custom accelerators
  remains a daunting task.
Developing a compiler
  for a custom accelerator
  requires significant
  \cref{thesis:devtime}.
Current frameworks for compiler generation
  generally require significant compilers expertise,
  and the sheer amount of effort
  required to build a compiler from the ground up
  generally limits bespoke compiler construction
  to teams
  at large companies,
  e.g. the TensorFlow 
  stack~\cite{abadi2016tensorflow}
  for Google's 
  TPU~\cite{jouppi2017tpu,jouppi2020tpu}.
Though projects
  such as 
  MLIR~\cite{mlir,
  lattner2021mlir,
  eldridge2021mlir}
  and
  Exo~\cite{ExoPldi22}
  have begun to prescribe
  a general framework
  for structuring a compiler,
  these tools are built for
  domain experts,
  and require significant time investment.
Even once a compiler
  is constructed
  for a piece of custom hardware,
  it may still miss crucial
  \cref{thesis:optimizations}---%
  in this case, taking the form of
  accelerator mapping opportunities.

Furthermore, the difficulty in building compilers
  for accelerators
  has a negative effect
  on accelerator 
  \cref{thesis:correctness}.
Core to ensuring correctness of
  accelerators
  is the process of end-to-end
  hardware \gls{validation}---%
  that is, testing the hardware design
  on full applications---%
  and thorough hardware validation
  requires a working compiler.
Some accelerator bugs 
  will only be caught
  during
  full-application validation,
  especially bugs in the
  microarchitectural optimizations common in accelerator design~\cite{chan2014itrs,fang2019understanding,lai2021programming}.
For example,
  a deep learning
  accelerator may use
  a custom numeric format
  tuned to be storage-efficient,
  while still providing enough
  precision
  to support effective inference.
Testing the numeric format
  on a single layer
  of the deep learning model
  may produce results
  well within the designer's
  error bounds.
However, if the designer
  neglects to test
  \textit{all}
  layers of the model,
  they may fail to discover that,
  even though individual layers
  behave well,
  the error may accumulate
  across layers,
  producing inaccurate application-level results~\cite{zorn2021rounding}.
Early end-to-end application level validation is thus essential
  for avoiding expensive and complex late stage hardware design changes.

\begin{figure}
\centering
\includegraphics[width=.6\textwidth]{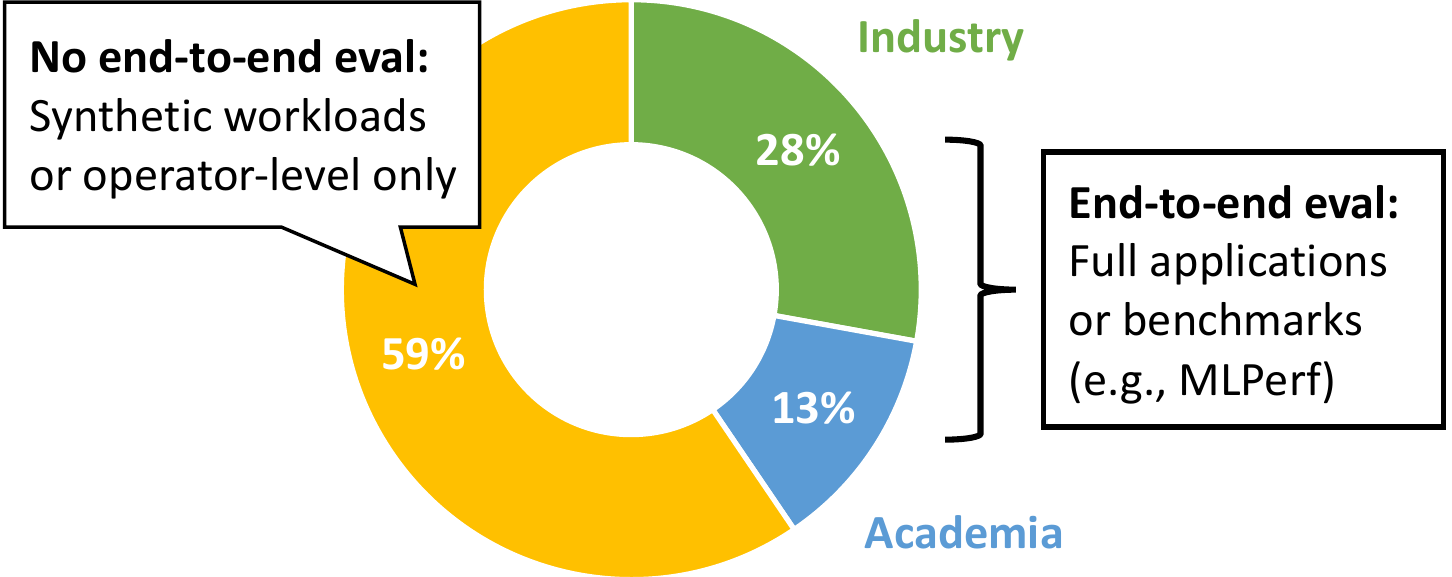}
\caption{
\textit{This figure is reproduced
  from Huang et al.~\cite{huang2024application}.}
\textbf{Gap in end-to-end evaluation of accelerators for neural network applications:} 
Survey of papers from ISCA, MICRO, VLSI, and ISSCC in 2021 and ICCAD, DAC in 2020.
The authors surveyed $79$ papers introducing new DL accelerator designs/methodologies and determined how each accelerator was evaluated. Only 41\% of the works reported end-to-end evaluation on non-synthetic applications, of which 68\% (28\% of the total) were from industrial teams.
}
\label{fig:3la-pie}
\end{figure}

The difficulty in building compilers
  is reflected in the literature.
\Cref{fig:3la-pie} presents a figure
  originally published in
  Huang et al.~\cite{huang2024application},
  which shows how few
  new accelerator designs
  were evaluated on end-to-end applications.
As we will see,
  this has practical consequences,
  as bugs in these accelerators
  may not be apparent
  until full end-to-end
  validation.

\begin{figure}
    \centering
    \includegraphics[width=\textwidth]{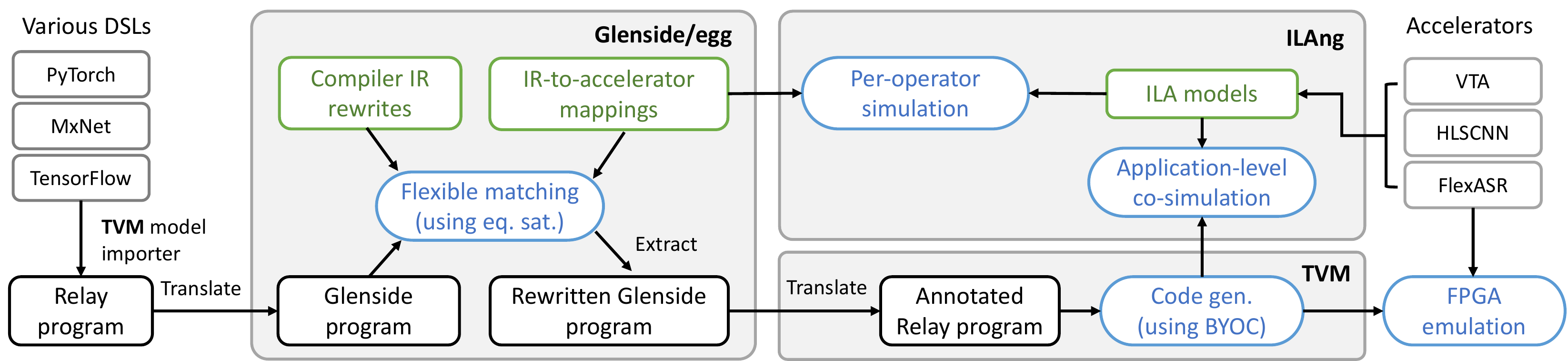}
    \caption{
\textit{This figure is reproduced from
  Huang et al.~\cite{huang2024application}.}
Diagram of the 3LA prototype's flow.
Note that only the ``Glenside/egg''
  portion of 3LA
  is considered a contribution
  of this dissertation;
  for more information on the other components,
  please see the original paper.
}
    \label{fig:3la-diagram}
\end{figure}

In response to the challenges
  of hardware validation
  to the common designer,
  Huang et al.~(including
    the author of this dissertation)
  developed
  3LA:
  a methodology
  to make testing
  easier for new accelerator designs~\cite{huang2022specialized,huang2024application}.
The \TLA flow
  is shown in 
  \cref{fig:3la-diagram}.
The primary contribution of 3LA
  is 
  a methodology to end-to-end evaluate accelerators on unmodified, full applications.
While 3LA as a whole
  is not a contribution of this dissertation
  (see the dissertations of Bo-Yuan Huang~\cite{huang2022specialized}
    and Steven Lyubomirsky~\cite{lyubomirsky2022compiler}),
  the motivations behind 3LA 
  and the story of its development
  are important for \cref{part:glenside-and-3la}
  of this dissertation.

3LA solves two major problems,
  the first being that
  generating simulators
  for hardware
  is difficult
  and time consuming.
3LA solves this problem
  using the Instruction-Level Abstraction (ILA)~\cite{huang2018instruction,huang2018formal}.
ILA is a specification system
  originally designed for
  system-on-chip verification purposes.
The authors of 3LA
  repurposed the ILA
  as a tool for capturing
  and simulating
  machine learning accelerator behavior,
  giving accelerator developers
  a framework for generating fast,
  high-level simulators
  for their accelerators.
The first contribution
  of 3LA
  we do not consider
  a contribution of this dissertation;
  again, please refer to the dissertations
  of the coauthors
  for more information about 3LA as a whole.

The second challenge 3LA solves
  is that,
  even given a simulator,
  compiling
  full programs to the simulator
  is difficult.
At the time of developing 3LA,
  there were few open-source, flexible 
  compiler frameworks
  for targeting custom hardware.
One open-source framework
  supporting mapping
  to custom accelerators
  is TVM's Bring Your Own Codegen (BYOC)~\cite{byoc,chen2021byoc}.
Initially, the 3LA authors
  attempted to use BYOC
  to address their second issue.
To use BYOC, the user
  provides syntactic patterns
  which BYOC then searches for
  in the workloads of interest.
However,
  exact syntactic pattern matching 
  (which we refer to simply as
    ``exact matching'')
  faces difficulties
  as there is often no canonical way
  to represent an operation,
  necessitating either the addition of more patterns
  or manual modifications to the input program
  to match the expected patterns.
Application code can vary greatly in structure,
  particularly in the case of compiler IRs,
  which may be produced after several iterations
  of program transformations.
Code variations are especially prevalent
  in machine learning compilers,
  where workloads are often imported
  from other languages via importers.
Even for the same machine learning model,
  importers from different languages
  can produce wildly different (but equivalent)
  imported programs.
Consider, for example,
  the compiler IR pattern for a linear layer
  in 
  LSTM-WLM:
\small
\[ \texttt{(bias\_add (nn\_dense \%a \%b) \%c)}. \]
\normalsize
However, 
  in another model (ResNet-20)
  the linear layers are equivalently expressed as: 
\small
\[ \texttt{(add (reshape (nn\_dense \%a \%b) \%s) \%c)} \]
\normalsize
when \instrInText{\%c} is a vector, for certain shapes \instrInText{\%s}.
Though they are \textit{semantically}
  equivalent,
  they are not \textit{syntactically}
  equivalent,
  and thus
  we cannot match both of these operations
  with
  a single syntactic pattern.
With an inflexible system like BYOC,
  we would be forced to add
  a new pattern for each possible
  way of writing
  the operator of interest.
To put this in the framing of my dissertation,
  BYOC is near the bottom-left corner
  on our model explicitness--algorithm adaptability spectrum
  in \cref{fig:intro:model-alg-spectrum}.
BYOC's \cref{thesis:models}
  of hardware are explicit:
  rewrites, capturing the functioning of hardware.
However, its \cref{thesis:algorithms}---%
  the underlying exact matching algorithm---%
  is inflexible,
  unable to find semantically equivalent
  but syntactically different
  matches.

This is where I was able to apply
  my dissertation.
When existing \cref{thesis:algorithms}
  (exact matching)
  proved inflexible,
  we introduced a new algorithm
  (\gls{equality-saturation})
  whose increased flexibility
  made it easier to find opportunities
  to invoke accelerators.
We developed a language
  called \g
  which allowed for the application
  of equality saturation
  to the task of accelerator mapping.
We then integrated \g in 3LA
  to solve the second challenge listed above.

Note that we do not improve upon
  model explicitness
  in \cref{part:glenside-and-3la};
  notice that the point for \g
  is above, but not further to the right of,
  the point for BYOC
  on \cref{fig:intro:model-alg-spectrum}.
Both \g and BYOC use similar patterns
  to capture the functioning of hardware.
In \cref{part:lakeroad},
  we will show how we can increase
  both algorithm adaptability
  \textit{and}
  model explicitness.

The rest of \cref{part:glenside-and-3la}
  proceeds as follows.
In \cref{chapter:part1-glenside},
  we introduce \g.
In \cref{chapter:part1-evaluation},
  we evaluate \g,
  primarily by demonstrating what
  \g is able to achieve
  when integrated into 3LA.
In \cref{chapter:part1-background},
  we present related work.

\chapter{Glenside}
\label{chapter:part1-glenside}

\textit{This chapter is derived from Smith et al.~\cite{smith2021pure}.}

\vspace{10mm}
\noindent
In the previous chapter,
  we laid out our need
  for tooling
  which can allow us
  to reason about tensor programs
  flexibly.
While developing the 3LA methodology~\cite{huang2024application},
  we needed a 
  tool 
  which would allow us to find opportunities
  to invoke accelerators
  within machine learning workloads.
TVM's Bring Your Own Codegen (BYOC)~\cite{chen2021byoc} framework
  was ostensibly designed for this purpose,
  but as we saw,
  BYOC was not flexible enough
  for the workloads we cared about.

To increase the likelihood of finding matches,
  pattern matching often relies on
  additional transformations
  to canonicalize intermediate representations (IRs)
  and massage data layouts into
  formats matching accelerator requirements~\cite{nvidia2020nhwc,newcomb2020halide-rewrite,
  hagedorn2020func-high-perf}.
Put in terms of our thesis,
  these transformations increase the
  \textit{flexibility}
  of our underlying matching \cref{thesis:algorithms},
  pushing us further up in our 
  algorithm adaptability--model explicitness spectrum
  (\cref{fig:intro:model-alg-spectrum}).
Once we start modifying the source program
  to find matches,
  the problem of mapping to accelerators
  becomes a \textit{term rewriting} problem,
  and thus
  we should be able to take advantage of
  the wealth of existing knowledge on 
  term rewriting techniques~\cite{baader1998term}.

Term rewriting is a well-known technique 
  for program transformations,
  with some compiler optimizations being implemented
  as term-rewriting systems~\cite{
    dershowitz1993taste,
    baader1999term,
    blindell2016instruction,
    regis-pact22}.
Given a set of syntactic rewrite rules ($\ell \rightarrow r$) that also preserve semantic equality, 
  a term-rewriting system 
  rewrites instances of pattern $\ell$ 
  in the input program with semantically equivalent pattern $r$ where applicable.

One term rewriting approach
  of particular interest
  is \gls{equality-saturation}.
In traditional term rewriting,
  applying one rewrite rule
  may prevent
  using other, potentially profitable, rewrite rules;
  this is referred to as the phase-ordering problem~\cite{whitfield1997approach}.
Equality saturation avoids
  phase-ordering issues 
  by searching over many equivalent rewritings of the same program~\cite{tate2011equality,joshi2002denali}.
Given an input program $p$, 
  equality saturation repeatedly applies 
  the given rewrite rules 
  to explore all equivalent ways to express $p$
using an \textit{e-graph} data structure
  to efficiently represent an exponentially large set of equivalent program expressions~\cite{nelson1980fast,nieuwenhuis2005proof}.
Upon reaching a fixed point, 
  i.e., when no application of any rewrite rule can introduce a new program expression,
  or upon hitting a predetermined resource limit,
  the optimal rewritten program
  can be extracted from an e-graph
  according to a given cost function.

To increase flexibility
  of accelerator mapping \cref{thesis:algorithms},
  {\TLA} sought to employ equality saturation;
  unfortunately, existing IRs in compilers for
  array/tensor programming DSLs 
  are not compatible with equality saturation.
Equality saturation is most easily applied in
  \textit{pure} (side effect--free) IRs
  that support equational reasoning.
Due to their purity, these IRs tend to be
  high-level.
However,
  mapping to accelerators requires considering
  low-level hardware details like data layout.
Existing pure IRs for ML frameworks are used
  primarily for high-level transformations
  (e.g., type elaboration and inlining)
  and do not expose low-level data layout details~\cite{relay}.
On the other hand,
  IRs used for crucial lower-level optimizations like
  operator fusion must support
  precise reasoning about memory use,
  and therefore are typically impure,
  hampering term rewriting.
In summary, for our purposes,
  existing IRs are either pure but too high-level,
  or low-level enough but impure.

To help mitigate such impedance mismatches,
  we present \textit{\g},\footnote{Publicly available at \url{https://github.com/gussmith23/glenside}.}
  a pure tensor program IR
  that enables hardware-level term rewriting.
\g is based on a simple
  \textit{access pattern} abstraction that
  supports expressing and reasoning about
  data layout transformations via
  syntactic rewrite rules.
When combined with standard arithmetic rewrites
  for per-tensor-element computations,
  access patterns enable implementing complex
  transformations for accelerator support as
  compositions of simple rewrites.

\begin{figure}
    \centering
    \includegraphics[width=.6\linewidth]{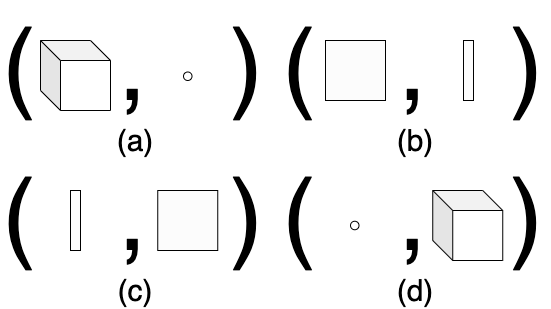}
    \caption{
      Four access patterns,
        representing different ways
        a
        tensor program
        (or \textit{kernel})
        might access
        the same 3D tensor. 
      For example, (c) represents
        accessing a 3D tensor as
        a vector of 2D matrices.}
    \label{fig:access-pattern-examples}
    \vspace{-1em}
\end{figure}

Tensors are traditionally characterized
  by their \textit{shape},
  an $n$-tuple 
  of positive integers
  indicating the size of each
  of a tensor's dimensions.
Access patterns instead characterize
  each tensor with two shapes, e.g.,
  \accesspatternshape{x}{y, z}, separating
  the dimensions which are \textit{iterated over} from
  the dimensions which are \textit{computed on.}
Figure~\ref{fig:access-pattern-examples}(c)
  depicts an example where a 3D tensor's
  first dimension is iterated over and
  some computation applied to each
  corresponding 2D matrix.


In the rest of this chapter,
  we will first walk through an example
  demonstrating the difficulty
  in developing a pure tensor IR
  \cref{sec:matmul}.
We will then describe  
  the implementation of \g in
  \cref{sec:glenside}.
Lastly, we describe how \g
  was incorporated into \TLA
  in \cref{sec:glenside-in-3la}.

\section{From Pure \texttt{matMul} to IR Design Goals}
\label{sec:matmul}

  
Applying functional techniques
  and term rewriting to tensor IRs
  requires careful design.
For example,
  we must ensure that operators be compositional
  with respect to tensor shapes
  and that the representation support
  generic rules within the
  target rewrite engine.
To highlight such constraints and
  motivate access patterns in \g,
  this section illustrates potential pitfalls
  with a simple matrix multiplication example.

\subsection{Pure Matrix Multiplication}
\label{subsec:pure-matmul}

We write
  \tcd{f64} for the type of 64-bit floats and
  \tcd{[A]} for vectors over type \tcd{A}.
Using this notation, we can specify operators like
  dot product and 2D matrix transpose as:
\begin{align*}
    \mcd{dotProd} &
    \mcd{ : [f64] * [f64] -> f64} \\
    \mcd{trans2} &
    \mcd{ : [[f64]] -> [[f64]]}
\end{align*} 






\noindent
Implementing 2D matrix multiplication
  on inputs $P$ and $Q$ requires computing
  an output matrix $R$ where
  $R_{ij} = \Sigma_k P_{ik} \, Q_{kj}
          =  P_i \cdot Q^{T}_{j}$. 
The need to compute \tcd{dotProd} for every pair
  of a row from $P$ and a column from $Q$
  suggests map and Cartesian product operators
  which we might specify with:
\begin{align*}
    \mcd{map} &
    \mcd{ : (A -> B) * [A] -> [B]} \\
    \mcd{cartProd} &
    \mcd{ : [A] * [B] -> [A * B]}
\end{align*}
Naively, we can almost implement matrix multiplication as:
{\color{red} \begin{align*}
  & \mcd{matMul(P, Q) :=} \\
  & \;\;\;\;\; \mcd{map(dotProd, cartProd(P, trans2(Q)))}
\end{align*} }
However, the result type will have been
  flattened to just {\color{red}\tcd{[f64]}},
  making it impossible to compose with other matrix
  operators that expect \tcd{[[f64]]} inputs.

Our first problem is that
  the \tcd{cartProd} specification above
  ``forgets'' the shape of its arguments.
We could change this specification to
  arrange the output as a matrix:
$$
  \mcd{cartProd2D : [A] * [B] -> [[A * B]]}
$$
But this result type prevents
  directly mapping \tcd{dotProd}.\footnote{
    This simple type does not specify how
    \tcd{cartProd2D} orders its output
    relative to its input vectors.
    We assume the order
    expected for matrix multiplication.}
Now the problem is that \tcd{map}
  only applies a computation by iterating
  over the first (outermost) dimension of a tensor.
If we specialize \tcd{map} to iterate
  over the second dimension:
$$
  \mcd{mapAt2 : (A -> B) * [[A]] -> [[B]]}
$$
then we can implement a compositional
  \tcd{matMul} operator that correctly produces
  results of type \tcd{[[f64]]} as:
\begin{align*}
  & \mcd{matMul(P, Q) :=} \\
  & \;\;\;\;\; \mcd{mapAt2(dotProd, cartProd2D(P, trans2(Q)))}
\end{align*}

While this gets us close
  to our goal of a pure, functional
  IR for tensor programs,
  as we'll see,
  this style also has its issues.

\subsection{\g Design Constraints and Goals}

This style of pure, higher-order functional
  program representation enables
  term rewriting and equational reasoning
  via rules like:
\begin{align*}
  \mcd{dotProd(P, Q)}
    & \leftrightsquigarrow
      \mcd{dotProd(Q, P)} \\[2pt]
  \mcd{trans2(trans2(P))}
    & \leftrightsquigarrow
      P \\[2pt]
  \mcd{map(f, map(g, P))}
    & \leftrightsquigarrow
      \mcd{map(f$\,\circ\,$g, P)} \\[2pt]
  \mcd{mapAt2(f, trans2(P))}
    & \leftrightsquigarrow
      \mcd{trans2(mapAt2(f, P))} 
\end{align*}


However, some of these rules depend on the
  shapes of dimension-specific operators aligning.
What happens when we need to support
  higher-dimensional tensors?
Without a mechanism to abstract
  which dimensions of a tensor
  are being iterated as opposed to computed over,
  we would have to generate versions of
  each rule for every combination of dimensions.
Worse, these problems
  do not only affect rewrite rules;
  they also lead to code blowup just to
  specify all the variants of tensor kernels
  that arise in practice---%
  e.g.~we would eventually need
  \texttt{mapAt3}, \texttt{mapAt4},
  and so on.

One strategy to address these challenges is
  adding support for anonymous functions (``lambdas''),
  currying, and closures to the 
  tensor program representation.
These features can provide sufficient
  flexibility to handle shape alignment
  issues that otherwise may require
  dimension-specific operators like
  \tcd{cartProd2D} and \tcd{mapAt2} above.
For example, given curried versions
  of \tcd{dotProd} and \tcd{map},
  we could have used such features
  to implement a curried \tcd{matMul} as:
\begin{align*}
  & \mcd{matMul' P Q :=} \\
  & \;\; \mcd{ \
      map' ($\boldsymbol\lambda\,$r =>\
        map' (dotProd' r) (trans2 Q)) P}
\end{align*}
Alternatively, some IRs rely on index notation
  for even pithier implementations like:
$$
  \mcd{matMul(P,Q)[i,j] := dotProd(P[i], trans2(Q)[j])}
$$

Unfortunately, these approaches all rely on some
  form of \textit{name binding} which can
  significantly complicate term rewriting.
Rewriting under binders,
  whether explicitly in the form of lambdas
  or implicitly with index notation,
  requires additionally analyzing the
  potential \textit{contexts}
  (what names are bound to)
  of every subexpression.
While it is still technically possible to
  apply state-of-the-art rewrite engines
  like \tcd{egg}~\cite{willsey2021egg}
  via explicit variable substitution rules and
  free variable analyses,
  we have found the additional complexity
  and rewrite search space blow up
  substantially eliminate the potential advantages
  of term rewriting in such IR designs.

All the above constraints inform \g's key design goal:
  providing an IR that flexibly supports specifying and
  composing higher-order tensor operators\footnote{
    As \tcd{map} and \tcd{mapAt2} in 
    Section~\ref{subsec:pure-matmul} illustrate,
    an IR can support higher-order operators without
    necessarily providing lambdas, currying, or closures.}
  over arbitrary dimensions while still enabling
  high-performance term rewriting techniques
  like equality saturation.
In the rest of \cref{part:glenside-and-3la},
  we show how \textit{access patterns} enable achieving
  these goals with a focus on applications to
  mapping application fragments down to
  specialized hardware accelerators.

\section{\g}
\label{sec:glenside}
  
\begin{table*}
\small
    \centering
    \caption{\g's access pattern transformers.}
    \label{tab:access-pattern-transformers}
    \begin{tabularx}{\linewidth}{lXX}
    Transformer 
    & Input(s)
    & Output Shape  \\
    \hline
    
    \texttt{access} 
    &
    \accesspatternshape{a_0,\dots}{\dots, a_n}
    and non-negative integer $i$
    & 
  \accesspatternshape
  {a_0, \dots, a_{i-1}}{a_i,\dots, a_n}
    \\
    
    \texttt{transpose} &
    \accesspatternshape{a_0,\dots}{\dots, a_n},  $\ell$ (a permutation of $(0, \dots, n-1)$) &
    \accesspatternshape{a_{\ell_0},\dots}{\dots, a_{\ell_n}}
    \\
    
    \texttt{cartProd} 
    &
    \accesspatternshape{a_0,\dots, a_n}{c_0, \dots, c_p},  \accesspatternshape{b_0,\dots, b_m}{c_0, \dots, c_p}
    & 
  \accesspatternshape
  {a_0, \dots, a_n, b_0,\dots, b_m}
  {2, c_0, \dots, c_p}
    \\
    
    \texttt{windows} 
    &
    \accesspatternshape{a_0, \dots, a_m}{b_0, \dots, b_n}, \newline
    window shape $(w_0, \dots, w_n)$,
    strides $(s_0, \dots, s_n)$
    &
    \accesspatternshape{a_0, \ldots, a_m, b'_0, \dots, b'_n}{w_0, \dots, w_n},\newline
    where $b'_i = \lceil (b_i - (k_i - 1)) / s_i \rceil $\\
    
    \texttt{slice} &
    \accesspatternshape{a_0, \dots }{\dots, a_n}, \newline
    dimension index $d$, bounds $[l, h)$
    &
    \accesspatternshape{a'_0, \dots }{\dots, a'_n} \newline
    with $a'_i = a_i$ except $a'_d = h - l$
    \\
    
    \texttt{squeeze} &
    \accesspatternshape{a_0, \dots }{\dots, a_n}, index $d$ where $a_d = 1$
    &
    \accesspatternshape{a_0, \dots }{\dots, a_n} with $a_d$ removed
    \\
    
    \texttt{flatten} &
    \accesspatternshape{a_0,\dots,a_m}{b_0,\dots,b_n} &
    \accesspatternshape{a_0 \cdots a_m}{b_0 \cdots b_n} \\
    
    \texttt{reshape} &
    \accesspatternshape{a_0,\dots,a_m}{b_0,\dots,b_n},\newline
    access pattern shape literal
    \accesspatternshape{c_0,\dots,c_p}{d_0,\dots,d_q}&
    
    \accesspatternshape{c_0,\dots,c_p}{d_0,\dots,d_q},\newline
    if $a_0 \cdots a_m = c_0 \cdots c_p$
    and $b_0 \cdots b_n = d_0 \cdots d_q$\\
    
    \texttt{pair}&
    two access patterns of shape
  \accesspatternshape
  {a_0, \dots}{\dots, a_n} &
  \accesspatternshape
  {a_0, \dots}{2, \dots, a_n}
    \\
    
    \end{tabularx}
\end{table*}






This section details \g's implementation,
  focusing on its core abstraction,
  \textit{access patterns}.
We use Section~\ref{sec:matmul}'s
  matrix multiplication as a
  running example throughout.

\subsection{Access Patterns}


Access patterns encode common
  tensor IR patterns where
  some tensor dimensions
  are \textit{iterated over} (accessed)
  while others are \textit{computed on}.\footnote{
    This is similar to NumPy's concept of \textit{universal functions.}}
Section~\ref{sec:matmul}'s \tcd{matMul} example
  \textit{iterates over} dimension 0 of input $P$,
  while \textit{computing on} dimension 1,
  effectively viewing $P$ as a 1D vector of 1D vectors.

Access patterns are specified by their \textit{shape} ---
  a pair of tuples of positive integers $(S_A, S_C)$.
An access pattern of shape $(S_A, S_C)$ is, in turn, a
  tensor $T$ whose shape is given by the
  concatenation of the access pattern shape tuples
  $S_A \,\mcd{++}\, S_C$; we refer to
  $S_A$ and $S_C$ as the \textit{access} and
  \textit{compute}
  dimensions of $T$, respectively.

Access patterns represent the view of an
  $(|S_A| + |S_C|)$--dimensional tensor
  as a tensor of shape $S_A$,
  each of whose elements has shape $S_C$.
For an access pattern $T$ of shape $(S_A, S_C)$
  where $|S_A| = n_A$, we use the syntax
  \tcd{(access $T$ $n_A$)} to represent $T$ in \g.
For example, if a 2D matrix $T$ has shape $(m, n)$,
  then the \g expression \tcd{(access $T$ 1)}
  yields an access pattern of shape $((m), (n))$.

  
The matrix multiplication example
  from Section~\ref{sec:matmul}
  directly accesses the rows of $P$,
  but uses \tcd{trans2} to iterate over
  the columns of $Q$.
Instead of requiring an explicit
  transpose operator, \g provides
  access pattern \textit{transformers}.
  
\subsection{Access Pattern Transformers}

Access pattern transformers 
  manipulate one
  or more access patterns
  to produce a new access pattern,
  allowing \g
  to support more complex patterns
  like
  slicing,
  transposing,
  and interleaving.
  Table~\ref{tab:access-pattern-transformers}
  lists \g's transformers.
  
To produce an access pattern
  representing
  the columns of $Q$
  for matrix multiplication,
  we employ
  the \texttt{transpose}
  transformer.
It takes an access pattern
  and a list of dimension indices,
  and rearranges
  the dimensions 
  of the access pattern
  in the order specified by the indices.
If $Q$ has shape $(N, O)$,
  \texttt{(transpose (access $Q$ 1) (list 1 0))}
  produces
  an access pattern
  of shape
  \accesspatternshape{O}{N}.
  
The \texttt{cartProd} transformer
  takes access patterns
  of shapes
  \accesspatternshape{a_0, \dots, a_n}{c_0, \dots, c_p}
  and 
  \accesspatternshape{b_0, \dots, b_m}{c_0, \dots, c_p}
  respectively, and produces 
  an access pattern of the shape
  \accesspatternshape
    {a_0, \dots, a_n, b_0,\dots, b_m}
    {2, c_0, \dots, c_p},
  where $(2, c_0, \dots, c_p)$
  represents a 2-tuple
  of the input access patterns'
  compute dimensions.
The access dimensions
  of the input access patterns
  are simply concatenated.
In the matrix multiplication example,
  the Cartesian product
  of the rows of $P$
  with the columns of $Q$
  is an access pattern
  of shape
  \accesspatternshape{M,O}{2, N},
  where the second shape
  represents a 2-tuple
  of a row from $P$
  with a column from $Q$.

We have nearly re-implemented
  matrix multiplication example
  in \g.
The final step
  is to implement the dot product, for which
  \g uses 
  access pattern \textit{operators}.
  
\subsection{Access Pattern Operators}

\begin{table}
    \centering
    \caption{\g's access pattern operators.}
    \label{tab:operators}
    \begin{tabularx}{\linewidth}{lXX}
    Operator & Type & Description\\
    \hline
    \texttt{reduceSum} & $(\dots) \rightarrow ()$ &
    sum values
    \\
    
    \texttt{reduceMax} & $(\dots) \rightarrow ()$&
    max of all values\\
    
    \texttt{dotProd} &
    $(t,s_0, \dots, s_n)\rightarrow ()$ &
    eltwise mul; sum
    \\
    

    \end{tabularx}
\end{table}

\textit{Operators}
  are the only \g
  constructs
  which
  perform computation.
They are invoked only
  in \texttt{compute} expressions,
  which map the operator
  over the compute dimensions
  of an access pattern.
For an input access pattern
  $A$
  of shape
  \accesspatternshape
  {s_0, \dots, s_{m-1}}
  {s_m, \dots, s_{n}},
  and an operator
  $f$
  with type
  $(s_m,\dots,s_n)
  \rightarrow
  (s'_{m'}, \dots, s'_{n'})$,
  the result of
  \texttt{(compute $f$ $A$)}
  will have shape
  \accesspatternshape
  {s_0, \dots, s_{m-1}}
  {s'_{m'}, \dots, s'_{n'}};
  that is, a \tcd{compute}
  expression
  cannot change
  the access dimensions
  of the input access pattern.
Table \ref{tab:operators}
  lists
  the operators
  in \g{}.
  
Recall where we are
  in converting
  our matrix multiplication
  example:
  we have accessed the rows of $P$
  and the columns of $Q$
  and taken their Cartesian product,
  resulting in an access pattern
  of shape
  \accesspatternshape
  {M, O}{2, N},
  and we need now
  to compute the dot product
  of these row-column
  pairs.
In \g,
  the \texttt{dotProd}
  operator
  (see Table~\ref{tab:operators})
  does just that.
To compute the dot product
  over our row-column pairs,
  we need only to apply
  \texttt{compute dotProd}
  to our access pattern,
  to produce an access pattern
  with final shape
  \accesspatternshape
  {M, N}{}.
The entire \g
  specification
  of matrix multiplication
  is shown in Figure \ref{fig:mat-mat-mult}.
  
\definecolor{gray}{Gray}{5} 
  
\begin{figure*}
\begin{minipage}{.54\textwidth}
\begin{subfigure}{\textwidth}
\begin{lstlisting}[basicstyle=\footnotesize,escapechar=!]
(transpose                   !\color{gray}; \hspace{2mm}\accesspatternshape{N, O, H', W'}{}!
 (squeeze                    !\color{gray}; \hspace{2mm}\accesspatternshape{N, H', W', O}{}!
  (compute dotProd           !\color{gray}; \hspace{2mm}\accesspatternshape{N, 1, H', W', O}{}!
   (cartProd                 !\color{gray}; \hspace{2mm}\accesspatternshape{N, 1, H', W', O}{2, C, K_h, K_w}!
    (windows                 !\color{gray}; \hspace{2mm}\accesspatternshape{N, 1, H', W'}{C, K_h, K_w}!
     (access activations 1)  !\color{gray}; \hspace{2mm}\accesspatternshape{N}{C,H,W}!
     (shape C Kh Kw)
     (shape 1 Sh Sw))
    (access weights 1)))     !\color{gray}; \hspace{2mm}\accesspatternshape{O}{C, K_h, K_w}!
  1)
 (list 0 3 1 2))
 
     \end{lstlisting}
       \vspace{-1.5em}
    \subcaption{2D convolution.
    }
    \label{fig:conv2d}
\end{subfigure}
\end{minipage}
\begin{minipage}{.45\textwidth}

\begin{subfigure}{\textwidth}
\begin{lstlisting}[basicstyle=\footnotesize,escapechar=!]
(compute dotProd          !\color{gray}; \hspace{2mm}\accesspatternshape{M, O}{}!
 (cartProd                !\color{gray}; \hspace{2mm}\accesspatternshape{M, O}{2, N}!
  (access activations 1)  !\color{gray}; \hspace{2mm}\accesspatternshape{M}{N}!
  (transpose              !\color{gray}; \hspace{2mm}\accesspatternshape{O}{N}!
   (access weights 1)     !\color{gray}; \hspace{2mm}\accesspatternshape{N}{O}!
   (list 1 0))))
  \end{lstlisting}
  \vspace{-1.5em} 
  \subcaption{Matrix multiplication.}
  \label{fig:mat-mat-mult}
\end{subfigure}

\begin{subfigure}{\textwidth}
\begin{lstlisting}[basicstyle=\footnotesize,escapechar=!]
(compute reduceMax       !\color{gray}; \accesspatternshape{N,C,H',W'}{}!
 (windows                !\color{gray}; \accesspatternshape{N,C,H',W'}{K_h, K_w}!
  (access activations 2) !\color{gray}; \accesspatternshape{N, C}{H, W}!
  (shape Kh Kw)
  (shape Sh Sw)))
\end{lstlisting}
  \vspace{-1em} 
  \subcaption{Max pooling.}
  \label{fig:maxpool-code}
\end{subfigure}

\end{minipage}
\caption{Common tensor kernels from machine learning expressed in \g. Lines containing access patterns are annotated with their access pattern shape.
$N$ is batch size; $H$/$W$ are spatial dimension sizes; $C$/$O$ are input/output channel count; $K_h$/$K_w$ are filter height/width; $S_h$/$S_w$ are strides.
}
\label{fig:all-kernels}
\end{figure*}

\section{\g in \TLA}
\label{sec:glenside-in-3la}

Now that we have presented
  \g,
  we will now briefly describe
  how \g was used within 3LA.

Operations which can be offloaded
  to accelerators
  can be captured via patterns,
  as discussed with 
  TVM's Bring Your Own Codegen
  framework.
Rather than attempt to enumerate all semantically equivalent patterns
 (a task that is tedious, error-prone, and likely to result in an incomplete enumeration), 
 or expect users to modify their application code to expose expected patterns (demanding knowledge of the model and patterns
  as well as engineering effort), 
   \TLA 
   increases the flexibility
   of compiler backend algorithms
  by utilizing term rewriting and \gls{equality-saturation} techniques to transform programs
  to expose the most  matching opportunities for accelerator operation selection. 
It is in this task
  that 
  \TLA uses \g.

Flexible matching uses two kinds of rewrite rules, both expressed in \g:
\begin{itemize}
\item Compiler IR rewrite rules: These are general-purpose \g-to-\g rules, independent of the accelerator, and are reusable and composable for various applications. We have developed a general set in \TLA 
including rules for, e.g., merging/splitting tensors, commutativity, associativity, and identities for common operators. 

\item \mapping rules: These rewrite rules are accelerator-specific,
  and translate from \g to black-box
  accelerator calls.
When targeting new accelerators, accelerator designers are expected to provide these mappings.
\end{itemize}

All rewrites in {\TLA} are polymorphic over tensor size, which requires specifying relationships between the input and output sizes for operations that merge, split, or broadcast over tensors. This also makes a given \mapping more general and provides support for applications using different block sizes, strides, etc., without changing any rules. 

In the extraction phase of equality saturation, the rewritten program optimizing the cost function is chosen. 
This provides flexibility in the criteria for selection among functionally equivalent candidates for accelerator offloads. In our evaluations where we focused on end-to-end functional testing, we used a simple cost function that maximizes the number of accelerator invocations. More sophisticated cost functions can incorporate information about performance or data movement costs, and thereby result in different offloads.

\textit{Compiler IR rewrite rules.} Here, we describe three examples of compiler IR rewrite rules to show different types of opportunities that can be exposed.
\footnotesize
\begin{eqnarray} 
  \texttt{(compute dot-product (reshape \%x \%s))} &\rightarrow& \texttt{(reshape (compute dot-product \%x) \%s)} \label{rr-reshape-bubbling} \\
  \texttt{(add (reshape (dense \%a \%b) \%s) \%c)} &\rightarrow& \texttt{(reshape (bias\_add (dense \%a \%b) \%c) \%s)} \label{rr-linear-layer} \\
  \texttt{\%x} &\rightarrow& \texttt{(reshape (flatten \%x) (shape-of \%x))} \label{rr-flatten}
\end{eqnarray}
\normalsize
The \texttt{reshape} operator takes a tensor and a shape vector as input and re-arranges the layout of the tensor to the given shape, and the \texttt{dot-product} operator takes a tensor as input and computes the inner product of vectors under the given axis~\cite{smith2021pure}. Rule~\ref{rr-reshape-bubbling} exploits the properties of the two operators and shows that rearranging the application order of \texttt{reshape} and \texttt{dot-product} operators
  preserves the semantics.
Rule~\ref{rr-linear-layer} shows that linear layer \glspl{mlkernel} can be expressed 
  using different arrangement and combinations of operators, e.g., $\texttt{bias\_add}$ (broadcasting) or the 
  elementwise $\texttt{add}$.
Rule~\ref{rr-flatten} shows that de-simplifying a computation 
  (e.g., flattening then unflattening) could expose more opportunities for matching 
  rewrites.
Moreover, combining these individual rewrite rules together enables more 
  sophisticated rewrites. 
For example, combining Rule~\ref{rr-reshape-bubbling} and Rule~\ref{rr-flatten} 
  allows for the emerging $\texttt{im2col}$ transformations for convolution kernels, 
  without needing to specify the transformation as a new rewrite rule.

\textit{IR-to-accelerator mapping rules.}
Similar to compiler IR rewrite rules, we specify IR-to-accelerator mapping 
  rules in Glenside.
These rules are accelerator-specific, mapping supported 
  operations to accelerator invocations.
We now describe three examples of IR-to-accelerator mapping rules.
\scriptsize
\begin{eqnarray}
  \texttt{(compute dot-product (cartesian-product ?x ?w))} &\rightarrow& \texttt{(vta-dense ?x ?w)} \label{rr-vta} \\
  \texttt{(conv2d ?input ?kernel ?group ...)} &\rightarrow& \texttt{(hlscnn-conv2d ?input ?kernel ?group)} \label{rr-hlscnn} \\
  \texttt{\{\{LSTM Relay Pattern\}\}} &\rightarrow& \texttt{(flexasr-lstm ?input ?hidden\_0 ...)} \label{rr-flexasr}
\end{eqnarray}
\normalsize
%
Rule~\ref{rr-vta} maps tensor-level computation of dense matrix multiplication 
  to VTA's \texttt{dense} operation. 
This allows matching decomposed coarse-grained operators and mapping to 
  fine-grained accelerator operations.
Rule~\ref{rr-hlscnn} maps kernel-level computation of a 2D convolution to 
  HLSCNN's \texttt{conv2d} operation---a common accelerator offloading for 
  deep learning kernels.
Rule~\ref{rr-flexasr} maps an LSTM computation to FlexASR's \texttt{lstm} operation.
Note that the LSTM computation (left-hand side) is specified using a pattern 
  compiled from a Relay program; this Glenside feature helps express complex
  operations.

\TLA utilizes equality saturation and the two types of rewrite rules to transform 
  programs, aiming to expose the most matching opportunities for accelerator 
  operation selection.
It was not clear \textit{a priori} whether flexible matching would be performant for accelerators with complex \mapping rules needed for available accelerator designs. Our evaluation results in the next chapter show that compiler IR rewrites can be combined effectively with a few \mapping rules in flexible matching, which finds more matches than exact matching in a reasonable time.

\chapter{Evaluation}
\label{chapter:part1-evaluation}

Thus far, we have described
  how we have applied
  the thesis of this dissertation---%
  that compiler backends
  should be generated
  from formal models of hardware---%
  in the realm of deep learning accelerators.
We first described the difficulties
  in developing compilers
  for deep learning accelerators.
We then described how these difficulties
  motivated the creation of
  3LA: a mostly-automated, end-to-end
  methodology
  for accelerator development.
(Note again that 3LA itself
  is not a contribution of this dissertation.)
We identified a specific problem
  within this domain
  as a problem of interest:
  mapping applications to accelerators.
To address this problem,
  this dissertation
  introduced
  \g,
  a tensor language
  which enables powerful rewriting techniques.
We then integrated \g 
  into \TLA,
  to bring the power of equality saturation
  to bear on the task
  of mapping to accelerators.
  
This evaluation will first demonstrate the utility of
  \g
  via a number of case studies
  in \cref{sec:case-studies}.
Then, 
  in \cref{secion:eval-3la}
  we will evaluate the specific claims
  of our thesis
    (improved \cref{thesis:optimizations},
     \cref{thesis:correctness}, and
     \cref{thesis:devtime})
  by evaluating the components of \TLA 
  to which \g was essential.

\section{Case Studies}
\label{sec:case-studies}

To demonstrate \g's utility,
  we first show how it enables
  concise specifications of several
  critical ML kernels
  (Section~\ref{section:representing-kernels}).
We then show how
  \g's pure, binder-free
  representation enables mapping kernels
  to an example accelerator via
  direct application of generic rewrite rules
  (Section~\ref{sec:case-study-tensorization}).
Finally,
  we highlight how \g
  enables the
  flexible mapping of
  larger, more diverse kernels
  to our accelerator,
  utilizing the power
  of equality saturation
  to automatically discover
  a variety of program transformations.
Specifically,
  we show how \g can automatically
  map convolutions to matrix multiplications
  (Section~\ref{sec:discovering-im2col})
  and automatically
  map large matrix multiplications into a
  sequence of smaller matrix multiplications
  (Section~\ref{sec:case-study-blocking}).
This portion of the evaluation is drawn from
  Smith et al.~\cite{smith2021pure}.

\subsection{Representation of Common ML Kernels}
\label{section:representing-kernels}

Figure~\ref{fig:all-kernels}
  lists the \g specifications
  of three common ML kernels:
  2D convolution,
  matrix multiplication,
  and max pooling.
Below, we discuss
  the specifications of 
  2D convolution
  and max pooling;
  see 
  Section~\ref{sec:glenside}
  for a description 
  of matrix multiplication.
  
\subsubsection*{2D Convolution}

2D convolution (\ctd{})
  is a core kernel
  in deep learning,
  defined element-by-element 
  over tensors storing
  activations $A$,
  strides $S$, and
  weights $W$ as: 
\begin{equation*}
\begin{split}
\mbox{out}&[n, o, x, y] =\\
\sum_{dx, dy, c}&
    (A[n, c, S[0] \cdot x  + dx, S[1] \cdot y + dy] \
    \cdot W[o, c, dx, dy])
\end{split}
\end{equation*}
where
  $n$ indexes the output batch,
  $o$ indexes output channels,
  $x$/$y$ index spatial dimensions,
  $dx$/$dy$ index
    the convolutional window spatial dimensions,
  and $c$ indexes input channels.
2D convolution
  slides each of the $o$
  filters
  of shape $(c, dx, dy)$
  through each possible
  $(c, dx, dy)$--shaped window
  of the input images.
At each of these locations,
  an elementwise multiplication
  and reduction sum
  is computed.

The \g specification
  of \ctd{}
  is shown in 
  Figure \ref{fig:conv2d}.
We access
  the \texttt{weights}
  as a vector of $O$ filters
  and the \texttt{activations}
  as a vector of $N$ images.
We leave the filters as they are,
  but form windows
  of shape
  $(C, K_h, K_w)$
  over the activations
  using the \texttt{windows}
  access pattern transformer
  (Table~\ref{tab:access-pattern-transformers}).
This produces an access pattern
  of shape
  \accesspatternshape
  {N, 1, H', W'}
  {C, K_h, K_w},
  i.e.,
  a batch of ``images''
  of new spatial shape
  $(H', W')$,
  where every location
  is a window of
  the original input.
Finally,
  we take the Cartesian product
  of the filters
  and the windows,
  compute their dot product,
  and \texttt{squeeze} and \texttt{transpose}
  the output
  into the correct layout.

\subsubsection*{Max Pooling}

Max pooling, commonly used in ML
  to condense intermediate activations (``act'' below),
  is defined as:
\begin{equation*}
\begin{split} 
\mbox{out}&[n, c, x, y] =\\
\max_{dx, dy}&
           (\mbox{act}[n, c,
                       \mbox{strides}[0] \cdot x  + dx,
                       \mbox{strides}[1] \cdot y + dy])
\end{split}
\end{equation*}

Max pooling 
  slides a window
  of shape $(dx, dy)$
  over all possible locations
  within the spatial (i.e.,~$x$ and $y$)
  dimensions.
At each window location,
  it reduces the window
  to a scalar
  with the $\max$ operator.
The \g specification merely applies
  \texttt{reduceMax} 
  over each two-dimensional window.
  
  
\subsubsection*{Discussion}\label{section:kernel-implementation-discussion}

\g separates
  the \textit{computation}
  from the \textit{data access patterns}
  in these kernels while exposing
  the simplicity of their computation---%
  and the relative complexity
  of their data access.
In all three kernels,
  the computation can be described
  with a single operator;
  most of the specification
  entails
  setting up the data access pattern.

Furthermore,
  \g exposes similar structure
  between kernels;
  for example,
  both \ctd
  and matrix multiplication
  feature the expression
  \tcd{(compute dotProd (cartProd ...))}.
  
At their core, these kernels
  are performing the same computation,
  but with different patterns
  of data access.
In Section~\ref{sec:discovering-im2col},
  we exploit this similarity in structure
  when mapping kernels to hardware.
  
These kernels highlight the expressive power
  of access patterns.
Consider the use of 
  \tcd{windows}
  in \ctd
   and max pooling.
Both kernels
  form windows
  differently:
  \ctd forms three-dimensional
  windows
  over the channels, height, and width
  dimensions,
  while max pooling forms two-dimensional windows
  over the height and width.
Rather than passing configuration parameters to \tcd{windows},
  \g attaches this information to the tensors themselves.
  
  
\begin{figure}
\begin{lstlisting}[escapechar=!]
(compute dotProd (cartProd ?a0 ?a1)) 
  !$\Longrightarrow$! (systolicArray ?rows ?cols ?a0 (access (transpose ?a1 (list 1 0)) 0))
!$\textrm{where \texttt{?a0} is of shape $((\mbox{\texttt{?batch}}), (\mbox{\texttt{?rows}}))$ and \texttt{?a1} is of shape $((\mbox{\texttt{?cols}}), (\mbox{\texttt{?rows}}))$}$!
\end{lstlisting}
\caption{Our rewrite rewriting matrix multiplication to a systolic array invocation.}
    \label{fig:systolic-array-rewrite}
\end{figure}
  
\begin{figure}
\begin{lstlisting}[escapechar=!]
?a !$\Longrightarrow$! (reshape (flatten ?a) ?shape) 

(cartProd (reshape ?a0 ?shape0) (reshape ?a1 ?shape1))
  !$\Longrightarrow$! (reshape (cartProd ?a0 ?a1) ?newShape)
  
(compute dotProd (reshape ?a ?shape)) 
  !$\Longrightarrow$! (reshape (compute dotProd ?a) ?newShape)
\end{lstlisting}
\caption{Rewrites enabling the discovery of the \itc transformation.}
\label{fig:im2col-rewrites}
\end{figure}

\subsection{Mapping \tcd{matMul} to Accelerators}
\label{sec:case-study-tensorization}

\g can be used to uncover opportunities
  to invoke accelerator components---%
  indeed, this is exactly how \g
  is used
  within \TLA.
Consider a 
  weight-stationary systolic array,
  a common matrix multiplication
  architecture.
A weight-stationary
  systolic array
  with $r$ rows
  and $c$ columns
  takes two lists
  of length-$r$ vectors
  (the activations
    and weights, respectively),
  pairing each vector
  from one list
  with each vector
  from the other,
  and computes a dot product
  over each pair.
The second list
  contains $c$ vectors,
  while the first
  can be of any length.

As discussed in \cref{sec:glenside-in-3la},
  \g's purity
  allows us to implement this hardware mapping task
  using a term rewriting system,
  in which we rewrite a matching program pattern
  to an invocation of our systolic array.
Our rewrite is shown in 
  Figure~\ref{fig:systolic-array-rewrite},
  mimicking
  \tcd{egg}'s rewrite syntax.
Tokens starting with a question mark
  (such as \texttt{?a0} in 
  Figure~\ref{fig:systolic-array-rewrite})
  are variables in the pattern,
  bound by the left-hand side (LHS),
  and then used on the right-hand side (RHS).
\tcd{egg} also allows for
  conditions on rewrites,
  which we print below our rewrites.

To design our rewrite,
  we first must design
  the LHS
  to match program patterns
  that resemble the data access pattern
  and compute pattern
  of our systolic array.
\g is eminently suitable for this task,
  as it can express
  exactly the data access
  and computation pattern
  we described
  for the systolic array.
Pairing all vectors from one list
  with all vectors from another
  and computing the dot product
  of the pairs
  is represented as
  \tcd{(compute dotProd (cartProd ?a0 ?a1))},
  binding
  \tcd{?a0}
  and \tcd{?a1}
  to the input access patterns.
We encode
  the systolic array's
  constraints
  on the input shapes
  as a condition on the rewrite.
Patterns which match the LHS
  are mapped to the RHS;
  in this case, we introduce a new
  \tcd{systolicArray} construct
  to represent the functioning of our systolic array.
The shape of the systolic array 
  is given by the \tcd{?rows} and \tcd{?cols}
  parameters,
  and the inputs are given
  as access patterns.
Note how we also transform
  the second access pattern
  to more accurately convey
  how the actual systolic array
  hardware
  accesses the weight tensor:
  it reads it all at once
  (hence, \tcd{(access ... 0)}),
  and expects it to be laid out
  in transposed form
  in memory.
This added information---%
  enabled by \g's access patterns---%
  provides richer data layout information,
  potentially helping future rewrites
  or code generation steps.
 
\subsection{Flexible Mapping: Discovering \itc{}}
\label{sec:discovering-im2col}

%

\begin{figure}
\begin{lstlisting}[escapechar=!]
(transpose                   
 (squeeze                    
  (reshape           !\color{gray}; \hspace{2mm}\accesspatternshape{N,1,H',W',O}{}!
   (compute dotProd  !\color{gray}; \hspace{2mm}\accesspatternshape{N \cdot 1 \cdot H' \cdot W',O}{}!          
    (cartProd                 
     (flatten        !\color{gray}; \hspace{2mm}\accesspatternshape{N \cdot 1 \cdot H' \cdot W'}{C \cdot K_h \cdot K_w}!
      (windows (access activations 1)  
               (shape C Kh Kw) (shape 1 Sh Sw)))
     (flatten        !\color{gray}; \hspace{2mm}\accesspatternshape{O}{C \cdot K_h \cdot K_w}!
      (access weights 1))))
   ?shape) 1) (list 0 3 1 2))
     \end{lstlisting}
     \vspace{-1em}
    \caption{An \tcd{im2col}-transformed 
      \ctd,
    after the application of the rewrites
    in Figure~\ref{fig:im2col-rewrites}
    and just before the application
    of the systolic array rewrite.}
    \label{fig:conv2d-im2col-rewritten}
\end{figure}
  
The \tcd{im2col} transformation
  is a data layout optimization
  which enables computing \ctd
  on matrix multiplication hardware.
The transformation
  involves instantiating
  the convolutional windows
  over the input activations
  directly in memory~\cite{im2col}.
This leads to data duplication,
  but the resulting speedup
  more than offsets that overhead.
In this case study,
  we show how a few
  general rewrites
  within \g
  lead to the 
  \textit{automatic rederivation}
  of the 
  \tcd{im2col} transformation.

\g's representation underscores
  the structural similarity
  between \ctd and matrix multiplication,
  reflected also by the shared
  \tcd{(compute dotProd (cartProd ...))}
  between \ctd 
  and the LHS of the systolic array rewrite
  in Figure~\ref{fig:systolic-array-rewrite}.
Using this rewrite 
  on \ctd would permit mapping it to the systolic array; 
  however, 
  the restrictions
  on the shape of
  \texttt{?a0}
  and \texttt{?a1}
  prevent its application.
The systolic array has 
  an activation access pattern
  of shape \accesspatternshape
  {a}{b}
  and a weight access pattern
  of shape \accesspatternshape
  {c}{d},
  while \ctd operates over
  access patterns
  of shape \accesspatternshape
  {N, 1,H',W'}{C, K_h, K_w}
  and
  of \accesspatternshape
  {O}{C, K_h, K_w},
  respectively.
Transforming the access pattern
  into a lower-dimensional form
  would enable the systolic array rewrite.
  
Figure~\ref{fig:im2col-rewrites}
  shows the rewrites
  which enable this transformation.
We call the first rewrite
  an 
  \textit{exploratory} rewrite
  as it
  optimistically matches
  any access pattern expression.
It flattens 
  an access pattern
  and immediately reshapes it
  back to its original shape, thus preserving equality
  (see Table~\ref{tab:access-pattern-transformers}
  for formal definitions).
This exploratory rewrite introduces the flattening
  necessary
  to convert the higher-dimensional access patterns
  of \ctd
  into the access patterns
  matched by the systolic array rewrite.
However, the \tcd{reshape} operator
  will still need to be moved 
  before we can fire 
  Figure~\ref{fig:systolic-array-rewrite}'s
  systolic array rewrite.
The second and third rewrites
  in Figure~\ref{fig:im2col-rewrites}
  take care of this;
  they implement \textit{composition commutativity}
  of \tcd{reshape}
  with \tcd{cartProd} and \tcd{compute dotProd},
  which ``bubble'' \tcd{reshape} operators
  up and out of expressions.
These rewrites
  express general properties of these operators
  and are not specific
  to this task.
  
These three rewrites work in concert 
  to map \ctd
  to a systolic array.
First,\footnote{
Since equality saturation 
  explores rewrites non-destructively, 
  the rewriting order here
  is purely for explanatory purposes.
}
  the exploratory rewrite
  flattens and reshapes
  all access pattern expressions.
This includes the inputs
  to \ctd's \tcd{cartProd}
  subexpression,
  which are flattened
  to shapes
  \accesspatternshape
  {N \cdot 1 \cdot H' \cdot W'}{C \cdot K_h \cdot K_w}
  and
  \accesspatternshape
  {O}{C \cdot K_h \cdot K_w}
  and reshaped
  back to their original shapes.
Next,
  the composition commutativity rewrites
  for \tcd{cartProd} 
  and
  \tcd{compute dotProd}
  fire in sequence,
  bubbling the \tcd{reshape} up
  through the 
  \tcd{cartProd} 
  and \tcd{dotProd} 
  expressions (shown in Figure~\ref{fig:conv2d-im2col-rewritten}).
Finally,
  the systolic array rewrite
  completes the \tcd{im2col} transform.
\g's equality saturation based rewrite engine
  discovers these rewrites
  as the exploratory rewrite 
  fires on every term
  and no rewrites are missed
  due to the absence of phase ordering.

This example highlights how,
  \textit{with straightforward,
  generally applicable rewrites
  defined over \g},
  equality saturation
  can emergently discover useful transformations
  that previously required
  expert insights to apply.

\subsection{Flexible Mapping: \tcd{matMul} Blocking}
\label{sec:case-study-blocking}


%

\begin{figure}
\begin{lstlisting}[escapechar=!]
?a !$\Longrightarrow$! (concat (slice ?a ?dim ?b0 ?b1) (slice ?a ?dim ?b1 ?b2) ?dim)
               
(cartProd ?a (concat ?b0 ?b1 ?dim)) 
  !$\Longrightarrow$! (concat (cartProd ?a ?b0) (cartProd ?a ?b1) ?newDim)
!$\textrm{if \texttt{?dim} is an access dimension}$!
 
(cartProd (concat ?a0 ?a1 ?dim0) (concat ?a2 ?a3 ?dim1)) 
  !$\Longrightarrow$! (concat (cartProd ?a0 ?a2) (cartProd ?a1 ?a3) ?newDim)
!$\textrm{if \texttt{?dim0} and \texttt{?dim1} are the same shape dimension}$!

(compute dotProd (concat ?a0 ?a1 ?dim)) 
  !$\Longrightarrow$! (concat (compute dotProd ?a0) (compute dotProd ?a1) ?dim)
!$\textrm{if \texttt{?dim} is an access dimension}$!

(compute dotProd (concat ?a0 ?a1 ?dim)) 
  !$\Longrightarrow$! (compute reduceSum (pair (compute dotProd ?a0) (compute dotProd ?a1)))
!$\textrm{if \texttt{?dim} is a shape dimension}$!
\end{lstlisting}
\vspace{-1em}
\caption{
Rewrites for blocking \tcd{matMul}.
}
\label{fig:all-blocking-rewrites}
\end{figure}

\begin{figure}
\begin{lstlisting}[escapechar=!]
(concat                          !\color{gray}; \hspace{2mm}\accesspatternshape{32,32}{}!
 (concat                         !\color{gray}; \hspace{2mm}\accesspatternshape{16,32}{}!
  (compute reduceSum             !\color{gray}; \hspace{2mm}\accesspatternshape{16,16}{}!
   (pair                         !\color{gray}; \hspace{2mm}\accesspatternshape{16,16}{2}!
    (compute dotProd             !\color{gray}; \hspace{2mm}\accesspatternshape{16,16}{}!
     (cartProd                   !\color{gray}; \hspace{2mm}\accesspatternshape{16,16}{2, 16}!
      (slice                     !\color{gray}; \hspace{2mm}\accesspatternshape{16}{16}!
       (slice (access activations 1) 0 0 16) 1 0 16)
      (transpose                 !\color{gray}; \hspace{2mm}\accesspatternshape{16}{16}!
       (slice 
        (slice (access weights 1) 0 0 16) 1 0 16)
       (list 1 0))))
    (compute dotProd             !\color{gray}; \hspace{2mm}\accesspatternshape{16,16}{}!
     (cartProd                   !\color{gray}; \hspace{2mm}\accesspatternshape{16,16}{2, 16}!
      (slice                     !\color{gray}; \hspace{2mm}\accesspatternshape{16}{16}!
       (slice (access activations 1) 0 16 32) 1 0 16)
      (transpose                 !\color{gray}; \hspace{2mm}\accesspatternshape{16}{16}!
       (slice                    !\color{gray}; \hspace{2mm}\accesspatternshape{16}{16}!
        (slice (access weights 1) 0 16 32) 1 0 16)
       (list 1 0)))))) ...
  \end{lstlisting}
  \vspace{-2em}
  \caption{A $32\times32$ \texttt{matMul}
  blocked into $16\times16$ \texttt{matMul}s.
  Only two of the eight total multiplications are shown.
  }
  \label{fig:matmul-rewritten}
  \vspace{-1em}
\end{figure}

Equality saturation 
  can also be used with \g 
  to emergently discover a
  matrix multiplication
  blocking scheme.
Matrix multiplication blocking
  is the common strategy
  of breaking up a single, large
  matrix multiplication
  into smaller multiplications,
  by multiplying subsets
  of the input matrices
  and assembling the results
  to form the output matrix.
This is essential in practice,
  as systolic arrays are small
  (often between $16\times16$ and $256\times256$)
  while matrices in ML and HPC applications
  can be much larger.

As in Section~\ref{sec:discovering-im2col},
  this transformation follows
  from an exploratory rewrite
  and associated ``cleanup'' rewrites.
The exploratory rewrite used for blocking
  is shown at the top of Figure~\ref{fig:all-blocking-rewrites}.
Given an access pattern,
  this rewrite slices the access pattern
  into two pieces
  along a dimension
  and then concatenates them back together.
The dimension
  as well as the division strategy
  are configurable.
For this example,
  we assume for simplicity
  that we run this rewrite
  on every available dimension,
  that we divide each dimension
  perfectly in half,
  and that all dimensions are powers of 2 in size.
Figure~\ref{fig:all-blocking-rewrites} gives rewrites
  for bubbling the introduced
  \texttt{concat}
  operators up through the expression,
  namely the compositional commutativity
  of \tcd{concat}
  with \tcd{cartProd}
  and \tcd{compute dotProd}.
Starting from the matrix multiplication
  in Figure \ref{fig:mat-mat-mult},
  assuming input shapes of $(32,32)$,
  the exploratory rewrite first slices and concatenates
  the access patterns
  at the input of \tcd{cartProd}.
Then, using the commutativity rewrites,
  the resulting \tcd{concat}s
  are bubbled up
  to produce the final expression
  in Figure~\ref{fig:matmul-rewritten}.
The effect of these rewrites
  is that the single 
  $32\times 32$ \tcd{matMul}
  becomes eight separate $16\times 16$ \tcd{matMul}s,
  which are
  summed
  and concatenated
  to form the full output matrix.
This case study demonstrates
  yet again
  that \g's expressiveness
  allows a small set
  of rewrites
  to produce interesting and useful
  emergent transformations.
  



\section{Evaluation as Part of 3LA}
\label{secion:eval-3la}

Next, we evaluate \g 
  by way of evaluating \TLA.
We first begin by describing
  the evaluation setup of \TLA:
  the accelerators we compile to in
  \cref{sec.eval-acc},
  and the workloads we compile in \cref{sec.eval-app}.
This section is drawn from Huang et al.~\cite{huang2024application}.

\subsection{Target Accelerators}
\label{sec.eval-acc}

We added support for three deep learning accelerators that provide hardware operators at different levels of granularity:
\begin{enumerate}[leftmargin=*]

\item \textbf{FlexASR} is an accelerator for speech and natural language processing (NLP) tasks that supports various RNNs~\cite{tambe20219}.
It uses a custom numeric data type called \textit{AdaptivFloat} for boosting the accuracy of quantized computations~\cite{tambe2020algorithm}.
%
%
%


\item \textbf{HLSCNN} is an accelerator optimized for 2D convolutions~\cite{whatmough201916nm}.
%
%
It operates on mixed 8/16-bit fixed point data
(8 bits for storing weights and 16 bits for computations).
%

\item \textbf{VTA} is a parameterizable accelerator for tensor operations featuring a processor-like design~\cite{moreau2019hardware}.
%
It supports element-wise arithmetic operations as well as generalized matrix multiplication,
operating on 8-bit integer data.
%

\end{enumerate}
%
For each accelerator, we defined an ILA model and a set of \mapping rules
  such as those described in
  \cref{sec:glenside-in-3la,sec:case-study-tensorization}.
The ILA models for FlexASR, HLSCNN, and VTA are approximately 5600, 1600, and 2100 lines of ILAng code (C++), respectively. 
The high-level synthesis (HLS) implementations
  of the accelerators are about 9300 (SystemC), 5100 (SystemC), and 6900 (Chisel) LoC, 
  respectively;
  the ILA specifications are thus of modest size,
  compared even to the relatively compact HLS implementations. 
For each \mapping rule, we represent the compiler side in Glenside IR, and the accelerator side 
as a program composed of ILA instructions (in a Python-embedded DSL). 
The total size of mapping rules (both the compiler and accelerator sides) for FlexASR (5 mappings), HLSCNN (1 mapping), and VTA (1 mapping) was 186, 22, and 49 LoC, respectively.
Recall that these mappings are polymorphic over tensor size on both sides, leading to general and compact representations. 
Additionally, the BYOC-based code generators and runtimes for these accelerators are approximately 450, 300, and 900 LoC of C++, 
respectively. 
These indicate the implementation of the code generation module in our prototype, as well as reusable utilities for data movement, handling custom numerics, and emitting the low-level MMIO code for each selected accelerator offload for end-to-end simulation of the application.
This reduction in total lines of code
  between \g and BYOC mapping rules
  represents a significant
  decrease in \cref{thesis:devtime}
  for those seeking to build a compiler
  for their accelerator.

%
%

\subsection{Target Applications}
\label{sec.eval-app}



We considered \AppNum DL applications corresponding to common neural network models for language and vision tasks that contain operators supported by the three target accelerators.
We selected applications with reasonable size for human inspection and in-depth analysis.
%
%
\begin{enumerate}[leftmargin=*]


\item \textbf{EfficientNet} is a recent convolutional neural network (CNN) designed for image classification~\cite{tan2019efficientnet}. 
It has convolutions that are supported by VTA and HLSCNN.



\item \textbf{LSTM-WLM} is a text generation application~\cite{pt2020wlm} implemented using an LSTM recurrent neural network architecture~\cite{graves2014towards}. The LSTM layer in this model is supported by FlexASR.


\item \textbf{MobileNet-V2} is a common CNN designed for mobile applications~\cite{howard2017mobilenets, sandler2019mobilenetv2}. We chose MobileNet-V2 due to its wide use, especially on embedded devices.

\item \textbf{ResMLP} is a recent residual network for image classification, comprised only of multi-layer perceptrons ~\cite{touvron2021resmlp}. Its linear layers could be accelerated by VTA and FlexASR.


\item \textbf{Transformer} is an NLP model comprised primarily of attention mechanisms~\cite{vaswani2017attention}. 
We chose Transformer as a representative of recent popular NLP models. 

\item \textbf{ResNet} is a popular CNN designed for image classification~\cite{he2016deep}. 
Besides ResNet-20, which we use in most of the evaluation, in \S\ref{sec.compilation-stats}, we additionally compare various implementations of ResNet-50 from MLPerf~\cite{mlperf} for its availability of different reference implementations.


\end{enumerate}
%
All applications were mapped to accelerators \emph{without any manual modifications}.
\begin{table*}
  \centering
  \caption{
  \textbf{End-to-end compilation statistics.}
  The total number of Relay operators (row 3) is given as a proxy
  for program complexity.
  In rows 4-6, we include rewrites for only one accelerator at a time; we do not offload to multiple accelerators at once like in \S\ref{sec.end-to-end}. Flexible matching identifies significantly more offloads than exact matching.
  Abbreviations: MN: MobileNet, Trans.: Transformer, and TF: TensorFlow.
  }
  \label{tab.compilation}
  \resizebox{\textwidth}{!}{
  \begin{tabular}{|clrrrrrr||rrr|}
  \hline
  \multicolumn{11}{|c|}{Application Statistics} \\ \hline
  \multicolumn{1}{|c|}{1} &
    \multicolumn{1}{|l|}{Application} &
    \multicolumn{1}{c|}{EfficientNet} &
    \multicolumn{1}{c|}{LSTM-WLM} &
    \multicolumn{1}{c|}{MN-V2} &
    \multicolumn{1}{c|}{ResMLP} &
    \multicolumn{1}{c|}{Trans.} &
    \multicolumn{1}{c||}{ResNet-20} &
    \multicolumn{3}{c|}{ResNet-50}
    \\ \hline
  \multicolumn{1}{|c|}{2} &
    \multicolumn{1}{|l|}{Source DSL} &
    \multicolumn{1}{c|}{MxNet} &
    \multicolumn{1}{c|}{PyTorch} &
    \multicolumn{1}{c|}{PyTorch} &
    \multicolumn{1}{c|}{PyTorch} &
    \multicolumn{1}{c|}{PyTorch} & 
    \multicolumn{1}{c||}{MxNet} &
    \multicolumn{1}{c|}{PyTorch} &
    \multicolumn{1}{c|}{ONNX} &
    \multicolumn{1}{c|}{TF}
    \\ \hline
  \multicolumn{1}{|c|}{3} &
    \multicolumn{1}{|c|}{\#Relay Ops} &
    \multicolumn{1}{c|}{232} &
    \multicolumn{1}{c|}{578} &
    \multicolumn{1}{c|}{757} &
    \multicolumn{1}{c|}{343} &
    \multicolumn{1}{c|}{872} &
    \multicolumn{1}{c||}{494} &
    \multicolumn{1}{c|}{709} &
    \multicolumn{1}{c|}{194} &
    \multicolumn{1}{c|}{609}
    \\ \hline \hline
  \multicolumn{11}{|c|}{Number of Static Accelerator Offloads Identified Using Exact Matching/Flexible Matching} \\ \hline
  \multicolumn{1}{|c|}{4} &
    \multicolumn{1}{|l|}{FlexASR} &
    \multicolumn{1}{r|}{0/35} &
    \multicolumn{1}{r|}{1/1} &
    \multicolumn{1}{r|}{0/41} &
    \multicolumn{1}{r|}{0/38} &
    \multicolumn{1}{r|}{0/66} &
    \multicolumn{1}{r||}{2/22} &
    \multicolumn{1}{r|}{0/54} &
    \multicolumn{1}{r|}{0/54} &
    \multicolumn{1}{r|}{0/54}
    \\ \hline 
  \multicolumn{1}{|c|}{5} &
    \multicolumn{1}{|l|}{HLSCNN} &
    \multicolumn{1}{r|}{35/35} &
    \multicolumn{1}{r|}{0/0} &
    \multicolumn{1}{r|}{40/40} &
    \multicolumn{1}{r|}{0/0} &
    \multicolumn{1}{r|}{0/0} &
    \multicolumn{1}{r||}{21/21} &
    \multicolumn{1}{r|}{53/53} &
    \multicolumn{1}{r|}{53/53} &
    \multicolumn{1}{r|}{0/53}
    \\ \hline 
  \multicolumn{1}{|c|}{6} &
    \multicolumn{1}{|l|}{VTA} &
    \multicolumn{1}{r|}{0/35} &
    \multicolumn{1}{r|}{36/36} &
    \multicolumn{1}{r|}{1/41} &
    \multicolumn{1}{r|}{38/38} &
    \multicolumn{1}{r|}{66/66} &
    \multicolumn{1}{r||}{0/22} &
    \multicolumn{1}{r|}{0/24} &
    \multicolumn{1}{r|}{0/24} &
    \multicolumn{1}{r|}{0/24}
    \\ \hline 
  \end{tabular}
  }
  \end{table*}
\subsection{Identifying Acceleration Opportunities}
\label{sec.compilation-stats}

%
We took the \AppNum DL applications, developed by different teams in different DSLs, and compiled them for the three target accelerators.
%
Our compiler successfully generated code that exploits the accelerators for supported computations---%
  thus improving over BYOC along the axis of \cref{thesis:optimizations}.
%


Table~\ref{tab.compilation} shows the compilation statistics of using exact matching and flexible matching.
%
%
%
Note that some accelerator operators correspond to multiple Relay operators; in particular, 
the LSTM RNN in LSTM-WLM corresponds to 566 Relay operators 
and maps to \textit{one} FlexASR operator, which shows \TLA effectively overcoming a dramatic granularity mismatch between the compiler IR and accelerator operators.
%
%
%
%

Our results demonstrate \TLA's viability across a range of DL applications and accelerators with the successful identification of acceleration opportunities
%
and provide
evidence for the utility of flexible matching. 
%
%
%
For example, the linear layer rewrite described in \cref{sec:glenside-in-3la}
  resulted in 66 invocations of FlexASR's linear layer
  in Transformer and 38 in ResMLP, in comparison to exact matching that produced no match. 
Furthermore, 
  the \texttt{im2col} transformation described in
  \cref{sec:discovering-im2col}
  rewrites 2D convolutions into matrix multiplications;
for VTA, this resulted in \emph{additional} 35 invocations in EfficientNet, 22 in ResNet-20, and 40 in MobileNet-V2.
Hence, flexible matching allowed us to support 2D convolutions on VTA even when 
there is no \mapping that maps 2D convolutions to VTA instructions.
Another rewrite
  that turns lone matrix multiplications
  into linear layers
  (by a zero-vector bias)
  works in concert with the \texttt{im2col} rewrites,
  resulting in offloads of 2D convolutions
  onto FlexASR  
  in EfficientNet, MobileNet-V2, and ResNet-20---%
  thus allowing an accelerator for NLP applications
  to also accelerate vision applications.
%
Note that these additional acceleration opportunities
  were identified automatically
  and are examples of \textit{emergent effects} resulting from simple, reusable (accelerator-agnostic) compiler IR rewrite rules.

We additionally
  evaluate the robustness
  of flexible matching
  by comparing
  the three implementations of ResNet-50
  from MLPerf~\cite{reddi2020mlperf}
  in Table~\ref{tab.compilation}, right.
Their Relay representations 
  differed in subtle ways (such as in reshaping operators)%
  \footnote{
For example, the TensorFlow implementation
  takes data in NHWC format rather than NCHW;
  Glenside can rewrite convolutions to use NCHW.}
  and are reflected in the difference in results of exact matching.
Flexible matching found the same (increased) number of matches for each accelerator, regardless of its source DSL.
All of these results speak to the increased
  \cref{thesis:optimizations}
  found by utilizing \g.
\subsection{Per-Operator Evaluation}



Having now evaluated \cref{thesis:devtime}
  and \cref{thesis:optimizations},
  we will evaluate \cref{thesis:correctness}.
We do so by demonstrating how \g, via \TLA,
  enables testing of accelerators.
We begin by evaluating single operators. 
Although evaluating individual operators
  does not suffice to characterize
  how an accelerator performs on a full application,
  it is a basic first step
  and provides 
  insights on the identified acceleration opportunities.
Here, we discuss both functional validation and performance evaluation at the operator level.

\subsubsection{Functional Validation}

\begin{table}
\caption{
\textbf{
Simulation-based validation results for checking {\mapping}s.
}
The average relative error (Avg. Err.) and the standard deviation (Std. Dev.) of errors are measured 
over 100 test inputs. 
For VTA, 
there was no error because the host supports 
8-bit integer operations.
}
\label{tab.layer-sim}
\centering
\begin{small}
\begin{tabular}{|c|l|l|r|r|}
\hline
  & Accel. & Operation & Avg.Err. & Std.Dev.  \\
  \hline \hline
  1 & VTA & All ops & 0.00\% & 0.00\% \\ 
  2 & HLSCNN & Conv2D & 1.78\% & 0.16\% \\ 
  3 & FlexASR & LinearLayer & 0.84\% & 0.29\% \\ 
  4 & FlexASR & LSTM & 1.21\% & 0.19\% \\ 
  5 & FlexASR & LayerNorm & 0.27\% & 0.20\% \\ 
  6 & FlexASR & MaxPool & 0.00\% & 0.0\% \\ 
  7 & FlexASR & MeanPool & 1.79\% & 0.28\% \\ 
  8 & FlexASR & Attention & 4.22\% & 0.09\% \\ 
\hline
\end{tabular}
\end{small}
\end{table}

The {\TLA} methodology readily enables operator-level validation through auto-generated ILA simulators.
In our experiments, we 
compared the outputs of the accelerator ILA simulator and those of TVM's runtime on host.
The accelerator ILA simulators precisely model the data types used by the accelerators.
%
%
For the reference results (TVM's runtime), we use 8-bit integer for comparing against VTA and 32-bit floating point for the other accelerators, as these are the closest host processor data types to those used by the accelerators. 
We measure the relative errors by using the standard Frobenius Norm~\cite{lapack} for the tensors based on the reference and accelerator generated output values as follows:
$Error = \| Out_{ref} - Out_{acc} \|_F / \| Out_{ref} \|_F$.

We validated all types of operators supported by the target accelerators.
Table~\ref{tab.layer-sim} shows a representative subset of the validation results: four {\mapping}s (Rows~1-4) that are used in the full application compilation (Table~\ref{tab.compilation}) and four additional mappings for non-trivial operations (Rows~5-8).
%
%
%
Note that some mappings introduce no numerical differences; e.g., the TVM runtime supports 8-bit integer execution, so the results for VTA match perfectly.
%
For other mappings, we see deviations caused by the custom numerics, especially for complex operators such as the attention operator on FlexASR.  
Such deviations should be carefully assessed
  in the context of application-level validation,
  as even small deviations could accumulate and affect the final accuracy.


\subsubsection{Performance Evaluation}

We also evaluated the performance gain of offloading operations from the host to accelerators 
using cycle counts as the performance metric, since we did not have clock frequencies for an SoC containing the host and accelerators.
For accelerators, we derived the cycle counts based on their cycle-accurate models
(VTA's Chisel model
and FlexASR's and HLSCNN's SystemC models).
For the host, we measured averaged cycle counts (1000 random inputs) in TVM's runtime
on one pinned EPYC-7532 core.

\begin{figure}
  \centering
  \includegraphics[width=.65\linewidth]{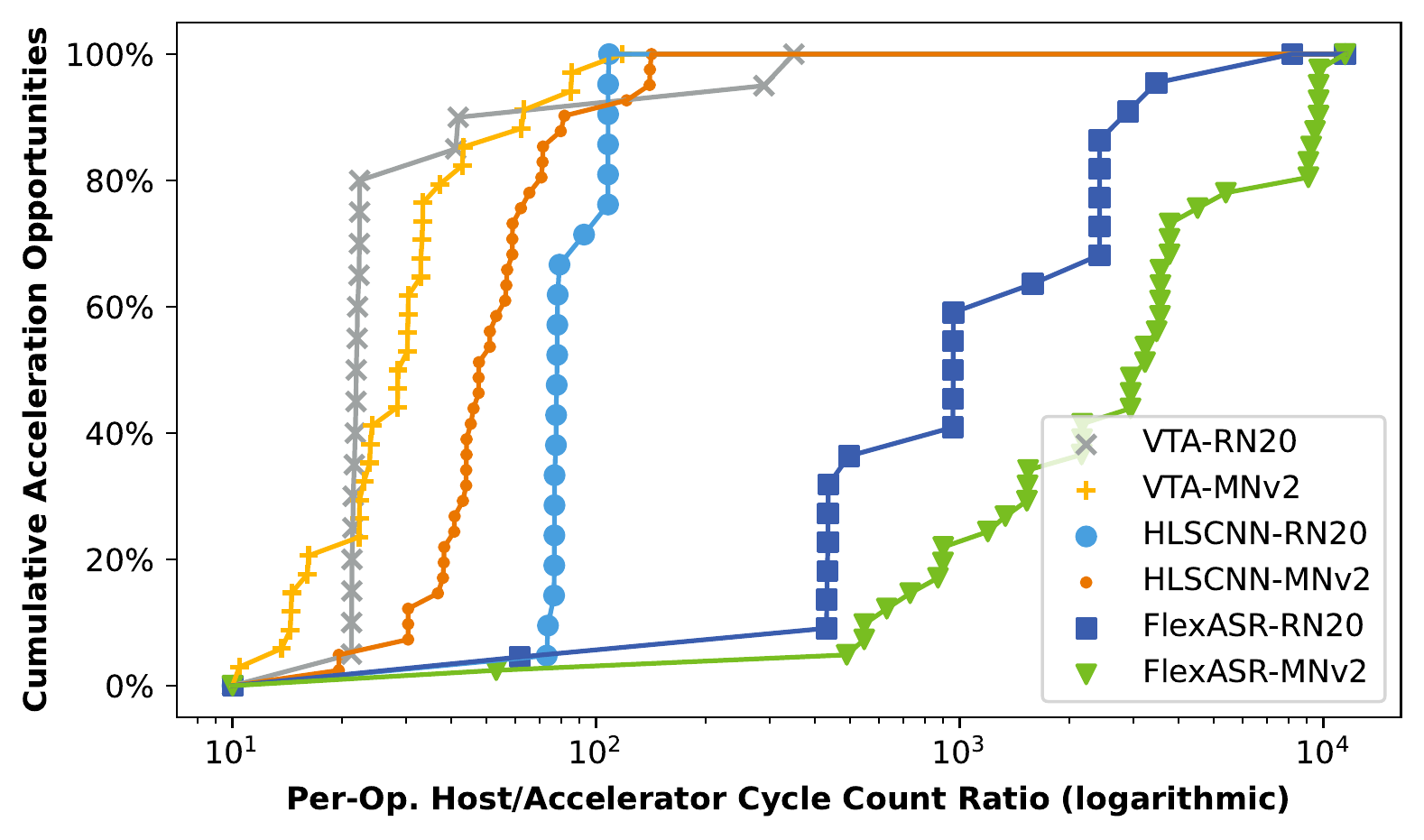}
  \caption{
\textbf{
Cumulative distribution of per-operator performance gains of all identified acceleration opportunities} in ResNet-20 (RN20) and MobileNet-V2 (MNv2) on the three accelerators.
Each point represents an operation offloaded from the host to the accelerator (as identified by flexible matching, Table~\ref{tab.compilation}).
The $x$-axis shows the host-to-accelerator cycle count ratio of each offloaded operation 
and the $y$-axis shows the cumulative distribution of offloaded operations. Points and plots more to the right are better; e.g., coarse-grained operators, supported with higher parallelism in FlexASR, offer greater speedup compared to the fine-grained operators in VTA. 
}
  \label{fig.performance}
\end{figure}

Fig.~\ref{fig.performance} shows the performance gains (ratio of host to accelerator cycles) of all identified acceleration opportunities in ResNet-20 and MobileNet-V2 when operations are offloaded from the host to VTA, HLSCNN, and FlexASR, respectively. 
Overall, as expected, all offloads resulted in performance gains relative to the host; we also see that accelerators providing coarser-grained operators (e.g., FlexASR), supported with higher parallelism, achieve higher performance gain per operator compared to finer-grained accelerators like VTA. 
\subsection{Application-Level Validation Through Co-Simulation}
\label{sec.end-to-end}

We performed application-level co-simulation
  by using the
  ILAng-generated simulators for accelerator computations
  and the host CPU for the rest of the computation.
%
%
We considered three applications,
  which between them provide opportunities to use
  each of the three accelerators: 
\begin{inlinelist}
  \item LSTM-WLM, where we accelerate linear layer and LSTM operations on FlexASR;
  \item ResNet-20, where we accelerate convolutions on HLSCNN and linear layers on FlexASR; and
  \item MobileNet-V2, where we accelerate convolutions and linear layers as in ResNet-20 and additionally accelerate both these operations on VTA (due to the \texttt{im2col} rewrites).
\end{inlinelist}
In ResNet-20 and MobileNet-V2,
  we were able to
  \emph{explore using HLSCNN and FlexASR together and separately}, simply by varying which 
  {\mapping}s
  we included in flexible matching.
Note again that this validation
  would not have been possible without the mapping power
  of \g.

We trained and validated the LSTM-WLM model using the WikiText-2 dataset~\cite{merity2016pointer}.
The image classification models (MobileNet-V2 and ResNet-20) were trained and validated using the CIFAR-10 dataset~\cite{cifar10}.
We additionally trained and validated a MobileNet-V2 model optimized for ImageNet using the ImageNet dataset~\cite{deng2009imagenet}.

Table~\ref{tab.verif-sim} shows the application-level co-simulation results.
%
%
%
%
%
%
For LSTM-WLM,
  the application-level results using the accelerators
  did not differ greatly
  from the reference results.
In the case of FlexASR,
  this was the \textit{first time}
  it had been run end-to-end on a full application---%
  this provided validation for its AdaptivFloat data type.
For VTA on MobileNet-V2,
  there was a small decrease in accuracy
  that may be attributed
  to quantization error.%
\footnote{We apply a form of uniform quantization~\cite{jacob2017quantization},
  which involves scaling the results
  based on the floating point reference.}

%

\begin{table*}
  \caption{
  \textbf{Application-level co-simulation results.}
  In each test, we evaluated 2000 CIFAR-10 images (for vision tasks) or 100 WikiText-2 sentences (for text generation) that were evenly sampled from the corresponding dataset.
  The reference results were obtained by running tasks in the original frameworks (MxNet for ResNet-20, PyTorch for the rest).
  The {results without numerical tuning} are 
  for the initial accelerator designs, modeled in ILA.
  The {result with numerical tuning}, where provided, were obtained by updating the ILA specifications
  according to 
  design revisions suggested by the accelerator developers.
  We measured the accuracy for image classification tasks (ResNet-20, MobileNet-V2) and perplexity for text generation (LSTM-WLM).}
  \label{tab.verif-sim}
  \centering
  \small
  \begin{tabular}{|l|c|c|c|c|}
  \hline
      \multicolumn{1}{|c|}{\multirow{2}{*}{Application}} &
      \multirow{2}{*}{Processing Platform} &
      \multirow{2}{*}{Reference Result$^\ast$} &
      {Result without} &
      {Result with}\\
    \multicolumn{1}{|c|}{} &
       &
       &
      {Numerical Tuning} &
      {Numerical Tuning} \\
    \hline \hline

  LSTM-WLM & 
    FlexASR & 
    122.15 & 
    121.97 & 
    N/A \\
    \hline


  ResNet-20 & 
    FlexASR &
    91.55\% &
    91.50\% &
    N/A  \\
    
   &
    HLSCNN &
    91.55\% &
    \cellcolor[HTML]{E9CECE}29.75\% &
    \cellcolor[HTML]{DDEFDE}92.10\%  \\
    
   &
    FlexASR \& HLSCNN & 
    91.55\% & 
    \cellcolor[HTML]{E9CECE}29.15\% & 
    \cellcolor[HTML]{DDEFDE}91.85\% \\
    \hline

  MobileNet-V2 &
    VTA &
    92.40\% &
    89.40\% &
    N/A \\
  
   &
    FlexASR &
    92.40\% &
    92.30\% &
    N/A \\

   & 
    HLSCNN & 
    92.40\% & 
    \cellcolor[HTML]{E9CECE}10.35\% & 
    \cellcolor[HTML]{DDEFDE}91.50\% \\
    
   & 
    FlexASR \& HLSCNN & 
    92.40\% & 
    \cellcolor[HTML]{E9CECE}10.35\% & 
    \cellcolor[HTML]{DDEFDE}91.20\% \\


    \hline
  \end{tabular}
  \begin{tablenotes}
    \item $\ast$ The reference result does not represent the best achievable accuracy/perplexity of the model on the given dataset. This table is intended for comparing the application-level results on different processing platforms.
    \item $\dagger$ Average simulation time of running one data point (e.g., an image or a sentence) on an AMD EPYC-7532 core.
  \end{tablenotes}
\end{table*}

However, the initial results for ResNet-20 and MobileNet-V2
  (both CIFAR-10 and ImageNet)
  using HLSCNN
  revealed a large loss in accuracy, 
  {as shown in Column~4 ``Results without Numerics Tuning'' in Table~\ref{tab.verif-sim}}.
We noticed that the linear layers 
  accelerated by FlexASR 
  did not impact the final accuracy,
  suggesting the issue stemmed from HLSCNN
  (for which this was also the first time it was run in an end-to-end application).
We then instrumented
  our {\TLA} prototype 
  to record additional information
  for each accelerator invocation,
  such as input and output ranges.
This helped 
  the accelerator developers
  determine that the loss of accuracy
  was due to a lack of dynamic range in the data type:
  weight data values 
  in HLSCNN's 2D convolutional layers
  were heavily quantized
  due to the narrow value range
  of their 8-bit fixed point representation.
After we updated the ILA specification (a much easier task than modifying the RTL implementation) based on the developers' suggestion to expand the fixed point representation to 16 bits and adjust the binary points in inputs' and accumulators' fixed point data types, the 
accuracy recovered.
{This is shown in Column~5 ``Results with Numerics Tuning'' in Table~\ref{tab.verif-sim}.}
This case study readily demonstrates
  how the {\TLA} methodology,
  enabled by \g,
  \textit{facilitates debugging and improving accelerator designs
  with rapid turnaround,} and thus improving overall
  \cref{thesis:correctness}.

The overall results in Table~\ref{tab.verif-sim} reaffirm the need for application-level validation, especially for accelerators utilizing custom numerics.
%
%
%
Thanks to formal ILA models
  and flexible compilation via \g, 
  \TLA provides quick design space exploration and numerics tuning without hardware engineering overhead 
in each design iteration.
%
%
Further, it provides handy debugging information and efficient simulation---%
for FlexASR, the ILA simulator yields a 30$\times$ speedup on average compared to RTL simulation.
\subsection{System Deployment and FPGA Emulation}
\label{sec.eval-fpga}

As an additional demonstration of {\TLA} and \g, we explored their use in 
  compiling workloads
  to a real hardware platform.
  Specifically, 
  we used our prototype to compile workloads
  to an \gls{fpga} emulation of FlexASR.%
\footnote{We synthesized and placed-and-routed the FlexASR accelerator on a Xilinx Zynq ZCU102 FPGA, which consumed 86\% of the available LUT resources.
Due to the significant engineering overhead of FPGA emulation, FlexASR is the only accelerator we deployed on an FPGA.}
%
We configured our prototype to lower
 FlexASR ILA instructions
 to the corresponding MMIO commands for FlexASR, 
 passing them to the FPGA using the Xilinx SDK~\cite{xsdk}.
%
Next, we compiled and executed synthetic workloads in which LSTM layers and linear layers were offloaded to the FlexASR accelerator.
The results matched those of the ILAng-generated simulator bit for bit, providing validation for 
the custom numerics.
This is a proof of concept for utilizing the {\TLA} methodology for an actual deployment, above and beyond simulation-based testing.
%


\chapter{Background and Related Work}
\label{chapter:part1-background}

In \cref{part:glenside-and-3la}
  of this dissertation,
  we have described an application
  of our thesis
  to the problem of
  generating compilers
  for deep learning accelerators.
We have described 
  the difficulties in building
  compilers
  for custom accelerators.
First, developing a compiler
  requires much
  developer effort
  and expertise
  (\cref{thesis:devtime}).
Second,
  even once a compiler is built,
  optimizations still often
  get left on the table
  (\cref{thesis:optimizations}).
And lastly,
  the difficulties in
  building compilers
  for accelerators
  often makes the process
  of validation
  of accelerators
  hard if not impossible
  for accelerator designers
  (\cref{thesis:correctness}).
We describe how
  we applied our thesis
  to produce
  \g,
  a domain-specific language
  which can be used to 
  automatically compile
  workloads to accelerators.
We used \g to build
  3LA, a new methodology
  for accelerator design.

We now present background
  and related work
  on the various topics
  involved in
  \cref{part:glenside-and-3la}
  of this dissertation.
We start from the basics, discussing
  existing
  machine learning accelerators
  (\cref{sec:part1:relatedwork:accelerators}),
  whose popularity
  is the core motivation
  behind building 3LA and \g.

\section{Machine Learning Accelerators}
\label{sec:part1:relatedwork:accelerators}

A variety of accelerators~\cite{
    jouppi2017tpu, chen2016eyeriss, moreau2018vta, markidis2018tensorcore, nvdla,
    genc2021gemmini}
  have been developed 
  to provide efficient implementations
  of tensor operators for ML applications.
These devices accelerate tensor operators 
  through hardware parallelism, 
  simultaneously applying related operations
  across many tensors in the accelerator's memory (which are often laid out according to custom rules that facilitate hardware optimization).
Tensor program compilers must translate
  expensive application code fragments
  down to accelerator invocations that
  adhere to these layout rules,
  which often involves both
  (a) higher-level transformations like
  tensor reshaping to match accelerator size bounds and
  loop unrolling to expose optimization opportunities, and
  (b) lower-level transformations like
  operator fusion and \tcd{im2col}
  to match accelerator calling conventions and
  even implement different operations
  using the same accelerator,
  e.g., on systolic arrays~\cite{im2col, jia2014semantic}.
  

\section{Validating Hardware Designs}

\Gls{validation} of hardware designs
  is an incredibly important task---%
  so important, in fact, that it often represents
  a majority of the cost
  for a new hardware design.
Validating a hardware design
  is the process of ensuring
  that it behaves as intended.
In the world of hardware design,
  this process is more commonly
  referred to as
  \textit{verification---}%
  however, in this dissertation,
  I prefer to use the definitions
  of \textit{validation} and \textit{verification}
  from the field of Programming Languages,
  in which \textit{validation}
  refers to non-formal-methods-based,
  (usually) non-exhaustive sanity checking,
  while \textit{verification}
  refers to formal, mathematical
  proving of properties (like correctness).
Hardware designers and validation engineers
  perform validation
  at many points in the hardware design process.
For example,
  they might validate that
  their initial prototype of the design,
  written in e.g.~Python or C++,
  behaves identically to their
  initial Verilog implementation
  using a simulator
  such as Verilator~\cite{verilator}.
Later in the design process,
  they might
  formally verify the equivalence
  of their initial Verilog implementation
  against a lower-level,
  backend-specific version of the Verilog
  using an equivalence checker
  like Cadence's Jasper~\cite{jasper},
  or similar tools from Mentor Graphics
  or Synopsys.

Simulation tools can operate
  at different levels of specificity.
Tools like Verilator~\cite{verilator} and Cuttlesim~\cite{pitclaudel2021cuttlesim} enable \textit{cycle-accurate RTL simulation}:
  i.e.~they simulate exactly what the hardware
  is doing
  at each clock cycle.
However,
  cycle-accurate simulation is slow,
  as it requires simulating
  every component within the hardware design.
Often, it is useful to run
  \textit{application-level co-simulation,} 
  in which a high-level software program
  (e.g.~a deep learning model)
  is simulated simultaneously with the hardware.
Co-simulation is integral
  to the 3LA methodology---%
  by running entire applications
  via co-simulation, 3LA helps designers
  find bugs in their hardware designs.
In general, cycle-accurate simulators
  are too slow to run full applications
  in any realistic amount of time,
  making co-simulation infeasible.
In these cases,
  higher-level, non-cycle-accurate simulation
  can enable fast simulation.
SystemC~\cite{SystemC}
  and ILA~\cite{huang2018instruction},
  are common tools for implementing
  these high-level models.
The 3LA framework relies on ILA, which,
  unlike SystemC,
  provides
  a clear formal verification path to RTL.
  
%

\section{Hardware--Software Co-Design}

\textit{Hardware--software co-design}
  refers to a loosely grouped
  set of techniques and ideas
  centered around one primary idea:
  rather than the well-defined separation
  between software and hardware design
  which existed in the past,
  hardware and software
  should instead be designed \textit{together.}
That is, to maximize
  the performance of hardware,
  hardware designers should be able to suggest
  changes to the software
  such that it will be more amenable to acceleration;
  similarly,
  software designers should be able to suggest
  hardware changes
  to enable new algorithms.
\g and the 
  \TLA methodology enable
  hardware--software co-design
  through the simple fact that
  they make
  hardware design iterations quicker.
By making it quicker and easier
  to simulate full end-to-end
  applications
  on prototype hardware,
  3LA and \g enable designers
  to see simulation results more quickly
  and adjust both their design
  and the software running on the hardware.

There exists much other work on
  hardware--software codesign
  for deep learning.
Work on accelerator generation and integration~\cite{
    bahr2020creating, truong2020fault}
  has explored adding support in the Halide~\cite{ragan2013halide}
  compiler flow for specialized Coarse-Grained Reconfigurable Array (CGRA) accelerators.
That work composes an
  impressive array of custom tools to
  generate and verify specialized CGRA accelerators
  and also map Halide program fragments
  down to accelerator invocations.
HeteroCL~\cite{lai2019heterocl} also provides
  a similar custom flow.

\section{Tensor IRs and Compilers}

\g, the primary contribution
  of \cref{part:glenside-and-3la}
  of this dissertation,
  is fundamentally an
  intermediate representation (IR)
  for tensor programs.
On top of the \g IR,
  we build the 3LA methodology,
  which includes a compiler
  utilizing \g
  to map deep learning workloads
  to accelerators.
In this section, we describe
  other IRs and compilers
  for tensor programs.

Machine learning workloads
 are generally viewed
 as a sequence of 
 of \textit{tensor kernel} invocations,
 where tensor kernels
 are large operations
 over multidimensional arrays (tensors)
 such as 2D convolution
 or dense matrix multiplication.
Machine learning frameworks
 (such as TensorFlow \cite{abadi2016tensorflow}
   and PyTorch \cite{pytorch})
 and machine learning compilers
 (such as TVM \cite{chen2018tvm})
 can optimize
 machine learning workloads
 at a number of levels,
 which often result
 in each framework
 having a number of different IRs.
TVM, for example,
 can optimize machine learning workloads
 at a high-level
 using its high-level IR
 Relax~\cite{lai2023relaxcomposableabstractionsendtoend}
 (and its former high-level
   IR Relay \cite{relay}),
 but
 low-level optimizations
 such as loop blocking and reordering
 must be done 
 in its lower-level IRs.


Tensor compilers for ML and HPC applications strive
  to balance clear, high-level operator semantics
  and support for the low-level optimizations
  necessary to target specialized accelerators.
Halide~\cite{halide}
  achieves this balance by separating
  operator \textit{specifications} (what is computed) from
  \textit{schedules} (how, when, and where
  each output element is generated).
This style of separation has proven
  highly effective across both
  application domains and hardware targets;
  numerous compilers including TVM~\cite{chen2018tvm},
  FireIron~\cite{hagedorn2020fireiron},
  LIFT~\cite{lift}, and Accelerate~\cite{accelerate}
  follow variations of this strategy.
  
The specification/schedule separation approach
  allows the same high-level program (specification)
  to be flexibly optimized for and mapped to
  different hardware targets by applying different schedules.
From this perspective,
  schedules represent different rewriting strategies
  to explore various loop ordering and memory layouts;
  in LIFT and Accelerate these
  take the form of functional combinators
  closely related to \g's approach.
As in classic term rewriting,
  experts must often carefully craft
  schedules for each target to achieve
  the best performance and mitigate
  phase ordering challenges~\cite{phase-ordering},
  though recent projects have produced promising results
  in automatic scheduling~\cite{
    chen2018autotvm, zheng2020ansor, anderson2020learning}.

Other tensor IRs like
  TACO~\cite{taco}, Keops~\cite{keops},
  and COMET~\cite{tian2021highperformance}
  rely on \textit{index notation}\footnote{
    Index notation is closely related to
    ``Einstein notation'', in which reduction
    indices are implicit.}
  to concisely express tensor operators
  and simplify optimization by
  uniformly representing
  per-output-element computations.
These approaches also rely on
  rewriting passes to generate
  kernel implementations specialized to
  tensor sparsity/density,
  operator combinations arising in
  the source application, and
  details of the target hardware.
In Section~\ref{sec:matmul} we discuss
  some of the tradeoffs of these approaches
  with respect to other rewriting strategies.
 
Polyhedral compilers~\cite{polyhedral-survey}
  like Tensor Comprehensions~\cite{vasilache2018tensor}
  and Tiramisu~\cite{tiramisu}
  optimize loop-filled programs
  by modeling loop nests as polyhedra
  and applying geometric transformations.
The polyhedral approach exploits
  regular loop structure,
  but is also restricted
  to geometrically affine transformations.
In contrast, term rewriting is
  neither guided nor restricted by
  geometric constraints, making
  these approaches broadly complementary.

Another key piece of background to note
  when it comes to deep learning compiler frameworks
  is interoperability of frontends.
Machine learning models can be expressed
  in many frontend languages,
  including MxNet~\cite{chen2015mxnet},
  PyTorch~\cite{paszke2019pytorch},
  TensorFlow~\cite{abadi2016tensorflow},
  ONNX~\cite{linux2019onnx},
  and CoreML~\cite{apple2022coreml},
  and generally,
  compilers should strive to support
  models expressed in all of these frontends.
As it is built on top of TVM,
  which supports importing from many
  frontend languages,
  our \TLA prototype
  supports all of these frontend languages.

\subsection{Term Rewriting and Equality Saturation}

Term rewriting is a classic
  program optimization technique~\cite{baader1998term}
  that relies on iteratively applying
  rewrite rules of the form $\ell \xrightarrow{} r$:
  when part of a program
  matches the pattern $\ell$
  under substitution $\sigma$,
  it is rewritten into $\sigma(r)$.
This approach is ubiquitous,
  frequently used in both mainstream and DSL compilers
  to implement features including preprocessing,
  strength reductions, and
  peephole optimizations~\cite{garavel2018rewrite-context}.
  
Classic term rewriting systems where
  rewrites are applied destructively suffer
  phase ordering problems~\cite{phase-ordering}:
  the order in which rewrites are applied can
  enhance or severely diminish performance.
Recent work has shown how program synthesis
  can help address this challenge in
  peephole optimizers like Halide's
  scalar expression rewriter~\cite{
    newcomb2020halide-rewrite}.
  
Advances in alternate rewriting techniques
  like equality saturation~\cite{tate2009equality}
  also mitigate phase ordering by exploiting
  the e-graph data structure from SMT solvers
  to repeatedly apply all rewrites simultaneously,
  thus obviating rule ordering considerations.
In particular,
  the \tcd{egg} library~\cite{willsey2021egg}
  has been used to develop new synthesis and
  optimization tools across diverse domains~\cite{
    herbie, szalinski, wang2020spores},
  including DSP compiler vectorization~\cite{
    vanhattum2021vectorization} and
  tensor computation graphs~\cite{yang2021equality}.

\g provides the first tensor IR
  amenable to equality saturation by
  introducing access patterns to
  provide pure, higher order tensor
  kernel combinators that support
  rank-polymorphism without the need
  for binding structures like
  anonymous functions or index notation.

\subsection{Pattern Matching Accelerator Calls}

The most closely related work to flexible matching is from 
   TVM BYOC~\cite{chen2021byoc}, which only provides exact syntactic matching.
Past work has also explored rewrite-based techniques for
  automatically inferring instruction selection passes
  between ISAs~\cite{
    ramsey2011resourceable,
    dias2010automatically}
  and in the context of superoptimization~\cite{
    bonsal-so,
    bonsal-so-translate}.
Rewriting in \TLA instead operates on a high-level IR
  to expose opportunities to invoke code generators,
  rather than performing low-level code generation directly.
Equality saturation has been
  used in the context of
  ML and DSP compilers for
  optimization~\cite{
    yang2021equality,
    alexa-dsp-eqsat,
    caviar-cc22}.
There has also been significant work on
  ML and HPC compiler frameworks with
  varying degrees of support for
  targeting custom accelerators~\cite{
    ragan2013halide,
    AtlPopl22,
    chen2018tvm,
    moreau2019hardware,
    lattner2021mlir}.
To the best of our knowledge,
  none of these frameworks provides support for
  testing prototype accelerators
  designs end-to-end on
  unmodified source applications.

\chapter*{\Cref{part:glenside-and-3la} Conclusion}

In \cref{part:glenside-and-3la}
  of this dissertation,
  I presented the first of two case studies
  giving evidence to my thesis.
In \cref{sec:part1-motivation}, 
  noting the gap in compiler and simulation tools
  for
  machine learning \glspl{accelerator}
  and how it was potentially affecting
  accelerator \cref{thesis:correctness},
  the authors of \TLA
  (including the author of this dissertation)
  set to developing a methodology for
  compiling to and testing
  designs.
As part of this flow,
  we needed a tool for finding places in
  machine learning workloads
  where we could invoke accelerators.
Existing
  tools missed mapping opportunities
  and were cumbersome to use,
  and thus poor with regards to \cref{thesis:optimizations}
  and \cref{thesis:devtime}.
In response,
  in \cref{sec:glenside}
  we introduced \g:
  a pure, binder-free tensor language
  which allowing for the use of
  more flexible \cref{thesis:algorithms}---%
  namely,
  \gls{equality-saturation}---%
  over machine learning workloads.
We incorporated \g into \TLA
  to automatically generate \gls{tensorization}
  routines
  in our compiler backend.
In \cref{chapter:part1-evaluation},
  we demonstrated how \TLA, with the power of \g,
  finds more accelerator mappings
  (greater \cref{thesis:optimizations})
  with less developer input
  (reduced \cref{thesis:devtime}).
Furthermore, we showed how we used \TLA
  to find and fix bugs in real
  accelerators (aiding in \cref{thesis:correctness}).

Note that, while we improve upon the state of the art 
  in algorithm flexibility (\cref{thesis:algorithms}),
  \cref{part:glenside-and-3la}
  does not improve upon the state of the art
  in model explicitness (\cref{thesis:models}).
In \cref{part:lakeroad}, I will
  more fully realize my thesis
  by utilizing \textit{both} more adaptable algorithms
  and more explicit models
  to automatically generate backends
  for FPGA compilers.

\part{Compilation to FPGAs}
\label{part:lakeroad}
\chapter*{\Cref{part:lakeroad} Abstract}

In \cref{part:lakeroad} of this dissertation,
  I apply my thesis
  to a fully separate domain:
  \gls{hardwaresynthesis} tools.
\gls{fpga} \gls{technology-mapping} is the process of
  implementing a hardware design expressed in 
  high-level HDL (hardware design language) code
  using the low-level, architecture-specific \glspl{primitive} of 
  the target FPGA.
As FPGAs become increasingly heterogeneous, 
  achieving high performance
  requires hardware synthesis tools 
  that better support mapping to complex, 
  highly configurable primitives 
  like digital signal processors (DSPs).
Current tools
  support DSP mapping via handwritten special-case mapping rules,
  which are laborious to write
  (poor \cref{thesis:devtime}),
  error-prone
  (poor \cref{thesis:correctness}),
  and often overlook mapping opportunities
  (poor \cref{thesis:optimizations}).
In \cref{part:lakeroad} of this dissertation,
  we introduce \lr,
  a principled approach to technology mapping via
  sketch-guided \gls{program-synthesis}
  (\cref{thesis:algorithms}).
A primary insight of \lr
  is to utilize vendor-provided
  simulation models
  (\cref{thesis:models})
  to generate the semantics
  needed by program synthesis.
\lr provides more
  extensible (\cref{thesis:devtime})
  technology mapping 
  with stronger correctness guarantees
  (\cref{thesis:correctness})
  and higher coverage of 
  mapping opportunities
  (\cref{thesis:optimizations})
  than state-of-the-art tools.
Across representative microbenchmarks,
  \lr produces
  1.4--3.6$\times$ the number of optimal mappings
  compared to proprietary state-of-the-art tools
  and
  6--30$\times$ the number of optimal mappings
  compared to popular open-source tools,
  while also providing correctness guarantees
  not given by any other tool.
\chapter{Introduction and Motivation}
\label{chapter:part2-intro}
\begin{figure}
\centering
\includegraphics[width=0.7\columnwidth]{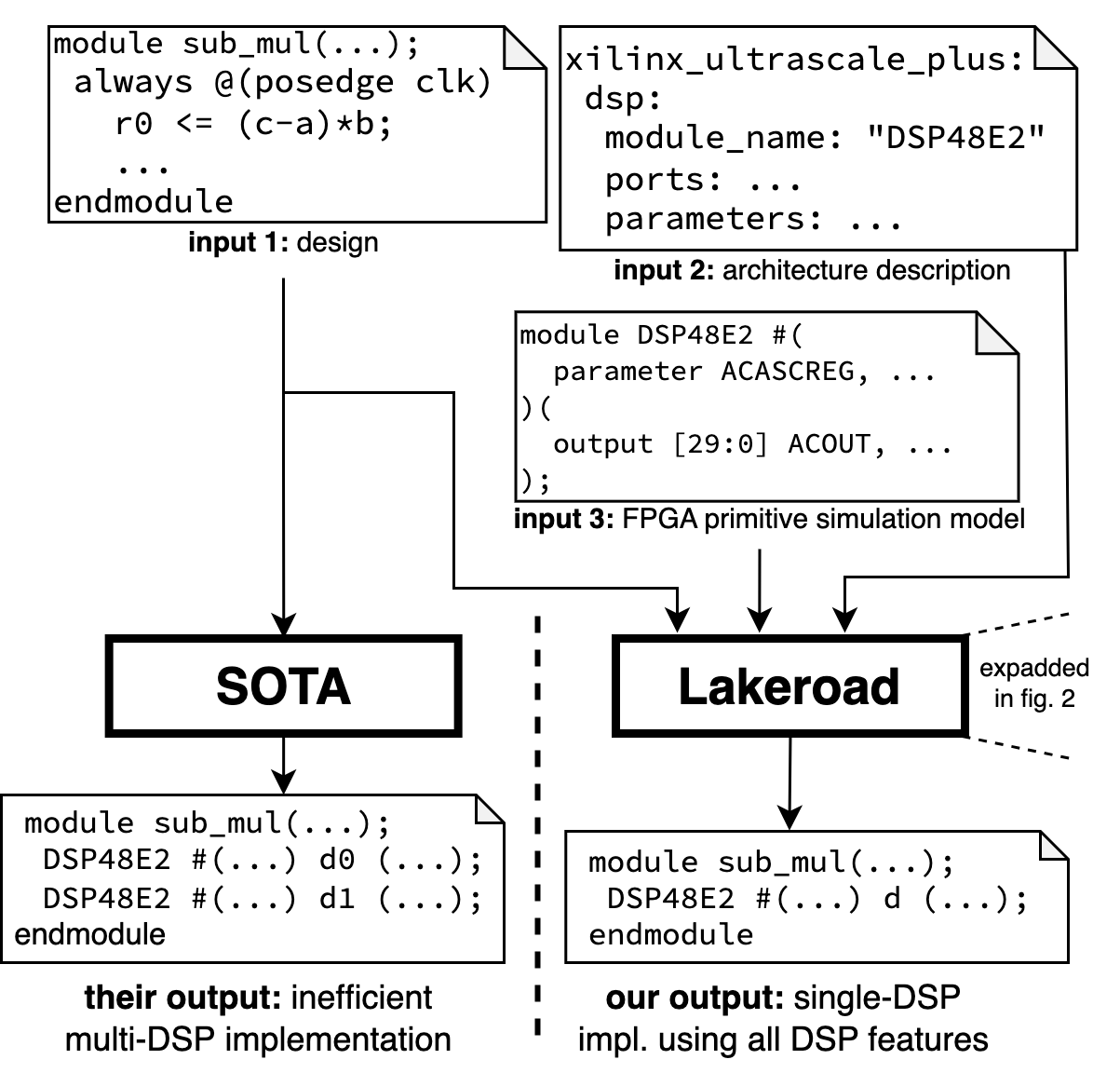}

\vspace{-3mm}
\caption{
Even given a simple input
  design (input 1),
  the state-of-the-art (SOTA)
  hardware synthesis tool
  for Xilinx FPGAs
  frequently
  fails to efficiently use 
  programmable primitives
  like DSPs.
\lr,
  on the other hand,
  can utilize all features
  of programmable primitives
  given just a short description
  of an FPGA architecture (input 2)
  and the vendor-provided 
  simulation models
  of the primitive (input 3).\tighten
}
\label{fig:firstpage}

\end{figure}

\textit{\Cref{part:lakeroad} is adapted from 
``FPGA Technology Mapping Using Sketch-Guided Program Synthesis'' by Smith, et al.~\cite{smith2024fpga}.
}

\vspace{10mm}

\noindent
Given a high-level hardware design specification
  (e.g., expressed in behavioral Verilog),
  FPGA technology mappers
  search for an equivalent
  low-level implementation
  in terms of the target FPGA's
  primitives.
See \cref{fig:firstpage} for an example, where
the high-level, behavioral \texttt{sub\_mul}
  module (``input 1'')
  is converted into FPGA-specific implementations
  (``their output'' and ``our output'')
  using the Xilinx-specific
  \texttt{DSP48E2} primitive.
  
Historically,
  FPGAs consisted of relatively simple
  primitives, such as
  lookup tables (LUTs) and carry chains.
Tools like
  ABC~\cite{ABC,abc2,brayton2010abc}
  \textit{automatically} 
  map to these basic primitives
  by translating designs
  to a library of simple logic gates
  and then packing those gates
  into LUTs.

However, FPGAs are becoming
  increasingly heterogeneous
  via
  the inclusion of specialized and diverse primitives
  such as digital signal processors (DSPs).
Utilizing these specialized primitives
  effectively
  is now
  crucial for achieving
  high performance~\cite{vega2021reticle}.
These specialized primitives
  make FPGA technology mapping far more challenging
  since technology mappers must now
  explore a much larger search space
  while also satisfying each primitive's
  complex set of restrictions and dependencies.
For example, Xilinx's DSP48E2
  is a multifunction 
  DSP
  with nearly
  100 ports and parameters,
  whose numerous configurations
  enable 
  support for a large variety of computations.
The manual for the DSP48E2 alone
  is 75 pages long,
  where considerable text details
  the complex restrictions
  between the settings of the nearly 100
  ports and parameters.

Existing technology mapping tools
  frequently fail to map designs
  to
  specialized primitives like DSPs,
  resulting in less-performant designs
  (that is, poor \cref{thesis:optimizations})
  and
  requiring manual work for the hardware designer
  to recover the performance of their design
  (that is, increased \cref{thesis:devtime}).
While existing toolchains
  have the ability to automatically infer
  locations where specialized primitives
  can be used in large designs,
  inference often fails~\cite{xilinxforum1,xilinxforum2,inferringreddit}.
In these cases, the designer can either
  accept lower performance and higher resource
  utilization,
  or they can perform
  what we call
  \textit{partial design mapping.}
During partial design mapping, 
  the designer
  manually identifies and separates out
  the module that should be mapped
  to a DSP.
They can attempt to re-run technology mapping
  on that module alone,
  in the hopes that mapping succeeds.
Yet existing toolchains often fail
  \textit{even in the partial design mapping case:}
  \cref{fig:firstpage} shows a
  simple module
  \texttt{sub\_mul}
  which \textit{should} fit on a single DSP48E2
  according to the DSP's manual,
  but is instead mapped to 
  two DSPs by current 
  state-of-the-art tools\footnote{
    Licensing restrictions forbid naming the
    specific proprietary tools, but they are familiar,
    standard packages used by many hardware designers.
  }---%
  a 100\% increase in resource utilization!
In the worst case,
  hardware designers are forced to manually
  instantiate complex primitives by hand,
  e.g., by looking through the 75-page
  DSP48E2 user manual
  to manually configure the DSP's dozens
  of ports and parameters.

Current state-of-the-art
  technology mappers 
  are implemented via
  ad hoc, handwritten pattern matching procedures,
  which
  fall short in three primary ways.
First,
  as we saw above,
  they are \textbf{incomplete:}
  they miss many mapping opportunities,
  even across microbenchmarks based on vendor documentation
  (\cref{thesis:optimizations}).
Second, they do not provide strong \cref{thesis:correctness} guarantees:
  recent work highlights the significant number of bugs found across 
  all major hardware synthesis tools~\cite{herklotz2020finding}.
Third, they are difficult to extend:
  \textit{each} new complex primitive requires
  supporting detailed semantics
  and adding hundreds of new, special-case
  syntactic pattern matching rules~\cite{wolf2013yosys}
  (hence increasing \cref{thesis:devtime}).

This chapter's
  key observation is that 
  technology mapping
  is well-suited for the application
  of automated reasoning procedures
  (\cref{thesis:algorithms})---%
  specifically,
  \textit{program synthesis}~\cite{gulwani2017program}.
We observe that 
  the configuration space of
  a programmable FPGA primitive
  is essentially a small, bespoke
  programming language,
  and that
  program synthesis
  could be applied
  to automatically generate
  primitive configurations.
We explore how
  program synthesis
  can simplify the design and implementation of
  FPGA technology mappers while providing
  \textbf{correct} (\cref{thesis:correctness}),
  \textbf{extensible} (\cref{thesis:devtime}), and
  \textbf{more complete}
  (\cref{thesis:optimizations})
  support for mapping to 
  diverse, highly configurable primitives
  like DSPs.
Program synthesis techniques rely on
  automated theorem provers like
  SAT and SMT solvers~\cite{de2008z3, barbosa2022cvc5}
  to automatically generate programs
  satisfying a given specification.
We demonstrate how
 \textit{sketch-guided program synthesis}~\cite{solar2008program}
  can be adapted
  for FPGA technology mapping,
  leveraging the
  Rosette~\cite{torlak2014lightweight} 
  program synthesis framework.


Sketch-guided program synthesis requires
  encoding the \textit{semantics}
  of the target language:
  in our case,
  a machine-readable,
  mathematical model specifying
  the behavior of each
  FPGA-specific primitive
  being mapped to.
In a typical synthesis tool,
  which generates programs
  for a single target language,
  this is a one-time cost.
However,
  in our setting,
  each new FPGA primitive
  introduces yet another new target language,
  which in turn requires
  extending the tool to encode
  yet another formal semantics.

To support
  correct, extensible, and more complete
  technology mapping, we propose
  automating this process with
  \textbf{semantics extraction from HDL}, 
  adapted from past work~\cite{daly2022synthesizing},
  to automatically extract
  complete primitive semantics
  from vendor-published HDL 
  \cref{thesis:models}
  (\cref{fig:firstpage}, ``input 3'').
Traditionally, such models have
   been used only 
  for
  simulation and validation
  \textit{after} technology mapping;
  we show that using the semantics
  to
  \textit{implement}
  technology mapping
  with a program-synthesis-based approach
  yields substantially more
  complete FPGA technology mapping.

Sketch-guided program synthesis also
  requires \textit{sketches}, which are 
  partially complete programs with ``holes'' to be filled in
  by the solver.
Sketches primarily serve to
  scale synthesis by
  constraining the set of programs that 
  solvers explore when searching for
  one that satisfies
  the given specification,
  i.e., performance at the cost of completeness.
In our setting,
  sketches correspond to
  arrangements of primitives,
  using holes
  as placeholders
  for some of the primitives' 
  ports and parameters.
This could be
  a single DSP with holes for
  its ports and parameters
  (as in the example in \cref{sec:overview-part-2}),
  or a number of LUTs with holes for their LUT memories,
  or even a mixture of LUTs, DSPs, and carry chains.
The synthesizer ``fills in the holes''
  as necessary for
  the low-level FPGA-specific primitive to implement
  a given high-level behavioral design fragment.
Unfortunately,
  developing effective sketches
  still requires synthesis expertise~\cite{10.1145/3140587.3062353,vanGeffenJITSynth}.
  Na\"ively, our approach would also
  require new sketches for each new
  FPGA primitive we target.


To address these challenges, 
  we introduce
  \textbf{architecture-independent sketch templates.}
Hardware designs are often implemented
  using high-level blueprints that are similar 
  across most FPGA architectures---%
  sketch templates
  capture these blueprints
  and make them reusable across architectures.
Therefore, by using sketch templates, we
  greatly reduce the overhead of supporting
  new architectures and
  diverse primitives.
Typically, when adding support for
  a new primitive or FPGA architecture in \lr,
  the hardware designer
  need not write or modify
  any sketch templates.


We leverage
  semantics extraction
  from HDL
  and architecture-independent
  sketch templates
  to build \lr,\footnote{
  \lr is publicly available at
  \url{https://github.com/uwsampl/lakeroad}.
  }
  a new FPGA technology mapper
  based on program synthesis.
  
\lr's prototype implementation automatically
  imports semantics for the LUTs, arithmetic carry chains,
  and DSPs of the Xilinx UltraScale+, Lattice ECP5,
  Intel Cyclone 10 LP,
  and SOFA~\cite{sofa} FPGA architectures. 
The only additional user input to \lr is a short 
  architecture description
  that lists the target FPGA's
  primitives (\cref{fig:firstpage}, ``input 2'').
Architecture descriptions
  only need to be written once per architecture,
  and \lr pre-supplies architecture descriptions
  for the aforementioned architectures.
With the automatically
  extracted primitive semantics
  and the user-provided architecture description,
  we demonstrate that \lr
  is more complete than proprietary tools on a variety
  of microbenchmarks
that are representative of program fragments
  implemented with DSPs during partial design mapping.
In particular,
  \lr maps up to 3.6$\times$ 
  more microbenchmarks than 
  state-of-the-art
  tools for Xilinx, Lattice, and Intel,
  and up to 30$\times$ 
  more microbenchmarks 
  than Yosys.\looseness=-1

\Cref{part:lakeroad} of this dissertation
makes
  the following key contributions:
\begin{itemize}[leftmargin=*]
\item The novel application of program synthesis
  to produce a technology mapper---\lr---that is
  more \textbf{correct, complete,} and \textbf{extensible} than state-of-the-art 
  tools.
\item A technique for applying
  \textbf{semantics extraction from HDL}
  to automatically generate models
  of hardware usable by 
  formal automated reasoning tools.
\item The concept of 
  \textbf{architecture-independent sketch templates,}
  which capture common patterns in hardware design
  in an architecture-independent way,
  plus \textbf{primitive interfaces} and \textbf{architecture descriptions}, the abstractions
  underlying these templates.
\item A formalization of the \lr
  toolchain and an argument
  for its correctness and sketch-completeness.
\item The first notion of \textbf{technology
  mapping completeness} for FPGA
  compilers.
\item \textbf{Empirical comparisons }of
  \lr and existing hardware synthesis
  tools to evaluate both their
  relative completeness and
  ease of extensibility.



\end{itemize}

In the following sections, we walk through a real-world example
  using both existing tools
  and \lr
  and 
  highlight \lr's design and key features (\cref{sec:overview}); 
  formalize \lr and
  demonstrate its correctness (\cref{sec:formalization}); 
  describe \lr's implementation (\cref{sec:implementation}); and 
  evaluate \lr on its
  completeness of mapping,
  extensibility,
  and expressiveness (\cref{sec:evaluation}) .
\cref{sec:background-and-related-work}
  discusses related work, and
  \cref{sec:lakeroad-conclusion} concludes.
\section{Overview}
\label{sec:overview}

We now walk through an 
  example
  of how current FPGA technology mapping tools can fail
  a hardware designer (\cref{sec:overview-part-1}) 
  and how \lr overcomes these limitations (\cref{sec:overview-part-2}).
In the process,
  we provide a high-level overview
  of \lr's main components.

\subsection{Compiling a Design to a DSP with Existing Tools}
\label{sec:overview-part-1}

Consider the following scenario:
A hardware designer
  is designing a large hardware block
  for the Xilinx UltraScale+ family of FPGAs.
The designer is specifically aiming to use 
  the UltraScale+'s specialized DSP48E2 units,
  which 
  can implement 
  combined multiplication, arithmetic,
  and logic operations, as 
  captured in this 
  simplified
  block diagram~\cite{userguide}:
\begin{center}
\includegraphics[width=.75\columnwidth]{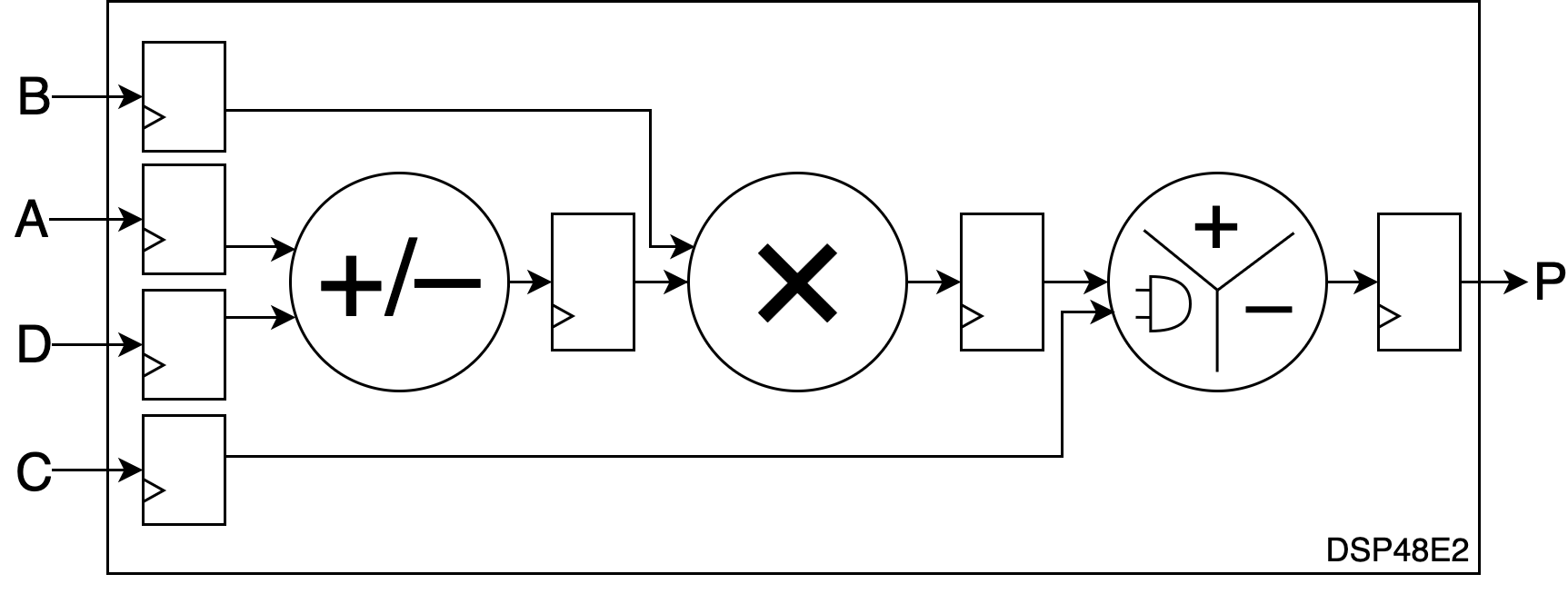}
\end{center}
The designer's hardware block
  involves the computation
  \texttt{(c-a)*b},
  which the manual states is implementable with a single DSP.
In particular, suppose the design
  consists of four separate instances of the following computation:
\begin{center}
\begin{minted}[baselinestretch=1]{verilog}
 for(i=0; i<4; i++) begin
  r[i] <= (c[i] - a[i]) * b[i];
 end
\end{minted}
\end{center}
It would be reasonable for the designer
  to expect the design
  to use a total of four DSPs.

\textit{\textbf{Current tools fail.}}
After compiling the design
  with existing tools,
  the designer is frustrated
  to find that the compiler returns a design
  that uses more
  resources than anticipated.
Instead of using the expected
  four DSPs, it uses eight---%
  a 100\% increase in resource consumption!
\textbf{The compiler has thus failed to 
  fully utilize
  the DSP}---%
  it has not configured a DSP48E2
  to implement
  \texttt{(c-a)*b} but has instead
  overflowed the
  computation
  onto
  an extra DSP.
The designer now faces a choice:
  either accept the result or attempt to coax the compiler
  into returning a more optimal design.

\textit{\textbf{Coaxing the compiler, to no avail.}}
Though many 
may choose to accept a less optimal result,
 this tenacious designer\footnote{
This may not be purely a personal preference. 
For example, a hardware design simply may not fit on an FPGA
  without manual optimizations!}
  tries to coax
  the compiler into giving the 
  expected results
  by 
  placing the computation
  of interest into a separate module:

\begin{minted}[baselinestretch=1]{verilog}
module sub_mul(input clk, input [15:0] a, b, c,
                   output reg [15:0] out);
 assign out = (c-a)*b;
endmodule
\end{minted}

\noindent
This allows the designer
  to  apply
  specific optimizations while mapping
  the module---%
  a process we call \textit{partial design mapping}.
They attempt various strategies,
  including
  annotating the module with
  Xilinx's \texttt{use\_dsp} Verilog attribute
  (to force the compiler to use a DSP where possible)
  and using a different
  synthesis directive
  (to apply a more 
  resource-intensive synthesis
  procedure).
\textbf{Despite these efforts,
  the compiler still cannot 
  map the design
  to a single DSP,}
  instead using two DSPs
  for the partial design.
Again, the designer must decide:
  give up and accept
  suboptimal results,
  or press on?

\textit{\textbf{Manual compilation.}}
The hardware designer presses on and
  now has only one option remaining:
  manually instantiating a DSP48E2
  with the desired behavior.
Skimming through the daunting 75-page DSP48E2's online user manual,
  the designer quickly discovers that
  configuring even
  the ``pre-sub'' \texttt{c-a}
  requires correctly setting 
  multiple ports and parameters
  (\texttt{INMODE}, \texttt{AMULTSEL}, \texttt{BMULTSEL}, and \texttt{PREADDINSEL}),
  whose descriptions span 10+ pages and multiple tables.
Correctly configuring the subsequent multiplier
  and logic unit proves even more difficult
  and time-consuming.
After configuring the computational units,
  the designer must still manually ensure
  correct pipelining
  of the 10+ pipeline registers.
After hours
  of frustration,
  a configuration is found that 
  seems to work, which the designer
  inserts into the design.
Precious time has been wasted,
  most of which will need to be repeated
  to configure the DSP again.
Making matters worse,
  \textbf{the designer has no formal guarantees
  about the correctness of this DSP configuration.}
It may work in a few simulated test cases,
  but are there corner cases
  that have been missed?\tighten



\subsection{Compiling a Design to a DSP with \lr}
\label{sec:overview-part-2}

\lr can save hardware designers
  the great effort involved
  in manual DSP configuration
  while also providing correctness guarantees.
Let us imagine how the designer 
  in this example,
  frustrated by conventional tools,
  can instead proceed using \lr
  during partial design mapping.
After putting 
  \texttt{sub\_mul}
  into its own module,
  the designer calls
  \lr from the command line:
\begin{minted}[baselinestretch=1]{bash}
$ lakeroad --template dsp \
           --arch-desc xilinx-ultrascale-plus.yml \
           sub_mul.v
\end{minted}
The \texttt{lakeroad} command outputs
  \texttt{sub\_mul\_impl},
  an implementation
  of \texttt{sub\_mul}
  that uses a single UltraScale+ DSP48E2:
\begin{minted}[baselinestretch=1]{verilog}
module sub_mul_impl(input clk, input [15:0] a, b, c, output [15:0] out);
 DSP48E2 #(
  .ACASCREG(32'd0), .ADREG(32'd0), .ALUMODEREG(32'd0),
  .AMULTSEL("AD"), .AREG(32'd0), .AUTORESET_PATDET("NO_RESET"), 
  // ...plus 30+ more parameters
 ) DSP48E2_0 (
   .A({ 14'h0000, a }), .ACIN(30'h00000000), .ALUMODE(4'hc),
   .B({ 2'h0, b }), .BCIN(18'h00000), .C({ 32'h00000000, c }),
   .CARRYCASCIN(1'h0), .CARRYIN(1'h0), .CARRYINSEL(3'h6),
   // ...plus 30+ more ports
 );
endmodule
\end{minted}
Unlike current compilers,
  \lr has produced an implementation
  using a single DSP48E2
  by utilizing more of the DSP's features.
Importantly, this compiled design
  is also formally guaranteed
  to implement
  the input \texttt{sub\_mul}
  design.
  
\begin{figure}
    \centering
    \hspace{-1cm}%
    \includegraphics[width=.5\columnwidth]{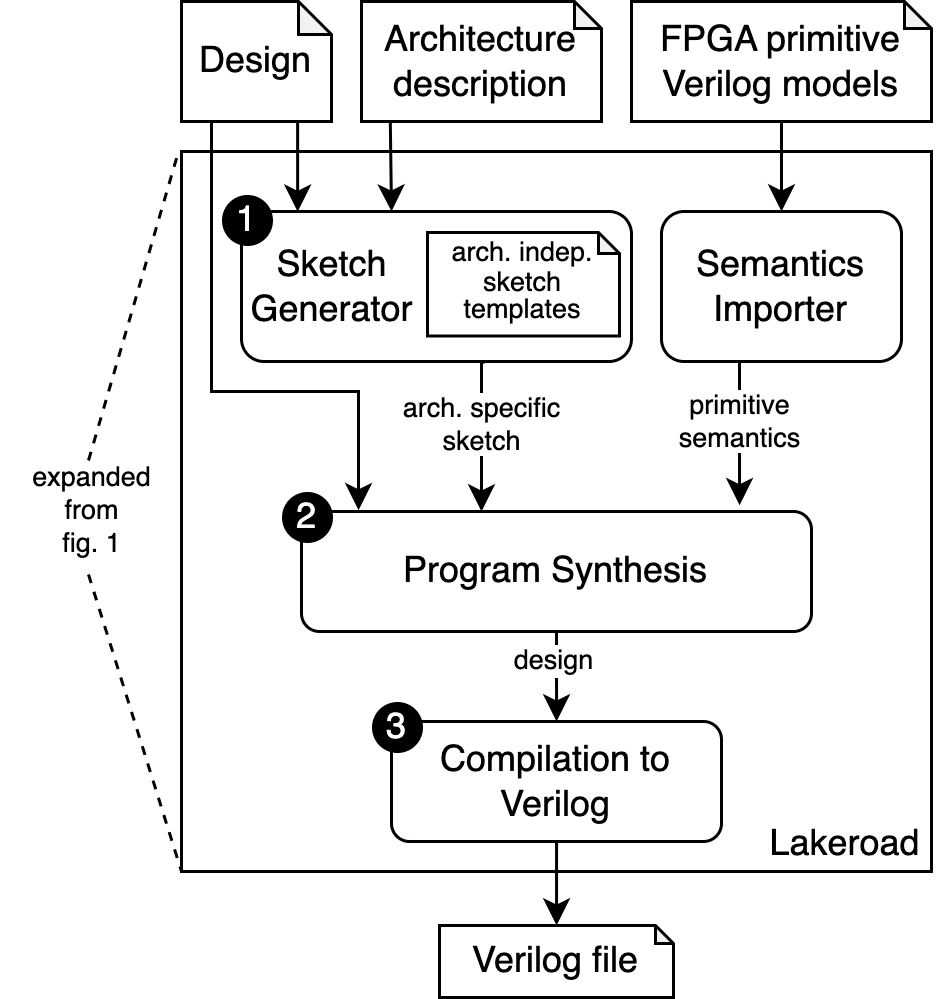}
   \caption{The components within \lr.}
    \label{fig:lakeroad-diagram}
\end{figure}

How does \lr provide 
  \textit{verified, more complete} support
  for the DSP48E2 over existing tools?  
At the core of \lr's
  correctness and completeness
  is
  \textit{sketch-guided program synthesis},
  a technique that  
  begins with a program \textit{sketch}, which
  captures a rough
  outline of a program
  and uses
  automated reasoning tools
  (e.g., SMT solvers)
  to fill in the sketch's \textit{holes.}
As shown in \cref{fig:lakeroad-diagram}, 
  \lr uses the following three-step process  
  to generate an efficient and correct
  DSP48E2 implementation
  of
  the \texttt{sub\_mul}
  design.

\textit{\textbf{Step 1: Generating a Sketch.}}
In
  the \texttt{sub\_mul}
  example,
  \lr
  generates
  the following sketch,
  which we refer to as \texttt{sketch}:\footnote{
Though this example is presented in a Verilog-like language,
  \lr's sketches are actually encoded in a Racket DSL that resembles structural Verilog.
  }
\begin{minted}[baselinestretch=1]{verilog}
module sketch(input clk, input [15:0] a, b, c, output [15:0] out);
 DSP48E2 #(
  .ACASCREG(??), .ADREG(??), .ALUMODEREG(??), .AMULTSEL(??), 
  .AREG(??), .AUTORESET_PATDET(??), ...
 ) DSP48E2_0 (
   .A({ 14'h0000, a }), .ACIN(??), .ALUMODE(??), 
   .B({ 2'h0, b }), .BCIN(??), .C({ 32'h00000000, c }),
   .CARRYCASCIN(??),  .CARRYIN(??), .CARRYINSEL(??), ...
 );
endmodule
\end{minted}
\noindent
This sketch consists of a single DSP48E2 instance
  with \textit{holes} 
  (represented by \texttt{??}) 
  serving as placeholders for most of its ports and parameters.
It is easy to see  
  the parallels
  between \texttt{sketch}
  and
  \texttt{sub\_mul\_impl};
  \texttt{sketch}
  is simply 
  \texttt{sub\_mul\_impl}
  with holes.
But how does \lr generate \texttt{sketch}
  in the first place?

To maximize portability across architectures,
  \lr does not store sketches 
  like \texttt{sketch}
  directly; 
  instead, it \textit{generates} sketches
  from architecture-independent
  \textbf{sketch templates.}
Instead of storing
  the preceding UltraScale+--specific sketch,
  \lr generates the sketch
  from the DSP sketch template, which
  the designer has chosen to use 
  with the \mbox{\texttt{-{}-template dsp}} flag.
A simplified form of this template looks like the following:
\begin{minted}[baselinestretch=1]{verilog}
module dsp_sketch_template(input clk,
                           input [n-1:0] a, b, c,
                           output [n-1:0] out);
 DSP dsp_instance(.clk(clk), .A(a), .B(b), .C(c), .D(d), .out(out));
endmodule
\end{minted}
Sketch templates
  capture hardware design patterns
that are common across FPGA architectures
  in an
  architecture-independent way.
\texttt{dsp\_sketch\_template},
  for example, 
  captures
  a basic pattern, i.e., 
  instantiating a single DSP.
\lr includes 
  sketch templates of varying complexity,
  from the simplicity of the one above 
  to the complexity of LUT-based multipliers.
Though new sketch templates
  can be added easily,
  in most cases
  (as in this example)
  users can simply apply
  \lr's provided templates.

To specialize \texttt{dsp\_sketch\_template}
  into \texttt{sketch},
  \lr translates
  the sketch template's generic \texttt{DSP}
  \textbf{primitive interface}
  into an UltraScale+--specific
  DSP48E2
  using the UltraScale+
  \textbf{architecture description.}
The generic \texttt{DSP}
  module is an instance of a
  \textbf{primitive interface:} 
  a \lr-introduced abstraction that 
  captures the similarities
  between primitives across 
  diverse FPGA architectures.
For example, 
  \lr's DSP primitive interface
  captures the facts that
  DSPs on all FPGA platforms
  generally have two to four data inputs
  (captured by \texttt{A}--\texttt{D};
  note that our example doesn't use
  input \texttt{D})
  and a clock input
  (captured by \texttt{clk}).
To convert the
  sketch template's
  DSP primitive interface instance
  into a DSP48E2,
  \lr utilizes the
  Xilinx UltraScale+ architecture description,
  which the designer has pointed to with the 
  \texttt{-{}-arch-desc xilinx-ultrascale-plus.yml}
  flag.
An \textbf{architecture description}
  specifies how \lr's various
  primitive interfaces
  are implemented for a given architecture.
The following simplified snippet of the UltraScale+
  architecture description, for example,
  tells \lr that,
  when generating a sketch for 
  UltraScale+,
  instances of the DSP
  primitive interface
  should be implemented with a DSP48E2:
\begin{minted}[baselinestretch=1]{yaml}
- interface: {name: DSP, params: { out-width: 48, a-width: 30, ...}} 
  holes: [?ACASCREG, ?ADREG, ?ALUMODEREG, ?AREG, ...]
  implementation:
    module: DSP48E2
    ports: [{ name: A, bitwidth: 30, value: A }, ...]
    parameters: [{ name: ACASCREG, value: ?ACASCREG }, ...]
    outputs: { O : P }
\end{minted}
Thus, while converting
  \texttt{dsp\_sketch\_template}
  into
  \texttt{sketch},
  \lr reads this architecture description
  and converts the single DSP instance
  into a DSP48E2,
  filling the ports and parameters with the concrete values
  and holes
  contained in the architecture description.
Architecture descriptions
  are usually short (100-400 LoC)
  and
  written only once per FPGA architecture; 
  \lr already contains such descriptions
  for Xilinx UltraScale+,
  Xilinx 7-series,
  Lattice ECP5,
  Intel Cyclone 10 LP,
  and the open-source FPGA SOFA~\cite{sofa}.

To generate a sketch,
  \lr takes an architecture-independent
  sketch template
  and specializes it using an
  architecture description.
Once the sketch is ready,
  the designer can move on to synthesis.

\textit{\textbf{Step 2: Program Synthesis.}}
  The next step 
  fills
  in the holes to generate
  a complete, correct
  hardware design,
  which is done automatically
  using a technique called
  \gls{program-synthesis}.
\textit{Program synthesis} is the process of
  using automated reasoning tools
  (like \gls{smt} solvers) 
  to generate correct programs
  by encoding program generation
  as a constraint solving problem.
In our \texttt{sub\_mul} example,
  \lr, aided  by
  Rosette~\cite{torlak2013growing,torlak2014lightweight}, 
  generates a query like the following:\footnote{
  We formalize this synthesis query and explain it precisely in \cref{sec:formalization}.
}
\footnotesize
$$\exists \ \mathtt{ACASCREG}, \mathtt{ADREG}, ...\ .\ \forall \ \mathrm{inputs} . \ 
  \texttt{sub\_mul}(\mathrm{inputs}) 
  = \texttt{sketch}(\mathrm{inputs}, \mathtt{ACASCREG}, \mathtt{ADREG}, ...)
$$
\normalsize
The query asks:
  are there concrete values for
  \texttt{ACASCREG}, \texttt{ADREG}, etc.,
  that will make our sketch's behavior
  equivalent to the input design's behavior
  on all inputs?
If the solver finds such values,
\lr can use them to fill the holes
  in the sketch
  and produce a compiled design.
However, if \lr tries to pass the preceding formula
  to an SMT solver,
  the solver will throw an error since
  the query is not expressed
  at a level 
 it understands, viz., %
  as equalities
  between bitvector expressions,
  using simple Boolean or arithmetic
  operations.
While it is conceivable that \texttt{sub\_mul}
  could be converted to
  a bitvector expression
  since its core computation is already
  expressed as
  \texttt{(c-a)*b},
it is unclear how to express
  \texttt{sketch}
  as an expression over bitvectors.
In particular, \lr must express
  bitvector-level semantics
  for Xilinx's DSP48E2 primitive.

To generate bitvector-level semantics
  for complex FPGA primitives,
  \lr introduces the concept of
  \textbf{semantics extraction}.
Rather than requiring
  manual effort to encode the semantics
  of the underlying hardware,
  which is notoriously difficult
  even for experts~\cite{Bernstein2021WhatAT},
  \lr's key insight is that these challenges
  can be avoided altogether
  by extracting low-level semantics
  directly from
  vendor-supplied simulation and verification models.
\lr builds on internal passes in 
  Yosys~\cite{wolf2013yosys} 
  to automatically extract
  solver-ready semantics from these vendor-provided HDL models,
  which we detail in \cref{sec:implementation:importing-semantics}.
For the
  \texttt{sub\_mul} example,
  the DSP48E2's semantics have already been imported into \lr.
Semantics need to be imported only when
  adding support for a new architecture, i.e., %
  about as infrequently
  as writing a new architecture description.
In most cases,
  \lr users can rely on already-imported semantics.
  
With the sketch generated
  and the DSP48E2's semantics imported,
  program synthesis can begin.
\lr utilizes Rosette to drive program synthesis,
  as detailed in \cref{sec:formalization}.
In our example, Rosette returns
  a configuration for the DSP48E2.
The last step, then, is to convert
  the compiled design
  to Verilog.
  
\textit{\textbf{Step 3: Compilation to Verilog.}}
Compiling \lr's internal representation into Verilog
  is a purely 
  one-to-one syntactic mapping;
  no optimizations are done at this stage,
  reducing the likelihood that bugs could be inserted.
In our example,
  the final Verilog produced
  results in the
  \texttt{sub\_mul\_impl}
  we saw at the start of \cref{sec:overview-part-2}.

\textit{\textbf{In summary.}}
\lr
  delivered 
  an implementation
  of the designer's
  \texttt{sub\_mul}
  module,
  improving upon both state-of-the-art compilers
  and manual approaches
  in multiple ways.
\lr's implementation
  is significantly more resource-efficient
  than the state-of-the-art compiler's.
\lr delivered its implementation
  in mere seconds,
  compared to the hours to days
  of work
  that manually instantiating
  a DSP might take.
Lastly,
  \lr's implementation
  is formally guaranteed to be correct.
Meanwhile, \lr did all of this
  while requiring no input from the user
  other than the Verilog to be compiled.

In the next chapter,
  we provide details on the concepts introduced in this overview.
We begin by formalizing \lr's compilation flow in \cref{sec:formalization}, 
  covering sketches and program synthesis in detail.
Then, in \cref{sec:implementation}, we give implementation details for sketch templates, primitive interfaces, semantics importing, and compilation to Verilog.

\chapter{Lakeroad}
\label{chapter:part2-lakeroad-detail}

Now that we have motivated
  the need for a tool like \lr,
  and have shown how it can improve
  a hardware designer's life in a simple example,
  in this chapter
  we provide the detailed information needed
  to reimplement \lr.
We begin by formalizing \lr
  in \cref{sec:formalization},
  providing a rigorous mathematical description
  of the entire compilation flow.
We then fill in implementation details
  not covered by formalization
  in \cref{sec:implementation}.
\section{Formalization}
\label{sec:formalization}

\newsavebox\boxlet
\newsavebox\boxassign
\newsavebox\boxin
\newsavebox\boxreg
\newsavebox\boxprim
\savebox{\boxlet}{\lstinline[language=thelang]!let!}
\savebox{\boxassign}{\lstinline[language=thelang]!:=!}
\savebox{\boxin}{\lstinline[language=thelang]!in!}
\savebox{\boxreg}{\lstinline[language=thelang]!Reg!}
\savebox{\boxprim}{\lstinline[language=thelang]!Prim!}

\newcommand{\Prim}[0]{\lstinline[language=thelang]{Prim}\xspace}

\newcommand{\Reg}[0]{\lstinline[language=thelang]{Reg}\xspace}

\newcommand{\Let}[0]{\lstinline[language=thelang]{let}\xspace}

\begin{figure*}

\begin{minipage}{.35\textwidth}
\begin{tabular}{l@{\hspace{.5em}}l}
$\SynProg$     & $\defn$ $\langle$ $\SynId$, $\seq{\SynId, \SynNode}*\rangle$\\[0.5em]
$\SynNode$       & $\defn \SynBv\ b$  | $\SynVar\ x$ \\
                 & \ \   | $\Op\ op\ \SynId$* \\
                 & \ \   | \IRReg{} $\SynId$ $(\SynBv~b)$ \\
                 & \ \   | \IRPrim{} $\mathsf{binds}$ $\SynProg$ \\
                 & \ \   | $\blacksquare_x$ \\[0.5em]
    
%
\end{tabular}
\end{minipage}%
\begin{minipage}{.25\textwidth}
\footnotesize{}
\begin{tabular}{l l}
$\SynId$  & $id  \in \mathbb{N}$ \\  
Bitvectors  & $b  \in \mathbb{BV}$ \\
Variables & $x  \in \text{LegalVarNames}$\\[0.5em]
Operators     & $op \in$ $\OpBv  \cup  \OpWire$\\[0.5em]
$\mathsf{binds}$ & $bs \in (\text{Variables} \rightharpoonup \SynId)$
\end{tabular}
\end{minipage}
\begin{minipage}{.4\textwidth}
\footnotesize{}
\begin{tabular}{l l}
\;\;\; Wire op & $\OpWire =$ \lstinline[language=thelang,mathescape]!$\{$concat$,$extract$, \ldots \}$! \\
\;\;\; Non-wire op & $\OpBv =$ \lstinline[language=thelang,mathescape]!$\{+, -, \times, \ldots \}$! \\
\end{tabular}
\end{minipage}

\caption{Syntax of $\UberLang$. $\blacksquare_x$ is a syntactic hole, labeled with variable $x$. $A \rightharpoonup B$ denotes the set of partial functions from $A$ to $B$.}
\label{fig:syntax}
\end{figure*}



\noindent
We claim that, through the use of
  program synthesis,
  we can improve the \cref{thesis:correctness}
  of FPGA compilers.
This section presents our core argument
  towards this point
  by formalizing \lr's semantics
  and arguing for its correctness.

We now formalize \lr{} 
  with functions $\lrfn$ and
  $\lrfnbmc$,
  and use these models
  to argue for the correctness
  and partial completeness of \lr{}. 
We first define  
  $\lrfn$ (\cref{subsec:the-lr-function}) and then 
motivate and define
  the language $\UberLang{}$,  
  specify its syntax and semantics,
  and define behavioral ($\SpecLang$), structural ($\ImplLang$), and sketch ($\SketchLang$) sublanguages (\cref{subsec:lrir-syntax-and-semantics}).
{
We next explain the
  underlying queries
  \lr{} uses to
  synthesize hardware programs
  that meet the desired specification
} (\cref{sec:formalization-program-synthesis}).
  We demonstrate the correctness
  and partial completeness of $\lrfn$,
  enumerate our Trusted Computing Base
  (\cref{subsec:lr-correctness-and-completeness}) and 
extend $\lrfn$ to $\lrfnbmc$,
    which ensures the generated program's
    semantics matches the design over multiple
    timesteps (\cref{subsec:lrfn-bmc}).
Finally,
 we highlight potential future
 applications that could be
 built on this section's formalization
 (\cref{sec:sem-beyond}).

\subsection{The \lr Function $\lrfn$}
\label{subsec:the-lr-function}

We model the execution of \lr
  with the partial function
  \small
\[\lrfn: \Sketch \times \SpecLang \times \Time \rightharpoonup \ImplLang,\]
\normalsize
where $\lrfn(\Psi, d, t)$
  invokes Rosette
  to synthesize a $t$-cycle
  implementation
  of behavioral design $d$ 
  using sketch $\Psi$ to guide
  the search,
  where a $t$-cycle
  implementation of $d$ is a program
  that is equivalent to $d$ at clock cycle
  $t$.
By not requiring program equivalence before
    clock cycle $t$ we allow the
    synthesized program's semantics to
    differ from the design during 
    an initialization period
    (e.g., as the pipeline is being filled).
To get guarantees beyond a single
    point in time $t$, we generalize
    $\lrfn$ to $\lrfnbmc$,
    which synthesizes a program
    that is equivalent to the design
    from time $t$ to $t + n$.
We formalize a sketch $\Psi \in \Sketch$  as a tuple
  $(\psi, h)$,
  where $\psi$ is a program in $\SketchLang$ 
  and $h$ is a map
  from the holes in $\psi$
  to a finite set of valid hole-free
  nodes in $\ImplLang$
  that can be used to fill
  the mapped hole.
This mapping $h$ is handled implicitly by Rosette's
  \texttt{choose} and \texttt{hole} constructs and
  need not be explicitly specified by the
  sketch writer.
We write $\lrfn(\Psi, d, t) = p$ to 
    indicate that synthesis succeeded and produced
    \lr{} program $p$.
However, it is possible that sketch $\Psi$
    cannot implement $d$, in which case
    the synthesis fails
    (i.e., returns UNSAT)
    and
    $\lrfn$ does not return anything.
Design $d$ belongs to 
  $\UberLang$'s behavioral fragment,
  $\SpecLang$ (see
  \cref{subsec:lrir-syntax-and-semantics}).
When $t = 0$, $\lrfn$ 
  synthesizes a 
  \textit{combinational design}; when 
$t > 0$, $\lrfn$ 
  synthesizes a \textit{sequential design}
  over $t$ clock cycles.
The rest of this section 
  considers sequential design synthesis 
  since its combinational counterpart is 
  a special case covered
  by our general approach.

\subsection{Defining $\UberLang$}
\label{subsec:lrir-syntax-and-semantics}

\lr uses the 
  $\UberLang$ language to
  translate behavioral
  HDL programs
  to structural, 
  hardware-specific
  HDL programs.
To facilitate this
  translation, we
  designed $\UberLang$
  to satisfy the
  following properties:
\begin{enumerate}[label=\textbf{P\arabic*}.]
    \item \textit{Easy translation to/from HDLs:}
        we must be able
        to translate
        designs from 
        a behavioral HDL
        to $\UberLang$
        and translate
        synthesized implementations
        to a structural HDL.
        
    \item \textit{Support parallel stateful execution:}
        FPGA designs
        consist of
        potentially stateful elements
        executing in parallel.
        $\UberLang$ must
        allow unambiguous
        parallel execution.

    \item \textit{Support graph-like program structures:}
        An FPGA component's outputs
        can be wired to
        multiple other components,
        including back
        to itself.
        This means that 
        FPGA programs can
        form arbitrary
        graphs, and $\UberLang$ must
        be able to express this.
        
    \item \textit{Support for sequential designs:}
        $\UberLang$ must handle designs
        that run over multiple clock cycles.
        
    \item \textit{Support for different architectures:}
        $\UberLang$ must handle FPGA components
        from different architectures.
\end{enumerate}
We describe how $\UberLang$ satisfies
  P1-P5 when we 
  define its syntax and semantics
  in the following subsections.

\subsubsection{$\UberLang$'s Syntax}
\label{subsubsec:syntax}

\Cref{fig:syntax} shows the $\UberLang$ syntax.
An $\UberLang$ program $\SynProg$
  consists of a root node ID
  and a graph of nodes,
  each of which is
  referred to by its ID.
A \textit{node} 
  can be 
  a constant bitvector,
  input variable,
  combinational (pure) operator,
  sequential (stateful) register,
  primitive,
  or hole.
Given a program 
  $p = (r, \langle id_1, {node}_1\rangle
           \ldots
           \langle id_n, {node}_n\rangle)$, 
  we use the notation 
  $p.root = r$, 
  $p.ids = \left\{id_1, \ldots, id_n\right\}$,
  and
  $p[id_i] = {node}_i$.
We define the free variables of a program $p.fv = \{x_i\}$ as the set of variable names occurring in $p$'s nodes of the form $(\SynVar\ x_i)$.\footnote{Note that this does not include variables of sub-programs occurring recursively inside of \IRPrim nodes.}  Finally, we use the notation $p.all\_ids$ for $p.ids$ together with $p'.all\_ids$ of any subprogram $p'$ of $p$ ($p'$ is a subprogram of $p$ if $\exists j, node_j = \textrm{\IRPrim}~bs~p'$).

Given a node $n$,
  we specify its inputs
  with the following function:
  \small
\begin{align*}
  &\textsc{inputs}(\SynBv\ b) = \{\}, \\
  &\textsc{inputs}(\SynVar\ x) = \{\}, \\
  &\textsc{inputs}(\Op\ op\ i_1 \ldots i_n)
     = \{i_1,\, \ldots,\, i_n\} \\
  &\textsc{inputs}(\text{\IRReg}\ i\ b_{init})
     = \{i\} \\
  &\textsc{inputs}(\text{\IRPrim}\ 
                     bs\ p')
                = \{bs[x]\ |\ x \in \textrm{domain}(bs)\}
\end{align*}
\normalsize
Note that we use 
    $A \rightharpoonup B$ 
    to denote the set
    of partial functions
    from $A$ to $B$;
    given 
    $bs \in A \rightharpoonup B$, 
    we write $\textrm{domain}(bs)$
    to denote
    the set of 
    $x\in A$ s.t.
    $bs[x]$ is defined.


A program $p$ is well-formed
  if and only if
  all the following
  conditions hold:

\begin{enumerate}[label=\bfseries{W\arabic*}.]
  \item $p.root \in p.ids$;

  \item All ids are unique and distinct. (i.e. for any sub-program $p'$, $p.ids$ and $p'.all\_ids$ are disjoint, and for any two sub-programs $p'$ and $p''$, $p'.all\_ids$ is disjoint from $p''.all\_ids$.)
  
  \item The inputs of all nodes in $p$ are ids of other nodes in $p$:
     $\forall id \in p.ids$, 
     $\text{inputs}(p[id]) \subseteq p.ids$;

  \item All primitive nodes contain 
      well-formed programs;
  \item All primitive nodes bind exactly their free variables; i.e., for $\text{\IRPrim}\ bs\ p'$, $\textrm{domain}(bs) = p'.fv$; and
        
  \item Program $p$ is free of combinational loops (formalized below in \cref{property:free-of-combinational-loops}).
        %
\end{enumerate}

\begin{property}[Free of Combinational Loops]
\label{property:free-of-combinational-loops}
Formally, a program $p$ is free of combinational
loops if there exists a function
$w : p.all\_ids \to \mathbb{N}$, that satisfies the following properties (collectively ``monotonicity''):
\begin{enumerate}
\item If $p[id] = \text{\IRReg{}}~\_~\_$, then $w(id) = 0$;
\item If $p[id] = \text{\IRPrim{}}~bs~p'$, then $w(id) > w(p'.root)$;
\item if $p[id] = \text{\IRPrim{}}~bs~p'$ and $p'[id'] = Var~x$, \\ then $w(id') > w(bs[x])$; and
\item Otherwise (e.g., $p[id] = \Op~op~ids^{*}$), \\
    if $id' \in \textsc{inputs}(p[id])$,
    then $w(id) > w(id')$. 
\end{enumerate}
\end{property}
\noindent The function $w$ acts as a witness to the 
absence of combinational loops because it is
impossible to define a strictly monotonic function without acyclicity.
We consider only well-formed
  $\UberLang$ programs.

%

$\SynBv$, $\SynVar$, and $\Op$ nodes
  encode bitvectors, variables, and
  operators.

\Reg $i_{data}\ b_{init}$ nodes
  let \UberLang implement
  sequential designs (P4).
$i_{data}$ is the 
  register's data input,
  which updates the stored
  value at the positive edge
  of each clock cycle,
  and $b_{init}$ is the
  register's initialization value.

\IRPrim{} $bs\ p$ nodes
  let \UberLang programs
  use hardware-specific components
  from different architectures (P5).
The $bs$ component is a \textit{variable map},
  mapping $\SynVar$s to input $\SynId$s.
The $p$ component is an $\UberLang$ program
  that defines the semantics of the
  hardware primitive.
A \Prim node also carries some metadata 
  used during compilation
  to a structural
  HDL, which we omit 
  for clarity.\tighten

$\SpecLang$ is the concrete
  \textit{behavioral}
  fragment of $\UberLang$ used for
  writing specifications; it 
  is formed by
  excluding \Prim
  nodes and holes
  from $\UberLang$.
  
$\ImplLang$ is the concrete
  \textit{structural}
  fragment of $\UberLang$ used
  for lowering $\UberLang$
  to structural HDLs; it 
  is formed
  by excluding \Reg nodes, $\Op$ nodes,
  and holes from $\UberLang$,
  with the following
  exception:
  the $p$ term
  in $\usebox{\boxprim}\ bs\ p$ must
  always be from the $\SpecLang$ since it 
  is used to specify the semantics
  of the \Prim node
  to the synthesis engine.
The behavioral node $p$
  is not used during
  compilation to HDL,
  and this behavioral
  expression does not 
  propagate to the
  structural HDL output.

$\SketchLang$ is another sublanguage of $\UberLang$ that is $\ImplLang$ but also including holes. 
Let $s$ be a program in $\SketchLang$
  with holes 
  $\blacksquare_{x_1}, \ldots, \blacksquare_{x_k}$.
These holes can be \emph{filled}
  with nodes $n_1, \ldots, n_k$
  in $\ImplLang$ by replacing
  each hole $\blacksquare_{x_i}$ 
  with its 
  corresponding node $n_i$
  to obtain a complete
  $\ImplLang$ program,
  denoted by $s[\blacksquare_{x_1} \mapsto n_1, \ldots]$.

The simplicity of this syntax
  makes translating to and from
  HDLs straightforward (P1).
\cref{sec:implementation}
  describes how \lr{} implements the
  translations to and from HDLs.

\subsubsection{$\UberLang$'s Semantics}
\label{subsubsec:semantics}

\begin{figure}


\small


\begin{flalign*}
  & \Time \hspace{0.2cm} t\in\mathbb{N} \hspace{1cm}  \textsf{Env} \hspace{0.2cm} e \in (\SynVar \rightharpoonup \Time \to \SynBv) \\
  &\vspace{0.5cm} \\
  &\textsc{Interp}\ :\ \SynProg \to \textsf{Env} \to \Time \to \SynNode \to \SynBv\\
  &\textsc{Interp}\ p\ e\ t\ (\SynBv\ b)\ = b&& \\
  &\textsc{Interp}\ p\ e\ t\ (\SynVar\ x)\ = e\ x\ t&& \\
  &\textsc{Interp}\ p\ e\ 0\ (\text{\IRReg}\ \_\ init)\ = init&& \\
  &\textsc{Interp}\ p\ e\ (t + 1)\ (\text{\IRReg}\ id\ \_)\ = \textsc{Interp}\ p\ e\ t\ p[id]&& \\
  &\textsc{Interp}\ p\ e\ t\ (\Op\ \mathsf{op}\ ids)\ = \llbracket op\rrbracket \ (\text{map}\ (\lambda id \ .\ \textsc{Interp}\ p\ e\ t\ p[id])\ ids)&& \\
  &\textsc{Interp}\ p\ e\ t\ (\text{\IRPrim}\ bs\ p')\ =&& \\
  &\quad \text{let}\ e' = \lambda x, t'\ .\ \textsc{Interp}\ p\ e\ t' \left(p[bs\ x] \right) \text{in}\\
  &\quad \textsc{Interp}\ p'\ e'\ t\ p'[p'.root]
\end{flalign*}



\caption{\lr's semantics as pseudocode.}
\label{fig:lr-interpreter-pseudocode}
\end{figure}

Before discussing the formal
  semantics of \UberLang,
  we present key 
  definitions.
We assume 
  a \textit{bitvector type}
  and, for simplicity,
  we elide bitvector
  widths.
We represent \textit{time} as a
  natural number.
A \textit{stream} is a function from $\Time$
  to bitvectors.
An \textit{environment} is
  a map from
  variable names
  to streams.

We give the semantics 
  for $\UberLang$ as an interpreter in 
  \cref{fig:lr-interpreter-pseudocode}.
We define the function \textsc{Interp} to interpret a program $p$ in environment $e$ at time $t$ and node $n$.
We do not define semantics for holes,
  as they are intended to be replaced
  by other constructs with well-defined semantics.

Most of the rules are straightforward.
A bitvector $\SynBv\ b$
  evaluates to
  its backing bitvector value $b$.
A variable node $\SynVar\ x$ 
  in an environment $e$
  at time $t$
  evaluates to
  the value returned 
  by the stream associated
  with $x$ in $e$
  at time $t$; 
using function notation, this is
  denoted by $e\ x\ t$.
A $k$-ary operator node
  $\Op{}\ op\ i_1\ldots i_k$ recursively
  interprets each operand in the current
  environment at the current time 
  and then applies $op$'s semantics,
  denoted $\llbracket op \rrbracket$,
  to the resulting values.
A register \IRReg $id\ b_{init}$ has two cases 
  depending on the current time: 
at time $t = 0$, a register evaluates
  to its initial bitvector value $b_{init}$; 
at nonzero times $t + 1$, a register evaluates
  to the value produced by the input $i$
  at the \textit{previous} timestep $t$.
A primitive \IRPrim{} $bs\ p'$
  in environment $e$ at time $t$
  is evaluated by interpreting
  the program $p'$ under
  the fresh environment $e'$
  formed by the binding map $bs$.

\subsection{Program Synthesis}
\label{sec:formalization-program-synthesis}


$\lrfn$ performs sketch-based program
  synthesis~\cite{solar2008program}.
Operationally, we implement
  the \textsc{Interp} function from 
  \Cref{fig:lr-interpreter-pseudocode}
  in Rosette, a solver-aided host
  language~\cite{torlak2014lightweight}.
Let sketch $\Psi = (\psi, h) \in \Sketch$, where
  $\psi \in \SketchLang$ has holes
  $\blacksquare_{x_{i}}$
  and $h$ maps $\psi$'s holes
  to the set of structural nodes
  that can legally fill the mapped hole.
Given a design $d$,
  we query Rosette if there
  are nodes 
  $n_1, n_2, \ldots n_k$ 
  such that 
  $n_i \in h[\blacksquare_{x_i}]$
  and
  $p = \Psi[\blacksquare_{x_1} \mapsto n_1, \ldots]$
   is well-formed and equivalent to $d$
   (i.e., we ask Rosette to fill
   each hole with a node associated with the node in $h$).
Program equivalence between well-formed
  programs $p$ and $d$ at time
  $t$, written $p\cong_t d$, is defined as
  \begin{align*}
    & p.fv = d.fv\ \wedge \\
    &\forall e\ s.t.\ \textrm{domain}(e) = p.fv, \\
    &\quad\textsc{Interp}\ p\ e\ t\ p.root =
   \textsc{Interp}\ d\ e\ t\ d.root.
  \end{align*}
In \cref{subsec:lrfn-bmc}, we
  use bounded model checking to
  extend $\lrfn$'s guarantees
  beyond the single timestep
  at clock cycle $t$.

\subsection{Correctness and Completeness of $\lrfn$}
\label{subsec:lr-correctness-and-completeness}

Recall that the synthesis
  function $\lrfn$ is partial.
We say that $\lrfn$
  is \emph{correct} if
  it 
  returns a program
  $\lrfn(\Psi,d,t) = p$  where
  $p$ is
  a well-formed
  completion of $\Psi = (\psi, h)$,
  meaning $p = \Psi[\blacksquare_{x_1} \mapsto n_1, \ldots]$
            such that $n_i \in h[\blacksquare_i]$ for all $i$
  and $p\cong_t d$.

Furthermore, we say
  that $\lrfn$ is
  \emph{sketch-complete}
  if $\lrfn(\Psi,d,t)$
  is defined whenever
  there exists a
  well-formed completion
  $p$ of $\Psi$
  such that $p\cong_t d$.
That is, synthesis is
  correct if it never
  returns an erroneous result
  and sketch-complete
  if it returns a
  correct result
  whenever one exists.

{
We have implemented $\lrfn$
    with Rosette 
    (see \cref{sec:formalization-program-synthesis}),
    which guarantees our system is correct and complete
    under the following assumptions:
\begin{enumerate}
    \item Correctness of Rosette and underlying SMT solvers;
    \item That our encoding of \lr{} is bug-free;
    \item That the lowering of \textsc{Interp}
    to SMT formulas by  Rosette always terminates.
    This is possible when partial evaluation of \textsc{Interp} on arguments $p$, $t$ and $n$ terminates (independently of the value of $e$).
\end{enumerate}

}

\begin{lemma}
\label{lemma:interp-is-primitive-recursive}
Let $p$ be a well-formed program, 
    $e$ an environment,
    $t$ a \Time,
    and $n$ be a node belonging to $p$.
Then \textsc{Interp} is primitive recursive 
    (i.e. terminates) in the arguments $p$, $t$, and $n$.\tighten
\end{lemma}

\begin{proof}[Proof of Lemma~\ref{lemma:interp-is-primitive-recursive}]
Recall that a function $f(x,y,z)$ is
    primitive recursive in arguments $x$ and $y$
    (under a lexicographic ordering) 
    if in the definition of $f$ every
    recursive call $f(x',y',z')$ is made
    with values $(x',y')$ such that $x' < x$ 
    or $x' = x \wedge y' < y$.
    If $x$ and $y$
    are drawn from the natural numbers 
    (or another well-ordered set),
    then the recursion is guaranteed to terminate.

Under what order is 
  \textsc{Interp} primitive recursive?
Because our program
    is well-formed, it must be free
    of combinational loops (see \cref{property:free-of-combinational-loops}).
Formally, this means we have an acyclicity
    witness function $w : p.all\_ids \to \mathbb{N}$
    that monotonically increases in the direction of
    dataflow in our circuit.
Each node $n$ argument passed to \textsc{Interp}
    has an \textsf{Id} that is unique and distinct
    from the \textsf{Id}s used in $p$ or any of $p$'s
    subprograms (\textbf{W2});
    we denote this \textsf{Id} as $id_n$.
We can associate each $n$ argument
    to a recursive call of \textsc{Interp}
    with a number $w(id_n)$.
We claim that \textsc{Interp} is
    primitive recursive under
    the lexicographic ordering on $(t, w(id_n))$.
    
To prove this claim we need to demonstrate that
    if \textsc{Interp} with time and node
    arguments $t'$ and $n'$ makes a recursive
    call to  \textsc{Interp} with time and
    node arguments $t''$ and $n''$, then the following
    condition holds:
    \small
\begin{equation}
\label{eqn:primitive-recursion-condition}
    t'' < t' \vee \left(t'' = t' \wedge w(id_{n''}) < w(id_{n'})\right).
\end{equation}
\normalsize
To do this it suffices to examine each case of \textsc{Interp}'s definition.

When $n'$ is a $\SynBv$ constant,
    \textsc{Interp} makes no recursive calls,
    and the condition in \cref{eqn:primitive-recursion-condition}
    holds vacuously.
    
When $n'$ is a \IRReg{} node \textsc{Interp} either terminates
  (when $t'=0$) or makes a
  recursive call with time value $t'' = t' - 1$,
  maintaining the condition in \cref{eqn:primitive-recursion-condition}.
  
When $n'$ is an operator node, 
  \textsc{Interp} recursively interprets
  the operands with time arguments $t'' = t'$.
However, each operand's id $id''$ belongs to $\textsc{inputs}(n')$,
    and, by~\cref{property:free-of-combinational-loops},
    $w(id_{n'}) > w(id'')$,
    so our condition holds.
  
This leaves us with the less obvious
  cases in which $n'$ is either a \IRPrim or $\SynVar$,
  which work together in tandem.
When $n' = \text{\IRPrim{}}~bs~p'$,
    \textsc{Interp} makes a recursive
    call with node argument $p'.root$
    and time argument $t$.
By ~\cref{property:free-of-combinational-loops},
    $w(p'.root) < w(id_{n'})$,
    and the condition in \cref{eqn:primitive-recursion-condition} holds.
\textsc{Interp} also defines a new environment for execution
    of $p'$ via $\lambda$-abstraction, and this in turn
    will recursively invoke \textsc{Interp}.
These environments are only invoked by the rule for variables,
    which we handle presently.
    
When $n' = \SynVar~x$, the environment is 
    invoked on variable $x$.
Here, there are two possible cases.
First, we are interpreting the
    top-level program $p$. 
As this is the initial, top-level environment, there is no further recursion.
Second, we are
    interpreting a sub-program $p'$
    and $e'~x~t = \textsc{Interp}~p~e~t~(p[bs~x])$
    is actually a recursive call into the
    program $p$ one level up,
    with its environment $e$.
In this latter case,
    note that $w$ is defined such that
     $w(id_{p[bs~x]}) = w(bs~x) < w(id_{\SynVar~x})$
    (item 3 of \cref{property:free-of-combinational-loops}),
    satisfying our property.
All cases are complete.
\end{proof}


From this, we conclude that
  all possible substitutions for $\Psi$
  are attempted, and $\lrfn$ is sketch-complete.


\paragraph{Trusted Computing Base.}

The \textit{trusted computing base} (TCB) of a
  system is the set of components
  it assumes to be correct~\cite{MacKenzieComputingTrust}.
A bug anywhere in the TCB
  could cause the guarantees
  made by that system to be violated.
\lr's  TCB includes:
  Rosette and the underlying
      SAT/SMT solvers that Rosette queries
      (Bitwuzla, cvc5, Yices2, and STP);
  the internal Yosys passes \lr
      uses to extract primitive semantics
      and translate design specifications
      from behavioral Verilog into
      $\SpecLang$;
  the semantics for $\UberLang$,
    which we assume conservatively
    models non-cyclic (DAG) designs;
  our code to translate from
      the $\ImplLang$ to
      structural Verilog; and
  the vendor-provided Verilog
    simulation models for FPGA primitives.
Each TCB component 
  has also been thoroughly tested,
  as described in \cref{sec:evaluation}.
Importantly,
  sketches and sketch generation
  are \textit{not} in \lr's TCB: %
  even if there were a
  bug in \lr's sketch-related components,
  it would not violate
  \lr's correctness guarantees.

\subsection{Multiple Clock Cycle Guarantees with $\lrfnbmc$}
\label{subsec:lrfn-bmc}

The preceding completeness and 
    correctness properties for
    $\lrfn$ 
    guarantee that
    running the
    synthesized program
    $p$ and the design $d$
    for $t$ clock cycles
    produces the same output.
To extend this guarantee, \lr supports
    a form of
    bounded model checking, 
    where
    synthesis ensures that
    $p$ is semantically equivalent
    to $d$ for $c$ additional clock cycles
    starting at time $t$.
We formalize this
    with the function $\lrfnbmc$,
    which takes a sketch $\Psi$,
    a behavioral design $d$,
    a number of clock cycles $t$,
    and a model checking
    time bound $c \geq 0$
    and returns an implementation
    $p \in \ImplLang$
    that is equivalent to $d$
    at time steps $t, t+1, \ldots, t + c$.

Our correctness and completeness guarantees are
    similar to those for $\lrfn$:

    \small
  \begin{align*}
    & p.fv = d.fv\ \wedge \\
    &\forall e\ s.t.\ \textrm{domain}(e) = p.fv, \\
    &\quad\bigwedge_{i=t}^{i=t+c}
      {\textsc{Interp}\ p\ e\ i\ p.root = \textsc{Interp}\ d\ e\ i\ d.root}.
  \end{align*}
  \normalsize

\subsection{Beyond \lr}
\label{sec:sem-beyond}

$\UberLang$, its semantics,
  and the synthesis approach we describe here 
  are useful for applying program synthesis
  to other hardware design problems.
For example,
  the synthesis problem detailed above could be ``flipped''
  to decompile structural designs back 
  to higher-level behavorial designs,
  i.e., synthesizing from $\ImplLang$
  to an expression in $\SpecLang$.
Such decompilation has seen recent
  interest for
  recovering equivalent but faster-to-simulate
  models and for porting models across
  different architectures~\cite{sisco2023loop}.
As another example,
  the synthesis approach could be
  adapted to help port designs by
  synthesizing expressions in
  $\ImplLang$ that use one set of primitives
  on one architecture from 
  other designs in $\ImplLang$ that use
  a different set of primitives from
  a different architecture.
Thus, the formalization in this section
  transcends the particular challenges
  of FPGA technology and provides
  a reusable foundation for exploring
  a much broader range of hardware design challenges
  from a program synthesis perspective.
  
\section{Implementation}
\label{sec:implementation}


\lr is composed of
  approximately 13K lines of Racket
  plus approximately 58K lines
  of Racket
  automatically generated from 
  vendor-supplied Verilog.
Vendor-supplied Verilog
  was obtained from Lattice Diamond,
  Intel Quartus,
  and Xilinx Vivado
  sources.
We used Vivado version v2023.1,
  Quartus 22.1std.1 Build 917 02/14/2023 SC Lite Edition,
  Diamond version 3.12,
  Yosys version 0.36+42 (commit \texttt{70d3531}),
  the cvc5~\cite{barbosa2022cvc5} and Yices2~\cite{dutertre2006yices,dutertre2014yices}
  solvers
  included in the 2023-08-06 release of \texttt{oss-cad-suite} from YosysHQ,
  the Bitwuzla solver at commit \texttt{b655bc0}~\cite{bitwuzla},
  the STP solver at commit \texttt{0510509a},
  Racket version 8.9~\cite{racket,racket:ref},
  and Rosette version 4.1~\cite{rosette4}.

\subsection{Primitive Interfaces}
\label{sec:impl-primitive-interfaces}

As described in \cref{sec:overview},
\textit{primitive interfaces}
  describe abstract versions
  of common FPGA primitives,
  which allow sketch templates
  to be architecture-independent.
To date, \lr declares primitive interfaces for
  $n$-input LUTs, $w$-width carry chains, 
  $n$-input muxes,
  and DSPs with up to four data inputs and one clock input.
The next section includes a concrete example
  of \lr's LUT4 primitive interface.\tighten


\subsection{Architecture Descriptions}
\label{sec:impl-arch-descs}

\begin{figure}
\begin{minted}[fontsize=\footnotesize]{YAML}
implementations:
  - interface: { name: LUT, num_inputs: 4 }
    internal_data: { sram: 16 }
    modules:
      - module_name: frac_lut4
        filepath: SOFA/frac_lut4.v
        ports:
          - { name: in, direction: in, width: 4, 
              value: (concat I3 I2 I1 I0) }
          - { name: mode, direction: in,
              width: 1, value: (bv 0 1) }
          - { name: lut4_out, direction: out,
              width: 1 }
        parameters: [{ name: sram, value: sram }]
    outputs: { O: lut4_out }
    \end{minted}
    \caption{
SOFA architecture description.
  }
    \label{fig:sofa-architecture-description}
\end{figure}

\noindent As described in \cref{sec:overview},
  \textit{architecture descriptions}
  convey the information
  required to convert
  each instance of a primitive interface
  into the corresponding
  architecture-specific module,
  which occurs while converting
  sketch templates
  into sketches.
The architecture description is
  the only additional input that
  may be required from a user
  to support a new architecture;
  it is a one-time effort 
  that is reusable
  for any designs in an architecture.
Architecture descriptions
  are simply lists
  (provided as YAML files)
  of the primitive interfaces
  that an architecture implements,
  but, crucially,
  also include 
  architecture-specific
  port and parameter values
  in a map called \texttt{internal\_data}.
Values in this map
  become symbolic values
  solvable by the SMT solver.
Additional constraints can also be specified
  in the architecture description 
  to rule out invalid configurations
  and minimize the solver's search space.\tighten

As an example,
  \cref{fig:sofa-architecture-description}
  shows the architecture description
  for the SOFA~\cite{sofa}
  FPGA architecture.
The description contains
  a single primitive interface implementation, i.e., 
  LUT4.
\lr's LUT4 primitive interface
  standardizes the names of a LUT4's inputs and outputs,
  naming the inputs
  \texttt{I0} through \texttt{I3}
  and the output
  \texttt{O}.
The SOFA implementation 
  of the LUT4 primitive interface
  uses
  the SOFA-specific
  \texttt{frac\_lut4}
  primitive.
Primitive interface inputs \texttt{I0} through \texttt{I3}
  are mapped to
  the actual
  input port of the \texttt{frac\_lut4},
  named \texttt{in}.
Likewise, the \texttt{frac\_lut4} output
  \texttt{lut4\_out}
  is mapped to the primitive interface output \texttt{O}.
The \texttt{internal_data} field
  declares \texttt{sram},
  the LUT's 16-bit internal memory,
  as an architecture-specific detail
  to be solved during synthesis.
 
If a sketch template uses a primitive interface
  not included in the architecture description
  (e.g., SOFA does not implement carries),
  \lr may still be able to implement the primitive interface
  based on primitive interfaces
  the architecture \textit{does} implement.
To date, \lr can implement any mux
  with LUTs, 
  a larger LUT from  smaller LUTs,
  a smaller LUT from a larger LUT,
  a carry from LUTs,
  and a smaller DSP from a larger DSP;
  it handles these conversions during sketch generation.

\subsection{Sketch Templates, Sketches, and Sketch Generation}

\label{sec:impl-sketch-templates}

As described in \cref{sec:overview},
  \lr 
  captures common FPGA implementation patterns
  in reusable, architecture-independent
  \textit{sketch templates.}
Thus far, we have described only 
  the relatively simple
  \texttt{dsp} sketch template,
  which instantiates a DSP.
As a more complex example
  of capturing common FPGA implementation patterns,
  consider
  the \texttt{bitwise-with-carry}
  sketch template, which
  uses $n$ LUTs and a carry chain
  to implement designs such as 
  addition or subtraction.
\lr currently provides 5 sketch templates: 
  \texttt{dsp},
  \texttt{bitwise}, \texttt{bitwise-with-carry},
  \texttt{comparison} (LUT- and carry-based arithmetic comparison), 
  and
  \texttt{multiplication} (LUT-based multiplication).
 
The process of converting sketch templates
  to sketches
  is implemented as described in
  \cref{sec:overview}
  and \cref{sec:impl-arch-descs}.
\lr iterates over every
  primitive interface instance
  in the sketch
  and replaces it with
  the concrete primitive
  in accordance with
  the architecture's 
  architecture description.
If the architecture description
  does not implement the requested
  primitive interface,
  \lr checks whether it can implement
  the primitive interface with other
  implemented interfaces
  (e.g., implementing a smaller LUT with
  a larger LUT)
  and raises an error otherwise.

Sketch templates and sketches alike 
  are written in a domain-specific language (DSL)
  embedded into Rosette,
  whose implementation closely mirrors the syntax
  and semantics of \UberLang. 
The only significant difference is that
  the interpreter implementation
  does not use bitvector streams natively.
Instead, each invocation of the interpreter
  represents a single timestep,
  and all intermediate values from the previous timestep
  are taken as input.
Streams are then built up
  using multiple invocations of the interpreter.

\subsection{Importing Semantics from Verilog Modules}
\label{sec:implementation:importing-semantics}

\lr uses 
  Yosys~\cite{wolf2013yosys}
  to convert Verilog modules
  into the \texttt{btor2} format~\cite{btor} 
  and then converts the resulting \texttt{btor2}
  to Rosette/Racket code.

Due to the semantics of the Verilog language
  and the internal implementation of Yosys,
  extracting semantics from Verilog modules
  may require the following manual modifications
  to accommodate semantics extraction and synthesis:
  
\begin{itemize}[leftmargin=*]
\item As Yosys converts
  parameters from variables
  to constant values
  immediately upon module import,
  module parameters should be converted to
  ports
  to ensure they remain variables
  (and thus solvable by the SMT solver).
Note that not all parameters 
  can always be converted to ports,
  meaning some parameters cannot be solved for.
\item Strings should be converted to bitvectors.
\item All registers should be initialized.
\item All instances of \texttt{x} and \texttt{z} values should be  
  converted to 2-state logic (0 or 1).
\end{itemize}
Note that these caveats
  apply only 
  to our prototype implementation,
  not the general technique
  of semantics extraction from HDL.
Once these manual modifications are made,
  the following series of Yosys passes
  can be used to convert the Verilog
  into suitable \texttt{btor2}:
\texttt{prep; flatten; pmuxtree; opt_muxtree; clk2fflogic; prep; write_btor}.

We implement
  the translation from \texttt{btor2}
  to Rosette bitvector expressions
  as a 1:1 translation 
  since both languages
  are simply operations over bitvectors.

\subsection{Program Synthesis and Compilation to Verilog}
\label{sec:implementation:program-synthesis}

We implement
  the synthesis procedure
  defined in \cref{subsec:lr-correctness-and-completeness}
  with Rosette.
Multiple clock cycle guarantees,
  as described in \cref{subsec:lrfn-bmc},
  are implemented simply by making $c+1$
  total assertions,
  asserting the output of the input design
  and the sketch are equal
  after each of the $c+1$ timesteps.
We use a portfolio solving
  method, running
  Bitwuzla~\cite{niemetz2020bitwuzla},
  cvc5~\cite{barbosa2022cvc5},
  Yices2~\cite{dutertre2006yices,dutertre2014yices},
  and STP~\cite{stp}
  in parallel
  and using results from the first
  solver to terminate.
To produce Verilog,
  \lr compiles the program from its internal DSL
  to the JSON format defined
  by Yosys using a straightforward translation 
  and then uses Yosys to output Verilog.



\chapter{Evaluation}
\label{sec:evaluation}

We now evaluate
  \lr in terms of completeness
  (related to \cref{thesis:optimizations})
  and extensibility
  (related to \cref{thesis:devtime}).
Note that we consider our formalization
  in \cref{sec:formalization}
  as evidence for our \cref{thesis:correctness}
  claim,
  and thus we don't include
  an evaluation of correctness
  in this chapter.
In the following experiments,
  we target four FPGA architectures:
  \textbf{Xilinx UltraScale+},
  commonly used
  for large, high-performance workloads;
  \textbf{Lattice ECP5},
  commonly used 
  in low-power, low-cost scenarios;
  \textbf{Intel Cyclone 10 LP},
  an FPGA designed for low-cost,
  high-volume use cases,
  and 
  \textbf{SOFA}~\cite{sofa},
  a recent, open-source
  FPGA developed by the
  research community.
We compare \lr to existing
  technology mappers.
For
  Xilinx Ultrascale+, Lattice ECP5, and
  Intel Cyclone 10 LP,
  we compare \lr against both
  the open source toolchain Yosys~\cite{wolf2013yosys}
  and the
  state-of-the-art,
  proprietary, closed source
  toolchains
  for each architecture.\footnote{
    Again, licensing restrictions 
    prevent our naming the specific 
    proprietary tools, but
    they are familiar, standard packages 
    used by many hardware designers.}\tighten
The experiments were conducted 
  on a system running Ubuntu 20.04.3 
  with an AMD EPYC 7702P 64-Core CPU.
The resident set size of a single \lr process
  did not exceed
  300MB while running our evaluation.
We use the software versions listed in \cref{sec:implementation}.

\section{\lr Completeness}
\label{sec:completeness}
The reliance of many technology mappers,
  including state-of-the-art tools,
  on hand-written patterns
  leads them to fail when attempting to map many workloads
  that \textit{should} be mapped to a single DSP.
In particular, 
  the process of partial
  design mapping 
  (illustrated in \cref{sec:overview}) 
  becomes
  a laborious endeavor because
  of this incompleteness:
  hardware designers
  hand-instantiate DSPs rather
  than rely on substandard
  automated tooling, repeating
  the work each time they identify a potential
  opportunity to use a DSP.
\lr's greater mapping completeness 
  significantly
  reduces the burden on hardware
  designers during partial
  design mapping and marks 
  the first step in automated
  mapping for full designs.
  We next evaluate how \lr's program synthesis approach
  enables it to achieve greater completeness
  for these program fragments.
In the context of this dissertation,
  this corresponds to providing evidence
  for our \cref{thesis:optimizations} claim:
  namely,
  through the use of
   more adaptable \cref{thesis:algorithms}
  (program synthesis)
  and more explicit \cref{thesis:models}
  (vendor-supplied simulation models),
  we can build a compiler,
  \lr,
  which finds more \cref{thesis:optimizations}
  in the form of mappings to
  specialized primitives.

\paragraph{\textnormal{\textit{\textbf{Evaluation Setup.}}}}

We highlight four particularly complex
  DSPs for the Xilinx Ultrascale+,
  Xilinx 7-series,
  Lattice ECP5,
  and Intel Cyclone 10 LP architectures: 
  the Xilinx DSP48E2,
  Xilinx DSP48E1,
  Lattice ALU54A--MULT18X18C 
  (a single DSP composed of 
  two primitives),
  and Intel cyclone10lp\_mac\_mult.
SOFA provides no DSP, and is not included
  in this part of the evaluation.
For each architecture's DSP,
  we enumerate a large subset 
  of the designs
  theoretically mappable
  to a single DSP 
  according to its
  configuration manual.
This microbenchmark set
  aims to capture
  the real-world designs
  which hardware designers
  would attempt to map
  to a platform's DSP.
For each architecture, we compare \lr
  to both the corresponding
  state-of-the-art toolchain for
  the architecture 
  as well as to Yosys.
For Xilinx Ultrascale+ and 7-series, 
  the DSP48E2 (resp.~DSP48E1) configuration manual details the
  structure of designs mappable
  to the primitive.
Our designs for Xilinx include 
  all permutations of the design form
  $((a \pm b) * c) \pm d$,
  as well as designs of the forms $(a * b)$, $(a\pm b) * c$ and $((a * b) \pm c)$.
We pipeline each of these
  workloads from zero to three stages 
  and use bitwidths from 8 to 18 bits.
For the DSP on Lattice, we similarly enumerate
  all designs of the form $(a * b) \odot c$, 
  where $\odot \in \{\&, |, \oplus, \pm\}$, and of the form $(a * b)$.
For each of these designs, 
  we use zero to two stages
  and bitwidths from 8 to 18 bits.
This results in
    792 microbenchmarks for Xilinx UltraScale+,
    396 for Lattice ECP5,
    and 66 for Intel Cyclone 10 LP.
Though \lr's output is correct by construction,
  we further validate its output
  by simulating each \lr-compiled design
  over thousands of consecutive cycles
  using Verilator.\tighten

\begin{figure}
    \centering
\includegraphics[width=0.7\textwidth]{lakeroad/assets/succeeded\_vs\_failed\_all.png}

\vspace{1em}
\footnotesize

    \begin{tabular}{|l|S[table-format=3.2]|S[table-format=3.2]|S[table-format=3.2]|}
    \hline
     Tool & {Median Time (s)} & \multicolumn{2}{c|}{Min / Max Time (s)} \\ \hline
    \multicolumn{4}{|c|}{\textbf{Xilinx UltraScale+}} \\ 
    \hline
         \lr & 14.99 & 2.99 & 127.70 \\  \hline
         SOTA Xilinx & 261.61 & 227.82 & 598.67 \\ \hline
         Yosys & 14.97 & 6.66 & 21.10 \\ \hline
    \multicolumn{4}{|c|}{\textbf{Xilinx 7-series}} \\ 
    \hline
         \lr & 5.63 & 3.13 & 62.68 \\  \hline
         SOTA Xilinx & 94.50 & 89.32 & 111.61 \\ \hline
         Yosys & 6.99 & 5.64 & 9.39 \\ \hline
    \multicolumn{4}{|c|}{\textbf{Lattice ECP5}} \\
    \hline
         \lr & 9.49 & 6.70 & 55.23 \\ \hline
         SOTA Lattice & 2.32 & 0.95 & 4.52 \\ \hline
         Yosys & 2.31 & 0.90 & 4.01 \\ \hline
    \multicolumn{4}{|c|}{\textbf{Intel Cyclone 10 LP}} \\
    \hline
         \lr & 2.92 & 2.12 & 4.13 \\ \hline
         SOTA Intel & 38.73 & 19.11 & 43.49 \\ \hline
         Yosys & 0.96 & 0.48 & 1.88 \\ \hline
    \end{tabular}


    \caption{
Results of 
  the completeness experiments
  described in \cref{sec:completeness},
  measuring the completeness 
  of technology mapping tools for DSPs on Xilinx UltraScale+, Xilinx 7-series, Lattice ECP5, and Intel Cyclone 10 LP,
  plus timing information.
A single bar in the bar chart
  communicates,
  for a given FPGA architecture
  and technology mapper,
  the proportion of the microbenchmarks
  that the given technology mapper could map to a single DSP.
In \lr's case, experiments can 
  either succeed 
  (\lr maps the microbenchmark
    to a single DSP),
  timeout,
  or return UNSAT.
For the other tools,
  experiments can either
  succeed
  or fail
  (i.e., the tool returns a mapping,
    but the mapping uses more than
    a single DSP).
There are a total of 
  792 experiments/microbenchmarks for Xilinx,
  396 for Lattice,
  and 66 for Intel.
    }
    \label{fig:xilinx-completeness}
\end{figure}

\paragraph{\textnormal{\textit{\textbf{Comparison to Existing Toolchains.}}}}
As demonstrated in Figure~\ref{fig:xilinx-completeness} (top),
  \lr maps $29\times$ more
  designs than Yosys
  and $1.4\times$ more designs
  than the proprietary,
  state-of-the-art toolchain on Xilinx Ultrascale+.
On Lattice ECP5,
  \lr maps $6.0\times$ more
  designs than Yosys 
  and $3.6\times$ more designs
  than the proprietary,
  state-of-the-art toolchain.
On Intel Cyclone 10 LP,
  \lr successfully maps all designs:
  $3\times$ more designs
  than the proprietary,
  state-of-the-art toolchain for Intel.
Yosys fails to map a single design
  on Intel.
State-of-the-art toolchains
  for all architectures fail
  to map more than half
  of the queried designs.
\lr times out on 
less than 20\% of designs.%
\footnote{We restricted Rosette
synthesis time to 
120 seconds, 40 seconds, and 20 seconds for
Xilinx, Lattice, and Intel
respectively, and
marked failure past that (though bitvector synthesis problems
are decidable).}
Note that \lr
  returns ``UNSAT'' on 
  a number of
  designs on Xilinx UltraScale+ and 7-series, i.e., 
  \lr claims there is
  \textit{no} possible mapping
  to a DSP48E2/DSP48E1 for the
  requested workload.
In all of these cases,
  both Xilinx SOTA
  and Yosys
  agree with \lr 
  and do not map the designs
  to a single DSP.
We conclude that 
  the set of designs we presented in 
  \textit{Evaluation Setup}
  must be overly broad;
  though the documentation implies
  that all of these designs are mappable
  to a single DSP,
  all three Xilinx synthesis tools 
  surveyed indicate that they are
  indeed not mappable.

\begin{figure}
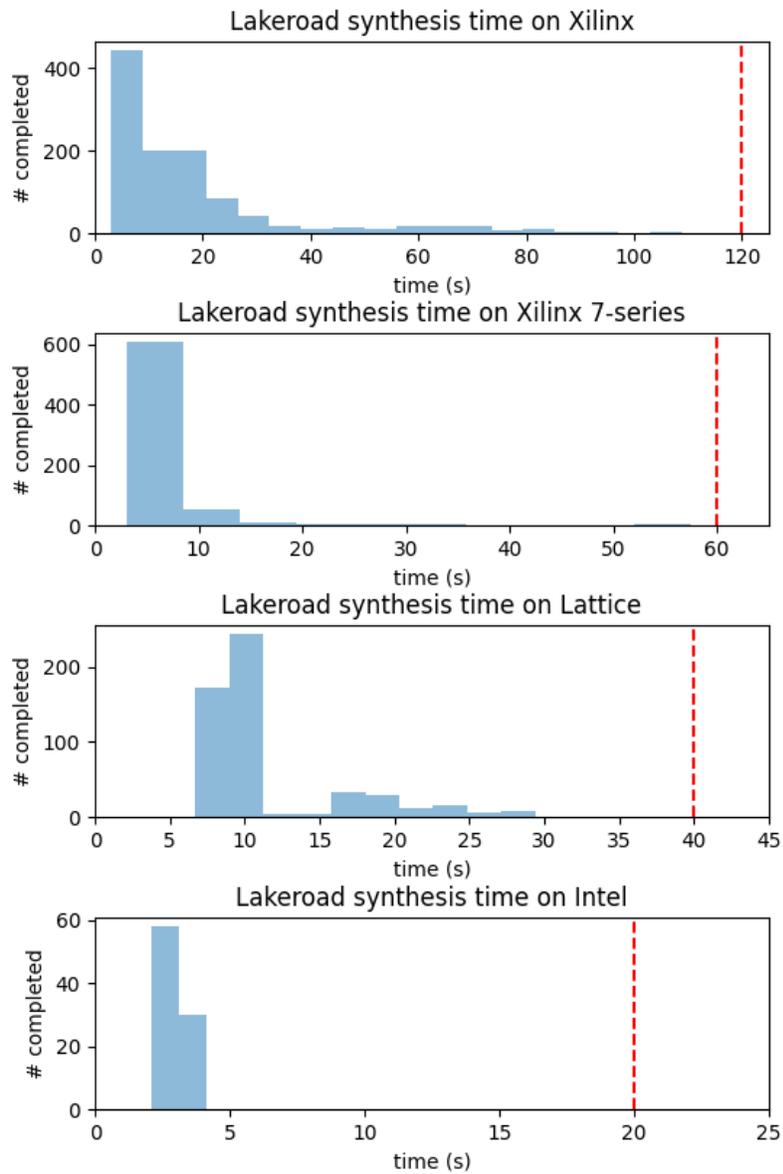

    \centering
    \includegraphics[width=0.7\textwidth]{lakeroad/assets/time/lakeroad\_time\_xilinx.png}
    \includegraphics[width=0.7\textwidth]{lakeroad/assets/time/lakeroad\_time\_xilinx\_7\_series.png}
    \includegraphics[width=0.7\textwidth]{lakeroad/assets/time/lakeroad\_time\_lattice.png}
    \includegraphics[width=0.7\textwidth]{lakeroad/assets/time/lakeroad\_time\_intel.png}
\caption{
Histograms of \lr program synthesis runtime
  for all terminating
  (success or UNSAT) \lr experiments
  described in \cref{sec:completeness}, with timeout thresholds indicated with a vertical
  dotted red line.
}
    \label{fig:lakeroad-synth-time}
\end{figure}

For timing, we compared the mapping time for each
  of the tools
  and report the results
  in Figure~\ref{fig:xilinx-completeness} (bottom).
The wide ranges for \lr
  show that solver time for different
  program synthesis queries
  is highly variable.
This is explored more deeply in
  \cref{fig:lakeroad-synth-time},
  which shows that most synthesis queries
  terminate quickly,
  with a long tail of slower queries.
Note that the state-of-the-art technology
mapper for Ultrascale+/7-series has a slow running time
  due to its long start-up process.\tighten

Regarding which solvers
  in the portfolio were
  most useful,
  of all terminating
  (success or UNSAT)
  \lr experiments,
  Bitwuzla was the first
  to complete
  for 806 of them,
  STP for 595,
  Yices2 for 538,
  and cvc5 for 54.

\paragraph{\textnormal{\textit{\textbf{Discussion.}}}}
Compared to Yosys,
  it is clear that
  \lr provides more complete support
  for
  programmable DSPs.
However, \lr's greater completeness
  over Yosys
  is perhaps not surprising since 
  Yosys is an open-source tool
  still under active development.
Part of the appeal
  of the Yosys toolchain
  is the diversity of backends
  it can target;
  these results show that, if incorporated
  into Yosys, \lr
  would further increase
  Yosys's flexibility and generality.
Perhaps most surprising
  is that \lr is more complete
  than
  specialized proprietary toolchains. 
Even the UNSAT results \lr produces 
  can be useful to designers 
  since they indicate
  potential flaws
  in the documentation or vendor-provided semantics.
In the context of a larger
  synthesis tool, \lr
  would provide stronger
  guarantees for mapping
  modules of larger designs.

\newcommand{\LatticeResourceTable}{add  & 1 & 1 & 2 &  &  & 1 &  &  &  \\
add  & 8 & 8 & 8 &  &  & 25 &  & 9 & 4 \\
add  & 32 & 32 & 32 &  &  & 135 &  & 55 & 28 \\
add  & 64 & 64 & 64 &  &  & 293 &  & 82 & 50 \\
and  & 1 & 1 &  &  &  & 1 &  &  &  \\
and  & 8 & 8 &  &  &  & 8 &  &  &  \\
and  & 32 & 32 &  &  &  & 32 &  &  &  \\
and  & 64 & 64 &  &  &  & 64 &  &  &  \\
not  & 1 & 1 &  &  &  & 1 &  &  &  \\
not  & 8 & 8 &  &  &  & 8 &  &  &  \\
not  & 32 & 32 &  &  &  & 32 &  &  &  \\
not  & 64 & 64 &  &  &  & 64 &  &  &  \\
mul  & 1 & 2 & 2 &  &  & 1 &  &  &  \\
mul  & 8 & 100 & 64 &  &  & 130 &  & 49 & 29 \\
mux  & 1 & 1 &  &  &  & 1 &  &  &  \\
mux  & 8 & 8 &  &  &  & 8 &  &  &  \\
mux  & 32 & 32 &  &  &  & 32 &  &  &  \\
mux  & 64 & 64 &  &  &  & 64 &  &  &  \\
sub  & 1 & 1 & 2 &  &  & 1 &  &  &  \\
sub  & 8 & 8 & 8 &  &  & 25 &  & 9 & 4 \\
sub  & 32 & 32 & 32 &  &  & 135 &  & 55 & 28 \\
sub  & 64 & 64 & 64 &  &  & 293 &  & 82 & 50 \\
}
\newcommand{\XilinxResourceTable}{add & 1 & 1 & 1 & \checkmark & 1 &  &  \\
add & 8 & 8 & 1 & \checkmark & 11 &  &  \\
add & 32 & 32 & 4 &  & 32 & 4 &  \\
add & 64 & 64 & 8 &  & 64 & 8 &  \\
add & 128 & 128 & 16 &  &  &  &  \\
and & 1 & 1 &  & \checkmark & 1 &  &  \\
and & 8 & 8 &  & \checkmark & 8 &  &  \\
and & 32 & 32 &  &  & 32 &  &  \\
and & 64 & 64 &  &  & 64 &  &  \\
and & 128 & 128 &  &  &  &  &  \\
not & 1 & 1 &  & \checkmark &  &  &  \\
not & 8 & 8 &  & \checkmark &  &  &  \\
not & 32 & 32 &  &  &  &  &  \\
not & 64 & 64 &  &  &  &  &  \\
not & 128 & 128 &  &  &  &  &  \\
mul & 1 & 2 & 1 & \checkmark & 1 &  &  \\
mul & 8 & 100 & 8 & \checkmark & 39 & 3 &  \\
mux & 1 & 1 &  & \checkmark & 1 &  &  \\
mux & 8 & 8 &  & \checkmark & 8 &  &  \\
mux & 32 & 32 &  &  & 32 &  &  \\
mux & 64 & 64 &  &  & 64 &  &  \\
mux & 128 & 128 &  &  &  &  &  \\
sub & 1 & 1 & 1 & \checkmark & 1 &  &  \\
sub & 8 & 8 & 1 & \checkmark & 11 &  &  \\
sub & 32 & 32 & 4 &  & 32 & 4 &  \\
sub & 64 & 64 & 8 &  & 64 & 8 &  \\
sub & 128 & 128 & 16 &  &  &  &  \\
}
\newcommand{\SofaResourceTable}{add  & 1 & 3 &  -  \\
add  & 8 & 24 &  -  \\
add  & 32 & 96 &  -  \\
add  & 64 & 192 &  -  \\
add  & 128 & 384 &  -  \\
and  & 1 & 1 &  -  \\
and  & 8 & 8 &  -  \\
and  & 32 & 32 &  -  \\
and  & 64 & 64 &  -  \\
and  & 128 & 128 &  -  \\
not  & 1 & 1 &  -  \\
not  & 8 & 8 &  -  \\
not  & 32 & 32 &  -  \\
not  & 64 & 64 &  -  \\
not  & 128 & 128 &  -  \\
mul  & 1 & 4 &  -  \\
mul  & 8 & 228 &  -  \\
mux  & 1 & 1 &  -  \\
mux  & 8 & 8 &  -  \\
mux  & 32 & 32 &  -  \\
mux  & 64 & 64 &  -  \\
mux  & 128 & 128 &  -  \\
sub  & 1 & 3 &  -  \\
sub  & 8 & 24 &  -  \\
sub  & 32 & 96 &  -  \\
sub  & 64 & 192 &  -  \\
sub  & 128 & 384 &  -  \\
}
\newcommand{\BigTable}{\rotatebox{\titletilt}{ Instr } & \rotatebox{\titletilt}{ Template } & \rotatebox{\titletilt}{ Runtime in sec } & \rotatebox{\titletilt}{ \#LUTs } & \rotatebox{\titletilt}{ \#CARRY8s } & \rotatebox{\titletilt}{ DSP? (sec) } & \rotatebox{\titletilt}{ Runtime } & \rotatebox{\titletilt}{ \#LUTs } & \rotatebox{\titletilt}{ \#MUXs } & \rotatebox{\titletilt}{ \#CARRY4s } & \rotatebox{\titletilt}{ \#INVs } & \rotatebox{\titletilt}{ DSP? } & \rotatebox{\titletilt}{ Runtime in sec } & \rotatebox{\titletilt}{ \#LUTs } & \rotatebox{\titletilt}{ \#CARRY8s } & \rotatebox{\titletilt}{ DSP? } & \rotatebox{\titletilt}{ Support } & \rotatebox{\titletilt}{ Runtime in sec } & \rotatebox{\titletilt}{ \#LUTs } & \rotatebox{\titletilt}{ \#CCU2Cs } & \rotatebox{\titletilt}{ DSP? } & \rotatebox{\titletilt}{ Runtime in sec } & \rotatebox{\titletilt}{ \#LUTs } & \rotatebox{\titletilt}{ \#MUXs } & \rotatebox{\titletilt}{ \#CCU2Cs } & \rotatebox{\titletilt}{ DSP? } & \rotatebox{\titletilt}{ Runtime in sec } & \rotatebox{\titletilt}{ \#LUTs } & \rotatebox{\titletilt}{ \#CCU2Cs } & \rotatebox{\titletilt}{ DSP? } & \rotatebox{\titletilt}{ Support } & \rotatebox{\titletilt}{ Runtime in sec } & \rotatebox{\titletilt}{ \#LUTs } \\\midrule
and8 & BW & 1.5 & 4 & -- & \checkmark (3.0) & 2.6 & 8 & -- & -- & -- &   & 32 & 4 & -- & ? & $\times$ & 1.6 & 8 & -- & ? & 1.5 & 8 & -- & -- &   & 1.1 & 8 & -- & ? & $\times$ & 1.8 & 8 \\
and16 & BW & 1.6 & 10 & -- & \checkmark (4.8) & 2.6 & 16 & -- & -- & -- &   & 36 & 10 & -- & ? & $\times$ & 1.6 & 16 & -- & ? & 1.5 & 16 & -- & -- &   & 0.6 & 16 & -- & ? & $\times$ & 2.2 & 16 \\
and32 & BW & 1.9 & 17 & -- &   & 2.6 & 32 & -- & -- & -- &   & 35 & 17 & -- & ? & $\times$ & 2.0 & 32 & -- & ? & 1.5 & 32 & -- & -- &   & 1.1 & 32 & -- & ? & $\times$ & 2.8 & 32 \\
or8 & BW & 1.7 & 4 & -- & \checkmark (7.4) & 2.5 & 8 & -- & -- & -- &   & 32 & 4 & -- & ? & $\times$ & 1.9 & 8 & -- & ? & 1.4 & 8 & -- & -- &   & 0.9 & 8 & -- & ? & $\times$ & 2.3 & 8 \\
or16 & BW & 1.8 & 10 & -- & \checkmark (7.1) & 2.6 & 16 & -- & -- & -- &   & 34 & 10 & -- & ? & $\times$ & 2.1 & 16 & -- & ? & 1.4 & 16 & -- & -- &   & 0.7 & 16 & -- & ? & $\times$ & 2.4 & 16 \\
or32 & BW & 1.7 & 17 & -- &   & 2.6 & 32 & -- & -- & -- &   & 31 & 17 & -- & ? & $\times$ & 2.2 & 32 & -- & ? & 1.5 & 32 & -- & -- &   & 0.6 & 32 & -- & ? & $\times$ & 3.1 & 32 \\
xor8 & BW & 1.7 & 4 & -- & \checkmark (5.1) & 2.6 & 8 & -- & -- & -- &   & 33 & 4 & -- & ? & $\times$ & 1.9 & 8 & -- & ? & 1.5 & 8 & -- & -- &   & 0.6 & 8 & -- & ? & $\times$ & 2.6 & 8 \\
xor16 & BW & 1.9 & 10 & -- & \checkmark (4.4) & 2.7 & 16 & -- & -- & -- &   & 32 & 10 & -- & ? & $\times$ & 2.0 & 16 & -- & ? & 1.5 & 16 & -- & -- &   & 1.0 & 16 & -- & ? & $\times$ & 2.4 & 16 \\
xor32 & BW & 2.2 & 17 & -- &   & 2.5 & 32 & -- & -- & -- &   & 33 & 17 & -- & ? & $\times$ & 1.8 & 32 & -- & ? & 1.5 & 32 & -- & -- &   & 0.7 & 32 & -- & ? & $\times$ & 2.4 & 32 \\
not8 & BW & 1.6 & 4 & -- & \checkmark (2.8) & 2.6 & -- & -- & -- & 8 &   & 34 & 4 & -- & ? & $\times$ & 1.8 & 8 & -- & ? & 1.4 & 8 & -- & -- &   & 0.7 & 8 & -- & ? & $\times$ & 1.9 & 8 \\
not16 & BW & 1.8 & 10 & -- & \checkmark (3.6) & 2.6 & -- & -- & -- & 16 &   & 32 & 10 & -- & ? & $\times$ & 1.8 & 16 & -- & ? & 1.5 & 16 & -- & -- &   & 0.6 & 16 & -- & ? & $\times$ & 2.6 & 16 \\
not32 & BW & 1.7 & 17 & -- &   & 2.6 & -- & -- & -- & 32 &   & 33 & 17 & -- & ? & $\times$ & 1.7 & 32 & -- & ? & 1.5 & 32 & -- & -- &   & 0.6 & 32 & -- & ? & $\times$ & 2.7 & 32 \\
mux8 & BW & 6.0 & 8 & -- & \checkmark (4.1) & 2.6 & 8 & -- & -- & -- &   & 35 & 8 & -- & ? & $\times$ & 2.0 & 8 & -- & ? & 1.6 & 8 & -- & -- &   & 1.1 & 8 & -- & ? & $\times$ & 2.7 & 8 \\
mux16 & BW & 8.5 & 16 & -- & \checkmark (6.3) & 2.6 & 16 & -- & -- & -- &   & 31 & 16 & -- & ? & $\times$ & 2.0 & 16 & -- & ? & 1.5 & 16 & -- & -- &   & 0.7 & 16 & -- & ? & $\times$ & 2.6 & 16 \\
mux32 & BW & 12.7 & 32 & -- &   & 2.8 & 32 & -- & -- & -- &   & 22 & 32 & -- & ? & $\times$ & 2.5 & 32 & -- & ? & 1.6 & 32 & -- & -- &   & 0.7 & 32 & -- & ? & $\times$ & 2.9 & 32 \\
add8 & Carry & 2.0 & 8 & 1 & \checkmark (3.7) & 2.6 & 8 & -- & 2 & -- &   & 33 & 11 & -- & ? & $\times$ & 2.3 & 8 & 4 & ? & 1.4 & -- & -- & 4 &   & 0.7 & 1 & 4 & ? & $\times$ & 2.6 & 24 \\
add16 & Carry & 2.3 & 16 & 2 & \checkmark (4.8) & 2.8 & 16 & -- & 4 & -- &   & 34 & 16 & 2 & ? & $\times$ & 2.4 & 16 & 8 & ? & 1.0 & -- & -- & 8 &   & 1.1 & 1 & 8 & ? & $\times$ & 2.9 & 48 \\
add32 & Carry & 2.3 & 32 & 4 &   & 2.7 & 32 & -- & 8 & -- &   & 34 & 32 & 4 & ? & $\times$ & 3.5 & 32 & 16 & ? & 1.5 & -- & -- & 16 &   & 0.7 & 1 & 16 & ? & $\times$ & 4.7 & 96 \\
sub8 & Carry & 2.0 & 8 & 1 & \checkmark (3.3) & 2.6 & 8 & -- & 2 & -- &   & 33 & 11 & -- & ? & $\times$ & 2.3 & 8 & 4 & ? & 1.4 & -- & -- & 4 &   & 0.8 & -- & 5 & ? & $\times$ & 2.6 & 24 \\
sub16 & Carry & 1.9 & 16 & 2 & \checkmark (3.9) & 2.7 & 16 & -- & 4 & -- &   & 34 & 16 & 2 & ? & $\times$ & 2.4 & 16 & 8 & ? & 1.4 & -- & -- & 8 &   & 0.8 & -- & 9 & ? & $\times$ & 3.2 & 48 \\
sub32 & Carry & 1.9 & 32 & 4 &   & 2.7 & 32 & -- & 8 & -- &   & 34 & 32 & 4 & ? & $\times$ & 3.3 & 32 & 16 & ? & 1.5 & -- & -- & 16 &   & 1.2 & -- & 17 & ? & $\times$ & 3.7 & 96 \\
mul8 & Mult & 2.0 & 91 & 8 & \checkmark (7.6) & 2.8 & 55 & 15 & 2 & -- &   & 36 & 33 & 3 & ? & $\times$ & 3.9 & 100 & 32 & ? & 1.9 & -- & -- & -- & \checkmark & 0.8 & -- & -- & ? & $\times$ & 6.2 & 228 \\
mul16 & DSP & 2.4 & -- & -- & \checkmark (2.4) & 3.4 & -- & -- & -- & -- & \checkmark & 32 & -- & -- & ? & $\times$ & -- & -- & -- & ? & -- & -- & -- & -- &   & -- & -- & -- & ? & $\times$ & -- & -- \\
mul32 & -- & -- & -- & -- &   & -- & -- & -- & -- & -- &   & -- & -- & -- & ? & $\times$ & -- & -- & -- & ? & -- & -- & -- & -- &   & -- & -- & -- & ? & $\times$ & -- & -- \\
eq8 & Comp & 1.8 & 8 & 1 &   & 2.6 & 3 & -- & -- & -- &   & 37 & 3 & -- & ? & $\times$ & 2.4 & 16 & 4 & ? & 1.4 & 5 & -- & -- &   & 0.8 & -- & 3 & ? & $\times$ & 4.6 & 32 \\
eq16 & Comp & 1.8 & 16 & 2 &   & 2.7 & 9 & 12 & -- & -- &   & 35 & 7 & -- & ? & $\times$ & 3.4 & 32 & 8 & ? & 1.5 & 12 & 1 & -- &   & 0.7 & -- & 5 & ? & $\times$ & 5.6 & 64 \\
eq32 & Comp & 3.3 & 32 & 4 &   & 2.6 & 24 & 1 & -- & -- &   & 36 & 11 & 2 & ? & $\times$ & 3.9 & 64 & 16 & ? & 1.6 & 25 & 2 & -- &   & 0.6 & -- & 9 & ? & $\times$ & 6.1 & 128 \\
neq8 & Comp & 2.1 & 8 & 1 &   & 2.6 & 3 & -- & -- & -- &   & 33 & 3 & -- & ? & $\times$ & 3.5 & 16 & 4 & ? & 1.5 & 5 & -- & -- &   & 0.7 & 1 & 3 & ? & $\times$ & 3.3 & 32 \\
neq16 & Comp & 2.6 & 16 & 2 &   & 2.7 & 9 & 12 & -- & -- &   & 23 & 7 & -- & ? & $\times$ & 2.6 & 32 & 8 & ? & 1.6 & 12 & 1 & -- &   & 1.1 & 1 & 5 & ? & $\times$ & 5.3 & 64 \\
neq32 & Comp & 3.5 & 32 & 4 &   & 2.7 & 24 & 1 & -- & -- &   & 33 & 11 & 2 & ? & $\times$ & 4.9 & 64 & 16 & ? & 1.6 & 25 & 2 & -- &   & 0.7 & 1 & 9 & ? & $\times$ & 7.0 & 128 \\
ugt8 & Comp & 2.7 & 8 & 1 &   & 2.6 & 6 & -- & 1 & -- &   & 37 & 4 & 1 & ? & $\times$ & 5.9 & 16 & 4 & ? & 1.5 & 1 & -- & 4 &   & 0.9 & 14 & -- & ? & $\times$ & 6.3 & 32 \\
ugt16 & Comp & 3.5 & 16 & 2 &   & 2.7 & 12 & -- & 2 & -- &   & 38 & 8 & 1 & ? & $\times$ & 7.4 & 32 & 8 & ? & 1.5 & 1 & -- & 8 &   & 1.0 & 35 & -- & ? & $\times$ & 6.1 & 64 \\
ugt32 & Comp & 3.4 & 32 & 4 &   & 2.7 & 22 & -- & 3 & -- &   & 32 & 16 & 2 & ? & $\times$ & 9.0 & 64 & 16 & ? & 1.5 & 1 & -- & 16 &   & 0.6 & 1 & 17 & ? & $\times$ & 10.0 & 128 \\
ult8 & Comp & 2.7 & 8 & 1 &   & 2.6 & 6 & -- & 1 & -- &   & 30 & 4 & 1 & ? & $\times$ & 5.8 & 16 & 4 & ? & 1.6 & 6 & 1 & 4 &   & 0.7 & 14 & -- & ? & $\times$ & 6.1 & 32 \\
ult16 & Comp & 3.3 & 16 & 2 &   & 2.7 & 12 & -- & 2 & -- &   & 32 & 8 & 1 & ? & $\times$ & 7.5 & 32 & 8 & ? & 1.6 & 16 & 2 & 8 &   & 1.0 & 35 & -- & ? & $\times$ & 6.2 & 64 \\
ult32 & Comp & 3.7 & 32 & 4 &   & 2.7 & 22 & -- & 3 & -- &   & 34 & 16 & 2 & ? & $\times$ & 9.1 & 64 & 16 & ? & 1.7 & 45 & 19 & 16 &   & 0.7 & 1 & 17 & ? & $\times$ & 10.3 & 128 \\
uge8 & Comp & 2.5 & 8 & 1 &   & 2.7 & 6 & -- & 1 & -- &   & 32 & 4 & 1 & ? & $\times$ & 4.6 & 16 & 4 & ? & 1.6 & 6 & 1 & 4 &   & 0.8 & 14 & -- & ? & $\times$ & 4.3 & 32 \\
uge16 & Comp & 2.4 & 16 & 2 &   & 2.7 & 12 & -- & 2 & -- &   & 34 & 8 & 1 & ? & $\times$ & 3.5 & 32 & 8 & ? & 1.6 & 16 & 2 & 8 &   & 1.0 & 35 & -- & ? & $\times$ & 6.7 & 64 \\
uge32 & Comp & 2.8 & 32 & 4 &   & 2.8 & 22 & -- & 3 & -- &   & 33 & 16 & 2 & ? & $\times$ & 11.4 & 64 & 16 & ? & 1.6 & 45 & 19 & 16 &   & 0.7 & -- & 17 & ? & $\times$ & 6.5 & 128 \\
ule8 & Comp & 2.6 & 8 & 1 &   & 2.7 & 6 & -- & 1 & -- &   & 34 & 4 & 1 & ? & $\times$ & 4.8 & 16 & 4 & ? & 1.4 & 6 & 1 & 4 &   & 0.6 & 14 & -- & ? & $\times$ & 3.8 & 32 \\
ule16 & Comp & 2.5 & 16 & 2 &   & 2.7 & 12 & -- & 2 & -- &   & 32 & 8 & 1 & ? & $\times$ & 3.6 & 32 & 8 & ? & 1.6 & 16 & 2 & 8 &   & 0.9 & 35 & -- & ? & $\times$ & 5.5 & 64 \\
ule32 & Comp & 2.5 & 32 & 4 &   & 2.8 & 22 & -- & 3 & -- &   & 32 & 16 & 2 & ? & $\times$ & 11.1 & 64 & 16 & ? & 1.8 & 45 & 19 & 16 &   & 0.9 & -- & 17 & ? & $\times$ & 5.7 & 128}
\newcommand{\LatticeTable}{\rotatebox{\titletilt}{ Instr } & \rotatebox{\titletilt}{ Template } & \rotatebox{\titletilt}{ Runtime in sec } & \rotatebox{\titletilt}{ \#LUTs } & \rotatebox{\titletilt}{ \#CCU2Cs } & \rotatebox{\titletilt}{ MUL? (sec) } & \rotatebox{\titletilt}{ Runtime in sec } & \rotatebox{\titletilt}{ \#LUTs } & \rotatebox{\titletilt}{ \#MUXs } & \rotatebox{\titletilt}{ \#CCU2Cs } & \rotatebox{\titletilt}{ \#MUL } & \rotatebox{\titletilt}{ Runtime in sec } & \rotatebox{\titletilt}{ \#LUTs } & \rotatebox{\titletilt}{ \#CCU2Cs } & \rotatebox{\titletilt}{ \#MUL } & \rotatebox{\titletilt}{ \#ALU } & \rotatebox{\titletilt}{ Support }  \\\midrule
and8 & BW    & 1.6  & 8 & --     &                   & 1.5 & 8 & -- & -- &         --      & 1.1 & 8  & -- &-- &-- & $\times$ \\
and16 & BW   & 1.6 & 16 & --    &                   & 1.5 & 16 & -- & -- &        --      & 0.6 & 16 & -- &-- &-- & $\times$ \\
and32 & BW   & 2.0 & 32 & --    &                   & 1.5 & 32 & -- & -- &        --      & 1.1 & 32 & -- &-- &-- & $\times$ \\
or8 & BW     & 1.9 & 8 & --       &                   & 1.4 & 8 & -- & -- &         --      & 0.9 & 8  & -- &-- &-- & $\times$ \\
or16 & BW    & 2.1 & 16 & --     &                   & 1.4 & 16 & -- & -- &        --      & 0.7 & 16 & -- &-- &-- & $\times$ \\
or32 & BW    & 2.2 & 32 & --     &                   & 1.5 & 32 & -- & -- &        --      & 0.6 & 32 & -- &-- &-- & $\times$ \\
xor8 & BW    & 1.9 & 8 & --      &                   & 1.5 & 8 & -- & -- &         --      & 0.6 & 8  & -- &-- &-- & $\times$ \\
xor16 & BW   & 2.0 & 16 & --    &                   & 1.5 & 16 & -- & -- &        --      & 1.0 & 16 & -- &-- &-- & $\times$ \\
xor32 & BW   & 1.8 & 32 & --    &                   & 1.5 & 32 & -- & -- &        --      & 0.7 & 32 & -- &-- &-- & $\times$ \\
not8 & BW    & 1.8 & 8 & --      &                   & 1.4 & 8 & -- & -- &         --      & 0.7 & 8  & -- &-- &-- & $\times$ \\
not16 & BW   & 1.8 & 16 & --    &                   & 1.5 & 16 & -- & -- &        --      & 0.6 & 16 & -- &-- &-- & $\times$ \\
not32 & BW   & 1.7 & 32 & --    &                   & 1.5 & 32 & -- & -- &        --      & 0.6 & 32 & -- &-- &-- & $\times$ \\
mux8 & BW    & 2.0 & 8 & --      &                   & 1.6 & 8 & -- & -- &         --      & 1.1 & 8  & -- &-- &-- & $\times$ \\
mux16 & BW   & 2.0 & 16 & --    &                   & 1.5 & 16 & -- & -- &        --      & 0.7 & 16 & -- &-- &-- & $\times$ \\
mux32 & BW   & 2.5 & 32 & --    &                   & 1.6 & 32 & -- & -- &        --      & 0.7 & 32 & -- &-- &-- & $\times$ \\
add8 & Carry & 2.3 & 8 & 4    &                   & 1.4 & -- & -- & 4 &         --      & 0.7 & 1  & 4  &-- &-- & $\times$ \\
add16 & Carry& 2.4 & 16 & 8  &                   & 1.0 & -- & -- & 8 &      --      & 1.1 & 1  & 8  &-- &-- & $\times$ \\
add32 & Carry& 3.5 & 32 & 16 &                   & 1.5 & -- & -- & 16 &     --      & 0.7 & 1  & 16 &-- &-- & $\times$ \\
sub8 & Carry & 2.3 & 8 & 4    &                   & 1.4 & -- & -- & 4 &      --      & 0.8  & --& 5  &-- &-- & $\times$ \\
sub16 & Carry& 2.4 & 16 & 8  &                   & 1.4 & -- & -- & 8 &      --      & 0.8  & --& 9  &-- &-- & $\times$ \\
sub32 & Carry& 3.3 & 32 & 16 &                   & 1.5 & -- & -- & 16 &     --      & 1.2 & -- & 17 &-- &-- & $\times$ \\
mul8 & Mult  & 3.9 & 100 & 32  & \checkmark (0.8)     & 1.9 & -- & -- & -- &     1       & 0.8 & -- & -- &-- &-- & $\times$ \\
mul16 & --  & -- & -- & --    & \checkmark (0.8)  & 1.3 & -- & -- & -- &      1      & 0.9  & -- & --  & 1 &-- & $\times$ \\
mul32 & --   & -- & -- & --     &                   & 1.1 & -- & -- & 14 &      3      & 2.8  & -- & --  & 4 & 2 & $\times$ \\
eq8 & Comp   & 2.4 & 16 & 4     &                   & 1.4 & 5 & -- & -- &      --      & 0.8 & -- & 3  &-- &-- & $\times$ \\
eq16 & Comp  & 3.4 & 32 & 8    &                   & 1.5 & 12 & 1 & -- &      --      & 0.7 & -- & 5  &-- &-- & $\times$ \\
eq32 & Comp  & 3.9 & 64 & 16   &                   & 1.6 & 25 & 2 & -- &      --      & 0.6 & -- & 9  &-- &-- & $\times$ \\
neq8 & Comp  & 3.5 & 16 & 4    &                   & 1.5 & 5 & -- & -- &      --      & 0.7 & 1  & 3  &-- &-- & $\times$ \\
neq16 & Comp & 2.6 & 32 & 8   &                   & 1.6 & 12 & 1 & -- &      --      & 1.1 & 1  & 5  &-- &-- & $\times$ \\
neq32 & Comp & 4.9 & 64 & 16  &                   & 1.6 & 25 & 2 & -- &      --      & 0.7 & 1  & 9  &-- &-- & $\times$ \\
ugt8 & Comp  & 5.9 & 16 & 4    &                   & 1.5 & 1 & -- & 4 &       --      & 0.9 & 14 & -- &-- &-- & $\times$ \\
ugt16 & Comp & 7.4 & 32 & 8   &                   & 1.5 & 1 & -- & 8 &       --      & 1.0 & 35 & -- &-- &-- & $\times$ \\
ugt32 & Comp & 9.0 & 64 & 16  &                   & 1.5 & 1 & -- & 16 &      --      & 0.6 & 1 & 17  &-- &-- & $\times$ \\
ult8 & Comp  & 5.8 & 16 & 4 &                      & 1.6 & 6 & 1 & 4 &           --      & 0.7 & 14 & -- &-- &-- & $\times$ \\
ult16 & Comp & 7.5 & 32 & 8 &                     & 1.6 & 16 & 2 & 8 &         --      & 1.0 & 35 & -- &-- &-- & $\times$ \\
ult32 & Comp & 9.1 & 64 & 16 &                    & 1.7 & 45 & 19 & 16 &      --      & 0.7 & 1 & 17  &-- &-- & $\times$ \\
uge8 & Comp  & 4.6 & 16 & 4 &                      & 1.6 & 6 & 1 & 4 &           --      & 0.8 & 14 & -- &-- &-- & $\times$ \\
uge16 & Comp & 3.5 & 32 & 8 &                     & 1.6 & 16 & 2 & 8 &         --      & 1.0 & 35 & -- &-- &-- & $\times$ \\
uge32 & Comp & 11.4 & 64 & 16 &   & 1.6 & 45 & 19 & 16 &     --      & 0.7 & -- & 17 &-- &-- & $\times$ \\
ule8 & Comp  & 4.8 & 16 & 4 &   & 1.4 & 6 & 1 & 4 &           --      & 0.6 & 14 & -- &-- &-- & $\times$ \\
ule16 & Comp & 3.6 & 32 & 8 &   & 1.6 & 16 & 2 & 8 &         --      & 0.9 & 35 & -- &-- &-- & $\times$ \\
ule32 & Comp & 11.1 & 64 & 16 &   & 1.8 & 45 & 19 & 16 &     --      & 0.9 & -- & 17 &-- &-- & $\times$}
\newcommand{\XilinxTable}{\rotatebox{\titletilt}{ Instr } & \rotatebox{\titletilt}{ Template } & \rotatebox{\titletilt}{ Runtime in sec } & \rotatebox{\titletilt}{ \#LUTs } & \rotatebox{\titletilt}{ \#CARRY8s } & \rotatebox{\titletilt}{ DSP? (sec) } & \rotatebox{\titletilt}{ Runtime } & \rotatebox{\titletilt}{ \#LUTs } & \rotatebox{\titletilt}{ \#MUXs } & \rotatebox{\titletilt}{ \#CARRY4s } & \rotatebox{\titletilt}{ \#INVs } & \rotatebox{\titletilt}{ \#DSP } & \rotatebox{\titletilt}{ Runtime in sec } & \rotatebox{\titletilt}{ \#LUTs } & \rotatebox{\titletilt}{ \#CARRY8s } & \rotatebox{\titletilt}{ \#DSP } & \rotatebox{\titletilt}{ Support } \\\midrule
and16 & BW    & 1.6  & 16 & -- & \checkmark (4.8) & 2.6 & 16 & -- & -- & -- &--   & 36 & 10 & -- &--   & $\times$ \\
and32 & BW    & 1.9  & 32 & -- &                  & 2.6 & 32 & -- & -- & -- &--   & 35 & 17 & -- &--   & $\times$ \\
or16  & BW    & 1.8  & 16 & -- & \checkmark (7.1) & 2.6 & 16 & -- & -- & -- &--   & 34 & 10 & -- &--   & $\times$ \\
or32  & BW    & 1.7  & 32 & -- &                  & 2.6 & 32 & -- & -- & -- &--   & 31 & 17 & -- &--   & $\times$ \\
xor16 & BW    & 1.9  & 16 & -- & \checkmark (4.4) & 2.7 & 16 & -- & -- & -- &--   & 32 & 10 & -- &--   & $\times$ \\
xor32 & BW    & 2.2  & 32 & -- &                  & 2.5 & 32 & -- & -- & -- &--   & 33 & 17 & -- &--   & $\times$ \\
not16 & BW    & 1.8  & 16 & -- & \checkmark (3.6) & 2.6 & -- & -- & -- & 16 &--   & 32 & 10 & -- &--   & $\times$ \\
not32 & BW    & 1.7  & 32 & -- &                  & 2.6 & -- & -- & -- & 32 &--   & 33 & 17 & -- &--   & $\times$ \\
mux16 & BW    & 8.5  & 16 & -- & \checkmark (6.3) & 2.6 & 16 & -- & -- & -- &--   & 31 & 16 & -- & -- & $\times$ \\
mux32 & BW    & 12.7 & 32 & -- &                  & 2.8 & 32 & -- & -- & -- &--   & 22 & 32 & -- & -- & $\times$ \\
add16 & Carry & 2.3  & 16 & 2 & \checkmark (4.8) & 2.8 & 16 & -- & 4 & -- &  -- & 34 & 16 & 2 & -- & $\times$ \\
add32 & Carry & 2.3  & 32 & 4 &                  & 2.7 & 32 & -- & 8 & -- &  -- & 34 & 32 & 4 & -- & $\times$ \\
sub16 & Carry & 1.9  & 16 & 2 & \checkmark (3.9) & 2.7 & 16 & -- & 4 & -- &  -- & 34 & 16 & 2 & -- & $\times$ \\
sub32 & Carry & 1.9  & 32 & 4 &                  & 2.7 & 32 & -- & 8 & -- &  -- & 34 & 32 & 4 & -- & $\times$ \\
mul16 & --   & 2.4  & -- & -- & \checkmark (2.4) & 3.4 & -- & -- & -- & -- & 1 & 32 & -- & -- & 1  & $\times$ \\
mul32 & --    & --   & -- & -- &                  & 3.6 & -- & -- & -- & -- & 3  & 22 & -- & -- & 3  & $\times$ \\
eq32  & Comp  & 3.3  & 32 & 4 &                   & 2.6 & 24 & 1 & -- & -- &  -- & 36 & 11 & 2 &--    & $\times$ \\
neq32 & Comp  & 3.5  & 32 & 4 &                   & 2.7 & 24 & 1 & -- & -- &  -- & 33 & 11 & 2 &--    & $\times$ \\
ugt32 & Comp  & 3.4  & 32 & 4 &                   & 2.7 & 22 & -- & 3 & -- &  -- & 32 & 16 & 2 &--    & $\times$ \\
ult32 & Comp  & 3.7  & 32 & 4 &                   & 2.7 & 22 & -- & 3 & -- &  -- & 34 & 16 & 2 &--   & $\times$ \\
uge32 & Comp  & 2.8  & 32 & 4 &                   & 2.8 & 22 & -- & 3 & -- &  -- & 33 & 16 & 2 &--   & $\times$ \\
ule32 & Comp  & 2.5  & 32 & 4 &                   & 2.8 & 22 & -- & 3 & -- &  -- & 32 & 16 & 2 &--  & $\times$}
\newcommand{\LatticeTable}{\rotatebox{\titletilt}{ Instr } & \rotatebox{\titletilt}{ Template } & \rotatebox{\titletilt}{ Runtime (s)} & \rotatebox{\titletilt}{ \#LUTs } & \rotatebox{\titletilt}{ \#CCU2Cs }  & \rotatebox{\titletilt}{ Runtime (s)} & \rotatebox{\titletilt}{ \#LUTs } & \rotatebox{\titletilt}{ \#MUXs } & \rotatebox{\titletilt}{ \#CCU2Cs } & \rotatebox{\titletilt}{ Runtime (s) } & \rotatebox{\titletilt}{ \#LUTs } & \rotatebox{\titletilt}{ \#CCU2Cs }  & \rotatebox{\titletilt}{ Support }  \\\midrule
and32 & BW   & 2.0 & 32 & --    & 1.5 & 32 & -- & -- &1.1 & 32 & -- & $\times$ \\
or32 & BW    & 2.2 & 32 & --    & 1.5 & 32 & -- & -- &0.6 & 32 & -- & $\times$ \\
xor32 & BW   & 1.8 & 32 & --    & 1.5 & 32 & -- & -- & 0.7 & 32 & -- & $\times$ \\
add32 & Carry& 3.5 & 32 & 16 & 1.5 & -- & -- & 16 & 0.7 & 1  & 16 & $\times$ \\
sub32 & Carry& 3.3 & 32 & 16 & 1.5 & -- & -- & 16 & 1.2 & -- & 17 & $\times$ \\
eq32 & Comp  & 3.9 & 64 & 16   & 1.6 & 25 & 2 & -- & 0.6 & -- & 9  & $\times$ \\
ugt32 & Comp & 9.0 & 64 & 16  & 1.5 & 1 & -- & 16 &  0.6 & 1 & 17  & $\times$ \\
ult32 & Comp & 9.1 & 64 & 16 & 1.7 & 45 & 19 & 16 &  0.7 & 1 & 17  & $\times$}
\newcommand{\SofaTable}{Instr & Template & Runtime in sec  & \#LUTs & Support & Support \\
\midrule
and8 & BW     & 1.8 & 8    & $\times$ & $\times$\\
and16 & BW    & 2.2 & 16   & $\times$ & $\times$\\
and32 & BW    & 2.8 & 32   & $\times$ & $\times$\\
or8 & BW      & 2.3 & 8    & $\times$ & $\times$\\
or16 & BW     & 2.4 & 16   & $\times$ & $\times$\\
or32 & BW     & 3.1 & 32   & $\times$ & $\times$\\
xor8 & BW     & 2.6 & 8    & $\times$ & $\times$\\
xor16 & BW    & 2.4 & 16   & $\times$ & $\times$\\
xor32 & BW    & 2.4 & 32   & $\times$ & $\times$\\
not8 & BW     & 1.9 & 8    & $\times$ & $\times$\\
not16 & BW    & 2.6 & 16   & $\times$ & $\times$\\
not32 & BW    & 2.7 & 32   & $\times$ & $\times$\\
mux8 & BW     & 2.7 & 8    & $\times$ & $\times$\\
mux16 & BW    & 2.6 & 16   & $\times$ & $\times$\\
mux32 & BW    & 2.9 & 32   & $\times$ & $\times$\\
add8 & Carry  & 2.6 & 24   & $\times$ & $\times$\\
add16 & Carry & 2.9 & 48   & $\times$ & $\times$\\
add32 & Carry & 4.7 & 96   & $\times$ & $\times$\\
sub8 & Carry  & 2.6 & 24   & $\times$ & $\times$\\
sub16 & Carry & 3.2 & 48   & $\times$ & $\times$\\
sub32 & Carry & 3.7 & 96   & $\times$ & $\times$\\
mul8 & Mult   & 6.2 & 228  & $\times$ & $\times$\\
mul16 & --   & -- & --    & $\times$ & $\times$\\
mul32 & --    & -- & --    & $\times$ & $\times$\\
eq8 & Comp    & 4.6 & 32   & $\times$ & $\times$\\
eq16 & Comp   & 5.6 & 64   & $\times$ & $\times$\\
eq32 & Comp   & 6.1 & 128  & $\times$ & $\times$\\
neq8 & Comp   & 3.3 & 32   & $\times$ & $\times$\\
neq16 & Comp  & 5.3 & 64   & $\times$ & $\times$\\
neq32 & Comp  & 7.0 & 128  & $\times$ & $\times$\\
ugt8 & Comp   & 6.3 & 32   & $\times$ & $\times$\\
ugt16 & Comp  & 6.1 & 64   & $\times$ & $\times$\\
ugt32 & Comp  & 10.0 & 128 & $\times$ & $\times$\\
ult8 & Comp   & 6.1 & 32   & $\times$ & $\times$\\
ult16 & Comp  & 6.2 & 64   & $\times$ & $\times$\\
ult32 & Comp  & 10.3 & 128 & $\times$ & $\times$\\
uge8 & Comp   & 4.3 & 32   & $\times$ & $\times$\\
uge16 & Comp  & 6.7 & 64   & $\times$ & $\times$\\
uge32 & Comp  & 6.5 & 128  & $\times$ & $\times$\\
ule8 & Comp   & 3.8 & 32   & $\times$ & $\times$\\
ule16 & Comp  & 5.5 & 64   & $\times$ & $\times$\\
ule32 & Comp  & 5.7 & 128  & $\times$ & $\times$   }
\newcommand{\SofaTable}{\rotatebox{\titletilt}{ Instr } & \rotatebox{\titletilt}{ Template } & \rotatebox{\titletilt}{ Runtime (s)} & \rotatebox{\titletilt}{ \#LUTs } & \rotatebox{\titletilt}{ Support } & \rotatebox{\titletilt}{ Support } \\\midrule
and32 & BW    & 2.8 & 32   & $\times$ & $\times$\\
or32 & BW     & 3.1 & 32   & $\times$ & $\times$\\
xor32 & BW    & 2.4 & 32   & $\times$ & $\times$\\
add32 & Carry & 4.7 & 96   & $\times$ & $\times$\\
sub32 & Carry & 3.7 & 96   & $\times$ & $\times$\\
eq32 & Comp   & 6.1 & 128  & $\times$ & $\times$\\
ugt32 & Comp  & 10.0 & 128 & $\times$ & $\times$\\
ult32 & Comp  & 10.3 & 128 & $\times$ & $\times$}
\newcommand{\FinalTable}{\rotatebox{\titletilt}{ Instr } & \rotatebox{\titletilt}{ Template } & \rotatebox{\titletilt}{ Runtime (s)} & \rotatebox{\titletilt}{ \#LUTs } & \rotatebox{\titletilt}{ \#CARRY8s } & \rotatebox{\titletilt}{ Runtime (s)} & \rotatebox{\titletilt}{ \#LUTs } & \rotatebox{\titletilt}{ \#MUXs } & \rotatebox{\titletilt}{ \#CARRY4s } & \rotatebox{\titletilt}{ \#INVs } & \rotatebox{\titletilt}{ Runtime (s)} & \rotatebox{\titletilt}{ \#LUTs } & \rotatebox{\titletilt}{ \#CARRY8s } & \rotatebox{\titletilt}{ Support } & \rotatebox{\titletilt}{ Runtime (s)} & \rotatebox{\titletilt}{ \#LUTs } & \rotatebox{\titletilt}{ \#CCU2Cs }  & \rotatebox{\titletilt}{ Runtime (s)} & \rotatebox{\titletilt}{ \#LUTs } & \rotatebox{\titletilt}{ \#MUXs } & \rotatebox{\titletilt}{ \#CCU2Cs } & \rotatebox{\titletilt}{ Runtime (s) } & \rotatebox{\titletilt}{ \#LUTs } & \rotatebox{\titletilt}{ \#CCU2Cs }  & \rotatebox{\titletilt}{ Support }  & \rotatebox{\titletilt}{ Runtime (s)} & \rotatebox{\titletilt}{ \#LUTs } & \rotatebox{\titletilt}{ Support } & \rotatebox{\titletilt}{ Support } \\\midrule
and32 & BW    & 1.9  & 32 & -- & 2.6 & 32 & -- & -- & -- & 35 & 17 & -- & $\times$ & 2.0 & 32 & --    & 1.5 & 32 & -- & -- &1.1 & 32 & -- & $\times$ & 2.8 & 32   & $\times$ & $\times$\\
or32  & BW    & 1.7  & 32 & -- & 2.6 & 32 & -- & -- & -- & 31 & 17 & -- & $\times$ & 2.2 & 32 & --    & 1.5 & 32 & -- & -- &0.6 & 32 & -- & $\times$ & 3.1 & 32   & $\times$ & $\times$\\
xor32 & BW    & 2.2  & 32 & -- & 2.5 & 32 & -- & -- & -- & 33 & 17 & -- & $\times$ & 1.8 & 32 & --    & 1.5 & 32 & -- & -- & 0.7 & 32 & -- & $\times$ & 2.4 & 32   & $\times$ & $\times$\\
add32 & Carry & 2.3  & 32 & 4  & 2.7 & 32 & -- & 8 & --  & 34 & 32 & 4 & $\times$  & 3.5 & 32 & 16 & 1.5 & -- & -- & 16 & 0.7 & 1  & 16 & $\times$ & 4.7 & 96   & $\times$ & $\times$\\
sub32 & Carry & 1.9  & 32 & 4 & 2.7 & 32 & -- & 8 & -- & 34 & 32 & 4 & $\times$ & 3.3 & 32 & 16 & 1.5 & -- & -- & 16 & 1.2 & -- & 17 & $\times$ & 3.7 & 96   & $\times$ & $\times$\\
eq32  & Comp  & 3.3  & 32 & 4 & 2.6 & 24 & 1 & -- & -- & 36 & 11 & 2 & $\times$ & 3.9 & 64 & 16   & 1.6 & 25 & 2 & -- & 0.6 & -- & 9  & $\times$  & 6.1 & 128  & $\times$ & $\times$\\
ugt32 & Comp  & 3.4  & 32 & 4 & 2.7 & 22 & -- & 3 & -- & 32 & 16 & 2 &$\times$ & 3.9 & 64 & 16   & 1.6 & 25 & 2 & -- & 0.6 & -- & 9  & $\times$ & 10.0 & 128 & $\times$ & $\times$\\
ult32 & Comp  & 3.7  & 32 & 4 & 2.7 & 22 & -- & 3 & -- & 34 & 16 & 2 &$\times$ & 9.1 & 64 & 16 & 1.7 & 45 & 19 & 16 &  0.7 & 1 & 17  & $\times$ & 10.3 & 128 & $\times$ & $\times$}
\newcommand{\titletilt}{90}

 


\section{\lr Extensibility and Expressiveness}

In addition to being
  correct by construction (\cref{sec:formalization}) and
  more complete 
  than existing FPGA technology mappers (\cref{sec:completeness}),
  \lr can also easily extend to new FPGA architectures.
Furthermore, automatic primitive semantics extraction 
  from vendor-provided HDL simulation models
  enables \lr to support diverse, highly configurable
  FPGA primitives.
In the context of this dissertation,
  this corresponds to providing evidence
  for our \cref{thesis:devtime} claim:
  namely,
  through the use of
   more adaptable \cref{thesis:algorithms}
  (program synthesis)
  and more explicit \cref{thesis:models}
  (vendor-supplied simulation models),
  supporting new FPGA architectures in
  \lr
  requires less \cref{thesis:devtime}
  compared to other tools.

The architecture descriptions
    vary in length from 20 to 240
    source lines of code (SLoC).
%
SOFA (20 SLoC) is the simplest, shown in full
  in \cref{fig:sofa-architecture-description}.
The descriptions for Xilinx UltraScale+ (185 SLoC), Xilinx 7-series (174),
  Lattice (240 SLoC), and Intel (178 SLoC)
  are longer since those
  FPGA architectures provide a
  wider range of configurable primitives.

As a point of comparison,
  the open-source Yosys toolchain,
  which has roughly 200 contributors on GitHub,
  provides technology mapping
  for Xilinx UltraScale+
  across over a dozen complex
  Verilog, C++, and Python files (about 1300
  lines of code).
We cannot provide similar numbers
  for state-of-the-art proprietary tools,
  but a developer
  of one such technology mapper
  shared that extending their tool to
  support new FPGA architectures
  was extremely difficult since it 
  ``spans millions of lines of low-level C.''
This is not surprising; Yosys aims to
  target a variety of vendor architectures, 
  while proprietary tools have teams of
  engineers to extract better mapping 
  (evident by Yosys' limitations
  in \cref{sec:completeness}).
By contrast,
  \lr supports
  diverse architectures and
  is easy to extend.
Even if a user
  wants to target a completely
  new architecture that
  \lr does not support,
  architecture-independent
  sketch templates allow reuse
  of previously implemented mapping
  strategies, and the user is
  only required to provide
  a few lines of
  high-level configuration
  for each primitive in 
  the architecture description.
  
\Cref{table:imported-primitves} further
  highlights \lr's expressiveness,
  i.e., its ability to support a diverse
  range of configurable primitives
  by automatically extracting semantics from
  vendor-provided HDL simulation models.
\lr can import the semantics
  of large configurable primitives, 
  such as the UltraScale+ DSP (896 lines of Verilog)
  or Lattice ECP5's ALU and multiplier units (1642 and
  795 lines of Verilog, respectively).
It is difficult and error-prone
  to manually formalize the full semantics for these primitives;
  partial support by ad hoc search procedures
  that rely on syntactic pattern matching
  leads to missing many mapping opportunities,
  as shown in \cref{sec:completeness}.




\begin{table}
\caption{
FPGA primitives imported 
  automatically by \lr from vendor-provided 
  Verilog models, with number of source lines of code 
  (excluding comments and empty lines) of 
  the original Verilog models.} 
\centering
\footnotesize
\label{table:imported-primitves}
\begin{tabular}{lrr}
 {\bf FPGA}   & \textbf{Primitive} & {\bf Verilog SLoC} \\\hline
 Xilinx Ultrascale+  & LUT6 & 88       \\
                     & CARRY8 & 23       \\
                     & DSP48E2 & 896       \\
                     \hline
 Xilinx 7-series
                     
                     & DSP48E1 & 1129       \\
                     \hline
 Lattice ECP5 & LUT2 & 5       \\
              & LUT4 & 7       \\
              & CCU2C & 60       \\
              & ALU54A & 1642 \\ 
              & MULT18X18C & 795 \\
              \hline
 Intel Cyclone 10 LP  & cyclone10lp\_mac\_mult   &  319       \\ 
              \hline
 SOFA      & frac\_lut4   &  69       \\ 
\end{tabular}
\end{table}

\chapter{Future Work: Churchroad}
\label{chapter:churchroad}

I wanted to briefly mention
  an ongoing extension to \lr.
A fundamental limitation
  of solver-aided approaches
  like \lr
  is problem size:
  the larger a constraint problem,
  the more likely solvers are to time out
  while trying to find a solution.
In \lr, this means that
  compiling larger hardware designs
  will present issues;
  we already saw in \cref{sec:evaluation}
  that \lr's underlying solvers
  time out
  on some benchmarks.
Furthermore, even some small tasks
  are 
  notably difficult for solvers---%
  namely, reasoning about multiplication~\cite{rath2024polysat,brain2021further}.
Scaling to large designs
  and compiling multipliers
  are core requirements
  for a hardware compiler,
  however.
So how do we ensure that
  we can use \lr
  on large designs?

In response,
  we have been developing Churchroad~\cite[section 4]{smith2024there},
  which combines many of the techniques from this dissertation.
Namely, we employ
  \gls{equality-saturation}---%
  our algorithm of choice in \cref{part:glenside-and-3la}.
Using equality saturation,
  we can capture an entire design
  in an \egr.
Once we have captured a full design, we can
  use equational reasoning via rewrites
  to achieve tasks that
  would otherwise be challenging
  for SMT solvers---%
  for example, splitting a wide
  multiplier
  into multiple multipliers and adders.
Similarly, we can use rewrites
  to discover places in the design
  where we might want to invoke \lr.
Consider the benchmarks
  we generated to evaluate DSP mapping in
  in \cref{sec:evaluation}---%
  we could similarly generate patterns
  to identify potential places where DSPs could be used
  in larger designs.
Once these locations are identified within the design,
  we can call \lr as a subroutine,
  and if \lr finds a mapping,
  we can then import it back into the \egr.

Churchroad has already shown promise as a method
  of scaling solver-aided hardware compilation.
Beyond that, however,
  Churchroad also presents new strategies
  for leveraging the strengths of
  of SMT solvers and equality saturation.

\chapter{Background and Related Work}
\label{sec:background-and-related-work}

To the best of our knowledge,
  \lr is the first work
  to apply the technique of program synthesis
  to FPGA technology mapping.
Indeed, as noted by Sisco \textit{et al.}~\cite{sisco2022synthesis},
  program synthesis has seldom been applied
  in the domain of hardware design although
  its underlying formal methods techniques
  are frequently used for
  the \textit{formal verification}
  of hardware designs rather than compilation,
  as in Bluespec SystemVerilog~\cite{nikhil2004bluespec},
  \koika~\cite{bourgeat2020essence},
  and Kami~\cite{choi2017kami}.
Sisco \textit{et al.} cite two examples
  of works that use program synthesis for hardware design,
  Verisketch~\cite{ardeshiricham19verisketch} and Sketchilog~\cite{becker14sketchilog},
  both of which apply program synthesis to produce HDL implementations from high-level designs.
Other works use program synthesis
  to generate \textit{software}
  that runs on low-powered hardware,
  like Chlorophyll~\cite{phothilimthana2014chlorophyll},
  which targets extremely memory-constrained
  power-efficient processors,
  Chipmunk~\cite{gao2019chipmunk},
  which targets programmable network switches,
  and Diospyros~\cite{vanhattum2021vectorization},%
  \footnote{Diospyros uses symbolic evaluation, which is related to program synthesis, to lift imperative programs for digital signal processors into a high-level mathematical representation that can then be used with the technique of equality saturation~\cite{tate2011equality} to generate optimized code for the target devices. This is also distinct from the program synthesis techniques referenced elsewhere in this dissertation.}
  which generates vectorized programs for standalone digital signal processors (more powerful and general-purpose devices than the DSP units in FPGAs).
These works demonstrate the utility of program synthesis 
  for generating code that handles
  specific wrinkles in hardware designs,
  as does the use of program synthesis in \lr
  to harness the programmability of FPGA DSPs.
Note that other types of solvers
  (beyond the SMT solvers used in Lakeroad)
  can be used within compilers,
  e.g.~using partitioned boolean quadratic problem
  (PBQP) solvers for instruction selection
  in LLVM~\cite{ebner2008generalized}

\lr is also related to past work in FPGA compilation and techmapping,
  much of which does not 
  entreaty to support
  programmable DSPs with as much generality.
ODIN~\cite{jamieson2005verilog} and ODIN-II~\cite{jamieson2010odin}
  are used in \textit{hard-block synthesis}
  for FPGAs,
  which is the task of mapping portions
  of hardware designs to specialized units (\textit{hard blocks}) like multipliers.
They operate purely over syntax (e.g., mapping \texttt{*} to a multiplier)
  and so are greatly limited in their ability
  to handle programmable DSPs.
The ABC~\cite{brayton2010abc} logic synthesis tool
  is used to lower hardware designs 
  into LUT and carry-chain configurations; it 
  is related to \lr in that it also uses constraint solvers
  to find configurations,
  though it is not general enough to handle
  a wide variety of programmable DSPs,
  unlike the program synthesis techniques used in \lr.
Note also that the use of
  configuration files in \lr to
  abstract away details of the FPGA architecture
  was inspired by past work in FPGA compilation,
  including OpenFPGA~\cite{tang2019openfpga}
  and the Verilog-to-Routing
  project (VTR)~\cite{rose2012vtr},
  both of which use abstract architecture descriptions
  to facilitate portability across designs,
  though these projects are limited in their support for DSPs.
Library-Parameterized Models~\cite{1993lpm, lpmaltera}
  define generic interfaces for common primitives and are also similar to \lr's primitive interfaces,
  though they are limited in their ability to represent configurable units like DSPs.\tighten

Virtual FPGA overlays~\cite{lysecky2005firm,brant2012zuma, landgraf2021compiler}
  are another approach to
  improving the mapping of hardware designs
  to hardware.
Overlays present a ``virtual''
  FPGA architecture;
  each actual architecture
  must then define a mapping
  from virtual to actual primitives.
This required translation is similar to
  \lr's requirement on users
  to implement primitive interfaces
  in an architecture description,
  though it requires more user effort.
The translation from virtual to actual architecture
  often comes with
  a steep resource
  and performance overhead.

\chapter*{\Cref{part:lakeroad} Conclusion}
\label{sec:lakeroad-conclusion}

\Cref{part:lakeroad} presented \lr,
  a novel approach to \gls{fpga} \gls{technology-mapping}.
\lr utilizes both more adaptable \cref{thesis:algorithms}---%
  sketch-guided \gls{program-synthesis}---%
  and more explicit \cref{thesis:models}---%
  vendor-supplied simulation models---%
  to provide greater \cref{thesis:correctness},
  completeness (i.e.~\cref{thesis:optimizations}),
  and extensibility (i.e.~reduced \cref{thesis:devtime})
  over state-of-the-art tools.
Because program synthesis tools
  can efficiently explore large search spaces, 
  \lr 
  can find mappings
  of hardware designs
  to FPGA DSPs
  in more cases
  than state-of-the-art tools,
  often finding more efficient implementations
  in the process.
With our techniques
  of semantics extraction
  from HDL
  and architecture-independent sketch templates,
  users must expend little manual effort 
  to apply \lr to
  a given FPGA architecture
  and extend it to handle further primitives.
Moreover, our formalization of \lr
  fosters greater confidence
  in its correctness.
\lr hence enables the 
  extensible, efficient, and correct 
  lowering of hardware designs to FPGAs,
  highlighting the effectiveness
  of program synthesis
  for FPGA technology mapping.

\lr cleanly and completely realizes
  my thesis set out at the start of this dissertation:
  utilize explicit, vendor-supplied \cref{thesis:models},
  apply state-of-the-art automated reasoning 
  \cref{thesis:algorithms},
  and you will produce a 
  powerful compiler backend when measured along the axes of
  \cref{thesis:optimizations},
  \cref{thesis:correctness}, and
  \cref{thesis:devtime}.
  

\addcontentsline{toc}{part}{Wrapping Up}
\chapter{Conclusion and Broader Thoughts}

In this dissertation,
  I have presented an argument
  for a certain method of building compiler backends:
  namely,
  automatically generating them
  from explicit \cref{thesis:models}
  of hardware
  using \gls{automated-reasoning} \cref{thesis:algorithms}.
I demonstrated how this method of backend generation
  leads to improved
  \cref{thesis:optimizations},
  greater 
  \cref{thesis:correctness}, and
  reduced
  \cref{thesis:devtime}
  in two case studies.
In \cref{part:glenside-and-3la},
  I introduced
  \g, a language and tool
  enabling the use of equality saturation
  on machine learning workloads.
When incorporated into the 
  \TLA methodology, \g
  enabled more flexible mapping
  of machine learning workloads
  to accelerators,
  ultimately enabling
  easier developer testing
  of hardware designs.
In \cref{part:lakeroad},
  I introduced \lr, which
  utilizes sketch-guided program synthesis
  and semantics extracted from vendor-supplied simulation models
  to generate more correct,
  more complete, and more extensible
  FPGA technology mappers.
While the examples presented in this dissertation
  are specific---%
  applying equality saturation to accelerator mapping,
  applying sketch-guided synthesis to technology mapping---%
  the underlying recipe is portable.
In the end, I hope it is this recipe you take away:
  to automatically generate better compiler backends,
  apply automated reasoning \cref{thesis:algorithms}
  to explicit, formal \cref{thesis:models} of the target hardware.

Though this dissertation
  may come to represent
  a substantial amount of my life's research output,
  I also consider it just a piece
  of a larger vision.
To conclude this dissertation,
  I will attempt to capture the
  the higher level thoughts,
  ideas, and inspirations
  underlying the projects in my dissertation.

This dissertation is simply
  two useful, testable
  implementations (\g and \lr)
  of a larger idea
  that's been bugging me
  since I was a Master's student
  at Penn State.
At Penn State,
  I worked on the problem of
  computing with emerging devices---%
  in our case, transistor which
  provided computing primitives other than just
  Boolean logic~\cite{raychowdhury2018computing,yoon2018fefet,yoon2019ferrofet}.
This is where the nagging feeling started:
  namely, the feeling that
  \textit{the hardware tells us what it does.}
Let me  explain what I mean.

Models of hardware are ubiquitous.
Any representation of hardware
  that isn't the literal silicon itself---%
  from a low-level GDSII file,
  to a mid-level \gls{rtl} representation,
  to a high-level \gls{hls} implementation or Python simulator---%
  is a \textit{model} of hardware.
All hardware in use
  likely has at least one model behind it,
  if not more.
Furthermore,
  models come in all shapes and sizes,
  and capture far more than just
  computational functionality,
  but also things like
  timing,
  area,
  and power.
The SkyWater PDK (recently open sourced
  by Google)
  or the ASAP7 PDK from Arizona State University
  are great examples
  of open-source process development kits,
  rich with models (simulation, timing, power, and area)
  of real, fabricatable
  hardware platforms.

There is currently a directionality
  associated with hardware models.
For example, the simulation
  models
  of an FPGA's primitives
  packaged inside \gls{hardwaresynthesis}
  tools
  are meant to be used
  to simulate a design
  after it has been compiled
  from its higher-level specification.
Alternatively,
  hardware designs intended for synthesis
  of an FPGA
  or ASIC
  are intended to be run through a synthesis tool
  and lowered to a netlist
  (to eventually become an ASIC
    or FPGA bitstream).
In both cases,
  the hardware model is only 
  \textit{lowered,} 
  such as when a synthesis-ready design
  is compiled,
  or \textit{lowered to,}
  such as when a design is compiled
  to simulation models.
It is not often the case that
  these models are used in the opposite direction---%
  i.e.~\textit{lifting} models
  to higher levels of abstraction,
  to generate higher-level outputs
  (like compiler backends).

It is surprising
  that models are only used in the lowering direction,
  as many of them are
  prime for lifting.
Models written in Verilog,
  for example,
  are readily usable
  by \gls{automated-reasoning} tools
  like \gls{smt} solvers.
For the most part,
  automated reasoning tools
  are only used to do post-lowering
  \gls{verification},
  but there is no reason that they
  can't be run ``in the other direction''---%
  in fact, this is exactly what \lr does.

With all of that in mind,
  let's return to the idea that
  \textit{the hardware tells us what it does.}
Rephrasing this idea
  in the terms presented in the paragraphs above,
  we are overflowing with rich,
  varied hardware models
  that are prime for use
  with automated reasoning tools,
  ready to have interesting,
  higher-level semantics
  lifted from them.
It's for these reasons
  that,
  for years,
  it's felt like hardware
  has been practically screaming at us,
  \textit{directly} providing all the information we need
  to understand how to compile to it optimally.
Yet we ignore it, choosing instead to use
  the models \textit{indirectly.}

When I say that tools use models ``indirectly'', 
  I'm referring to the current optimization loop
  standard in many optimization tasks.
Consider \gls{hardwaresynthesis} tools.
During synthesis, the tool will
  make some guesses during compilation
  about what optimizations will generate optimal output
  (often informed, indirectly, by the underlying hardware models).
The designer then
  simulates the results using the hardware models,
  to check whether the tool's guesses were correct.
The odd part of this loop, to me,
  is the \textit{indirect} use of the models
  to inform compiler construction,
  and the \textit{direct} use of the models
  only after the fact, to check compilation results.
Why not shorten this loop, and use the models \textit{directly}
   during compiler construction?
Hardware tells us what it does---%
  so why not listen?

\lr was the most concrete
  instantiation of this idea
  that I was able to achieve
  in the span of my PhD.
Rather than encoding DSP mapping rules
  as patterns (informed, indirectly, from models
    and DSP documentation),
  \lr determines how to map to primitives
  by ingesting models directly,
  utilizing \gls{program-synthesis}
  to map to primitives based on the ingested models.
However, \lr only scratches the surface
  of the broader idea 
  I'm discussing here.
\lr only maps based on functional behavior alone;
  it does not take advantage of, for example,
  the timing information
  also available in the models.

Furthermore,
  these ideas go far beyond digital hardware.
I've been lucky enough to apply these ideas
  in non-boolean, analog computing~\cite{yoon2018fefet,yoon2019ferrofet,raychowdhury2018computing}
  as well as DNA circuit design~\cite{wathieu2023fridge}.
It isn't just digital hardware
  that is telling us what it does.
Any computing substrate has \glspl{primitive};
  most primitives have models.
These models can be similarly used to generate compilers.

\vspace{10mm}

\noindent
With that, I will conclude this dissertation.
If nothing else, I hope you take with you
  the idea that
  \textit{hardware is telling us what it does}---%
  it's on us to listen!

\clearpage
\addcontentsline{toc}{chapter}{Bibliography}
\singlespacing
\bibliographystyle{plain}
\bibliography{references}

\begin{thebibliography}{100}
\bibitem{1993lpm}
Eia-is-103 : Library of parameterized modules (lpm).
\end{thebibliography}

\end{document}